\documentclass[a4paper,11pt]{article}
\pdfoutput=1 

\usepackage{jheppub} 
                     
\usepackage{amsmath}
\usepackage{graphicx}
\usepackage{amssymb}
\usepackage{subfigure}
\usepackage{cancel}
\usepackage{relsize}
\usepackage{bbold}
\usepackage{lineno}
\usepackage{braket}
\usepackage{slashed}
\usepackage{multirow}
\usepackage{placeins}

\newcommand{\av}[1]{\left\langle #1\right\rangle}

 \newcommand{\HEP}{\texttt{HEPfit}}

\usepackage[T1]{fontenc} 

\title{$B\to K^* \ell^+ \ell^-$ decays at large recoil in the Standard Model: a
  theoretical reappraisal}

\author{Marco Ciuchini$^a$,} 
\author{Marco Fedele$^{b,c}$,} 
\author{Enrico Franco$^c$,} 
\author{Satoshi Mishima$^d$,}
\author{Ayan Paul$^c$,} 
\author{Luca Silvestrini$^c$}
\author{and Mauro Valli$^{e,f}$}

\affiliation{$^a$INFN, Sezione di Roma Tre, Via della Vasca Navale 84,
  I-00146 Roma, Italy} \affiliation{$^b$Dipartimento di Fisica,
  Universit\`a di Roma ``La Sapienza'', P.le A. Moro 2, I-00185 Roma,
  Italy} \affiliation{$^c$INFN, Sezione di Roma, P.le A. Moro 2,
  I-00185 Roma, Italy} \affiliation{$^d$Theory Center, IPNS, KEK, Tsukuba 305-0801, Japan }
\affiliation{$^e$SISSA, via Bonomea 265, I-34136 Trieste, Italy}
\affiliation{$^f$INFN, Sezione di Trieste, via Valerio 2, I-34127
  Trieste, Italy}

\emailAdd{marco.ciuchini@roma3.infn.it}
\emailAdd{enrico.franco@roma1.infn.it}
\emailAdd{marco.fedele@uniroma1.it}
\emailAdd{satoshi.mishima@kek.jp}
\emailAdd{ayan.paul@roma1.infn.it}
\emailAdd{luca.silvestrini@roma1.infn.it}
\emailAdd{mauro.valli@sissa.it}

\abstract{We critically reassess the theoretical uncertainties in the
  Standard Model calculation of the $B \to K^* \ell^+ \ell^-$
  observables, focusing on the low $q^2$ region. We point out that
  even optimized observables are affected by sizable uncertainties,
  since hadronic contributions generated by current-current operators
  with charm are difficult to estimate, especially for
  $q^2 \sim 4 m_c^2\simeq 6.8$ GeV$^2$.  We perform a detailed
  numerical analysis and present both predictions and results from the
  fit obtained using most recent data. We find that non-factorizable
  power corrections of the expected order of magnitude are sufficient
  to give a good description of current experimental data within the
  Standard Model. We discuss in detail the $q^2$ dependence of the
  corrections and their possible interpretation as shifts of the
  Standard Model Wilson coefficients. }

\begin{document} 
\maketitle
\flushbottom
\allowdisplaybreaks

\section{Introduction}
\label{sec:intro}

Flavour-Changing Neutral Current (FCNC) processes are very sensitive
probes of New Physics (NP). Within the Standard Model (SM) they can
only arise at the loop level, and they are further suppressed by the
GIM cancellation mechanism, so that even very heavy new particles can
give rise to sizable contributions, especially if they carry new
sources of flavour violation.  In particular, the semileptonic decays
$B\to K^* \ell^+ \ell^-$ have been advocated to be among the cleanest
FCNC
processes~\cite{Eilam:1986at,Hou:1986ug,Godbole:1987tc,Hou:1987kf,
  Deshpande:1988mg,Grinstein:1988me,He:1988tf,Ciuchini:1989qw,Colangelo:1995jv,Altmannshofer:2008dz}.
Indeed, the dilepton invariant mass spectrum is accessible over the
full kinematic range allowing to cut the theoretically challenging
resonance-dominated regions. The description of the remaining part of
the spectrum is simplified using the heavy quark expansion at low
dilepton invariant mass $q^2$ \cite{Beneke:2000wa,Beneke:2001at},
while an Operator Product Expansion (OPE) can be used at large $q^2$
\cite{Grinstein:2004vb,Bobeth:2010wg,Beylich:2011aq,Bobeth:2011gi}.
In particular, heavy quark symmetry allows to reduce the number of
independent form
factors~\cite{Isgur:1990kf,Charles:1998dr,Grinstein:2002cz}, while
non-factorizable corrections are power suppressed.\footnote{This is
  not the case for charmonioum resonant contributions. They need to be
  controlled using experimental cuts
  \cite{Beneke:2009az,Lyon:2014hpa}.}  Experimentally, the full
angular analysis can be performed allowing for the extraction of
twelve angular coefficients (plus twelve more for the CP-conjugate
decay) in several $q^2$ bins.\footnote{The number of independent
  angular coefficients reduces to eight neglecting the lepton masses.}
From these coefficients, exploiting the symmetries of the infinite
mass limit, one can define ``optimized'' observables in which the soft
form factors cancel out, drastically reducing the theoretical
uncertainties~\cite{Kruger:2005ep,Egede:2008uy,
  Descotes-Genon:2013vna}. For these observables, very precise
predictions can be found in the
literature~\cite{Egede:2010zc,Becirevic:2011bp,Matias:2012xw,
  Das:2012kz,DescotesGenon:2012zf,Descotes-Genon:2013wba,Altmannshofer:2013foa,Beaujean:2013soa,Hurth:2013ssa,Matias:2014jua,Descotes-Genon:2014uoa,Altmannshofer:2014rta,Mandal:2014kma,Altmannshofer:2015sma,Mandal:2015bsa}.
Some deviation from these predictions has been observed in recent LHCb
data~\cite{Aaij:2013iag,Aaij:2013qta,LHCb:2015dla,Aaij:2015oid}.  In
this article we argue that no deviation is present once all the
theoretical uncertainties are taken into account. Among those, the
most important is a conservative evaluation of the deviation from the
infinite mass limit. This kind of corrections is known to be important
in other $b\to s$
decays~\cite{Ciuchini:2001gv,Ciuchini:2008eh,Duraisamy:2009kb}. In
fact, nonperturbative contributions from non-leptonic operators with
charm, although power suppressed, can compete with the contribution of
semileptonic and radiative operators even below the $c\bar c$
threshold.  A first estimate at small $q^2$ of this effect has been
provided by ref.~\cite{Khodjamirian:2010vf}, showing indeed that these
contributions are non-negligible. Furthermore, $c\bar{c}$ resonances
at threshold give a contribution to the rate that is two orders of
magnitude larger than the short-distance one
\cite{Beneke:2009az,Lyon:2014hpa}; indeed, no OPE can be performed in
this kinematical region, and quark-hadron duality is expected to hold
only for $q^2 \ll 4 m_c^2$.  At present, the effect of power
corrections and nonperturbative contributions cannot be fully computed
from first principles. Unfortunately this is the main limiting factor
in searching for NP in those amplitudes where these contributions are
present. Indeed, underestimating them might lead to too early claims
of NP.  First steps towards a careful assessment of the hadronic
uncertainties have been taken in
refs.~\cite{Jager:2012uw,Jager:2014rwa}.

In this work we show that, given the above arguments, present data do
not unambiguously point to the presence of NP in
$B\to K^* \ell^+ \ell^-$. We will discuss below what kind of NP
contributions can be disentangled from hadronic contributions; those
which cannot be disentangled are hindered by the hadronic
uncertainties.

This paper is organized as follows. In section~\ref{sec:pow} we discuss
power corrections to factorized formul\ae\ at low $q^2$.
In section~\ref{sec:fit}
we present results and predictions, discussing the size and role
of nonfactorizable terms. 
Our conclusions are drawn in section~\ref{sec:conclusions}.
Appendices \ref{sec:FF}--\ref{sec:other} contain some technical details.

\section{Power corrections to factorization at low \boldmath$q^2$}
\label{sec:pow}
Both $\bar{B} \rightarrow {\bar K}^*\ell^+\ell^-$ and
$\bar{B} \rightarrow {\bar K}^* \gamma$ can be described
by means of the $\Delta B = 1$ weak effective Hamiltonian
\begin{equation}
\label{eq:Heff}
\mathcal{H}_\mathrm{eff}^{\Delta B = 1} =
\mathcal{H}_\mathrm{eff}^\mathrm{had} +
\mathcal{H}_\mathrm{eff}^\mathrm{sl+\gamma}, 
\end{equation}
where the first term is the hadronic contribution 
\begin{equation}
\label{eq:H_had}
\mathcal{H}_\mathrm{eff}^\mathrm{had} =
\frac{4G_F}{\sqrt{2}}\Bigg[\sum_{p=u,c}\lambda_p\bigg(C_1 Q^{p}_1 +
C_2 Q^{p}_2\bigg) -\lambda_t \bigg(\sum_{i=3}^{6} C_i P_i +
C_{8}Q_{8g} \bigg)\Bigg] \,,  
\end{equation}
involving current-current, QCD penguin and chromomagnetic dipole
operators \cite{Chetyrkin:1997gb}
\begin{eqnarray}
  Q^p_1 &\:\:=\:\:& (\bar{s}_L\gamma_{\mu}T^a p_L)(\bar{p}_L\gamma^{\mu}T^ab_L)\,,\nonumber \\
  Q^p_2 &=& (\bar{s}_L\gamma_{\mu} p_L)(\bar{p}_L\gamma^{\mu}b_L)\,, \nonumber \\
  P_3 &=& (\bar{s}_L\gamma_{\mu}b_L)\sum\phantom{} _q(\bar{q}\gamma^{\mu}q)\,, \nonumber \\
  P_4 &=& (\bar{s}_L\gamma_{\mu}T^ab_L)\sum\phantom{}
          _q(\bar{q}\gamma^{\mu}T^aq) \,,\nonumber \\ 
  P_5 &=&
          (\bar{s}_L\gamma_{\mu1}\gamma_{\mu2}\gamma_{\mu3}b_L)\sum\phantom{}
          _q(\bar{q}\gamma^{\mu1}\gamma^{\mu2}\gamma^{\mu3}q)
          \,,\nonumber \\  
  P_6 &=&
          (\bar{s}_L\gamma_{\mu1}\gamma_{\mu2}\gamma_{\mu3}T^ab_L)\sum\phantom{}
          _q(\bar{q}\gamma^{\mu1}\gamma^{\mu2}\gamma^{\mu3}T^aq)
          \,,\nonumber \\  
  Q_{8g} &=& \frac{g_s}{16\pi^2}m_b\bar{s}_L\sigma_{\mu\nu}G^{\mu\nu}b_R \,,
\end{eqnarray}
while the second one, given by
\begin{equation}
\mathcal{H}_\mathrm{eff}^\mathrm{sl+\gamma} = -
\frac{4G_F}{\sqrt{2}}\lambda_t\bigg( C_7Q_{7\gamma} + C_9Q_{9V} +
C_{10}Q_{10A} \bigg) \,, 
\label{eq:H_sl}
\end{equation}
includes the electromagnetic penguin plus the semileptonic operators
\begin{eqnarray}
Q_{7\gamma} &\:\:=\:\:&
                        \frac{e}{16\pi^2}m_b\bar{s}_L\sigma_{\mu\nu}F^{\mu\nu}b_R\,,
                        \nonumber \\  
Q_{9V} &=&
           \frac{\alpha_{e}}{4\pi}(\bar{s}_L\gamma_{\mu}b_L)(\bar{\ell}\gamma^{\mu}\ell)\,,
           \nonumber \\  
Q_{10A} &=&
            \frac{\alpha_{e}}{4\pi}(\bar{s}_L\gamma_{\mu}b_L)(\bar{\ell}\gamma^{\mu}\gamma^5\ell)
            \,, 
\end{eqnarray}
where $\lambda_{i} \equiv V^{}_{ib} V^{*}_{is} $ with $i=u,c,t$.

Considering the matrix element of
$\mathcal{H}_\mathrm{eff}^{\Delta B = 1}$ in eq.~(\ref{eq:Heff})
between the $\bar B$ initial state and $\bar K^*\ell^+\ell^-$ final
state, the contribution of
$\mathcal{H}_\mathrm{eff}^\mathrm{sl+\gamma}$ in eq.~(\ref{eq:H_sl})
clearly factorizes into the product of hadronic form factors and
leptonic tensors at all orders in strong interactions.  On the other
hand, the matrix elements of $\mathcal{H}_\mathrm{eff}^\mathrm{had}$
in eq.~(\ref{eq:H_had}) factorize only in the infinite $m_b$ limit
below the charm
threshold~\cite{Beneke:2000wa,Beneke:2001at,Beneke:2009az}.  Moreover,
in this regime, heavy quark symmetry reduces the number of independent
form factors from seven to two soft form
factors~\cite{Isgur:1990kf,Charles:1998dr,Grinstein:2002cz}. Therefore,
in this limit, the amplitudes have simpler expressions so that
optimized observables dominated by short distance physics can be
defined~\cite{Kruger:2005ep,Egede:2008uy,Descotes-Genon:2013vna}. The
main issue however is how important departures from the infinite
mass limit are, in particular when $q^2$ is close to $4m_c^2$.

Concerning factorized amplitudes, these can be described using the
full set of form factors, which have been estimated using QCD sum
rules at low
$q^2$~\cite{Colangelo:1995jv,Ball:2004ye,Ball:2004rg,Khodjamirian:2006st,Straub:2015ica}.
In particular we use the very recent results of
ref.~\cite{Straub:2015ica} with the full correlation matrix.  While
the form factor calculation is a difficult one, we think that QCD sum
rules provide a reasonable estimate of low $q^2$ values and
uncertainties, compatible with the lattice estimate at high
$q^2$~\cite{Horgan:2015vla}. Using full QCD form factors reintroduces
some hadronic uncertainties into optimized observables which have been
estimated in refs.~\cite{Egede:2010zc,Becirevic:2011bp,Matias:2012xw,
  Das:2012kz,Jager:2012uw,DescotesGenon:2012zf,Descotes-Genon:2013wba,Matias:2014jua,Descotes-Genon:2014uoa,
  Jager:2014rwa}. In this respect, it has been suggested in
ref.~\cite{Descotes-Genon:2014uoa} that including some
power-suppressed terms in the definition of the soft functions could
reduce the uncertainty on some optimized observables. Since observables
cannot depend on arbitrary scheme definitions, their deviation
from the infinite mass limit cannot be reduced in this way. 

The main point of our paper concerns the non-factorizable contribution
present in the matrix element of the Hamiltonian in
equation~(\ref{eq:H_had}) involving a $c\bar c$ loop.
In the infinite mass limit, this term can be computed using QCD
factorization including $\mathcal{O}(\alpha_s)$
corrections~\cite{Beneke:2001at,Bosch:2001gv}. Beyond the leading
power, the contribution of $Q^c_{1,2}$ to the
$\bar{B} \rightarrow {\bar K}^*\ell^+\ell^-$ amplitude at $q^2\sim 1$
GeV$^2$, as well as the contribution to the
$\bar{B} \rightarrow {\bar K}^*\gamma$ amplitude, has been estimated
using light-cone sum rules in the single soft-gluon
approximation~\cite{Khodjamirian:2010vf}.  This approximation worsens
as $q^2$ increases and breaks down at $q^2\sim 4m_c^2$, as each
additional soft gluon exchange is suppressed by a factor
$1/(q^2-4m_c^2)$.  In ref.~\cite{Khodjamirian:2010vf} the authors
proposed also a phenomenological model interpolating their result at
$q^2\sim 1$ GeV$^2$ with a description of the resonant region based on
dispersion relations. While this model is reasonable, clearly there
are large uncertainties in the transition region from $q^2\sim 4$
GeV$^2$ to $m_{J/\psi}^2$.  Therefore, we consider the result of
ref.~\cite{Khodjamirian:2010vf} at $q^2\lesssim 1$ GeV$^2$ as an
estimate of the charm loop effect, but allow for larger effects as
$q^2$ grows and reaches values of $\mathcal{O}(4 m_c^2)$. 

While $Q_{1,2}^c$ are expected to dominate the
$\langle \bar{K}^* \gamma^* \vert
\mathcal{H}_\mathrm{eff}^\mathrm{had} \vert \bar{B} \rangle$ matrix
element, the effect of all operators in the hadronic Hamiltonian can
be reabsorbed in the following parameterization, generalizing the one
in ref.~\cite{Jager:2012uw}:\footnote{Since $h_\lambda$ is a smooth
  function of $q^2$ in the range considered, the first hadronic
  threshold being at $q^2 = m_{J/\psi}^2 \sim 9.6$ GeV$^2$, we are
  using a simple Taylor expansion. While the expansion might have
  significant corrections in the last bin considered, with current
  experimental uncertainties this is not problematic. We have also
  checked that using a parameterization with an explicit singularity
  at $m_{J/\psi}^2$ one obtains compatible results.}
\begin{eqnarray}
 h_\lambda(q^2) \;\;&=&\;\; \frac{\epsilon^*_\mu(\lambda)}{m_B^2} \int d^4x e^{iqx} \langle \bar K^* \vert T\{j^{\mu}_\mathrm{em} (x) 
 \mathcal{H}_\mathrm{eff}^\mathrm{had} (0)\} \vert \bar B \rangle \nonumber \\
 \;\;&=&\;\; h_\lambda^{(0)} + \frac{q^2}{1\,\mathrm{GeV}^2}
         h_\lambda^{(1)} + \frac{q^4}{1\, \mathrm{GeV}^4} h_\lambda^{(2)} \,,
 \label{eq:hlambda}
\end{eqnarray}
where $\lambda=+,-,0$ represents the helicity. Notice that
$h_\lambda^{(0)}$ and $h_\lambda^{(1)}$ could be reinterpreted as a
modification of $C_7$ and $C_9$ respectively, while the term
$h_\lambda^{(2)}$ that we introduce to allow for a growth of
long-distance effects when approaching the charm threshold cannot be
reabsorbed in a shift of the Wilson coefficients of the operators in
eq.~(\ref{eq:Heff}). We notice here the crucial point regarding NP
searches in these processes: one cannot use data to disentangle
long-distance contributions such as $h_\lambda^{(0,1)}$ from possible
NP ones, except, of course, for NP-induced CP-violating effects and/or
NP contributions to operators other than $C_{7,9}$.  Thus, in the
absence of a more accurate theoretical estimate of $h_\lambda(q^2)$
over the full kinematic range it is hardly possible to establish the
presence of NP in $C_{7,9}$, unless its contribution is much larger
than hadronic uncertainties. In this work we show that hadronic
contributions are sufficient to reproduce the present data once all
the uncertainties are properly taken into account.  We conclude that,
given the present hadronic uncertainties, the NP sensitivity of these
decays is washed out. In order to recover it, a substantial reduction
of these uncertainties is needed. This however requires a theoretical
breakthrough in the calculation of the hadronic amplitude in
eq.~(\ref{eq:hlambda}).

The $h_\lambda(q^2)$ are related to the $\tilde{g}^{\mathcal{M}_i}$
functions defined in ref.~\cite{Khodjamirian:2010vf} as follows:
\begin{eqnarray}
\tilde{g}^{\mathcal{ M}_1} \;\;& = &\;\; -
                                     \frac{1}{2C_1}\frac{16m_B^3(m_B +
                                     m_{K^*})\pi^2}{\sqrt{\lambda(q^2)}V(q^2)q^2}
                                     \left(h_-(q^2) - h_+(q^2)\right)\,, \nonumber\\
\tilde{g}^{\mathcal{ M}_2} \;\;& = &\;\; -
                                     \frac{1}{2C_1}\frac{16m_B^3\pi^2}{(m_B
                                     + m_{K^*})A_1(q^2)q^2}
                                     \left(h_-(q^2) + h_+(q^2)\right)\,, \label{eq:gtildetohlambda}\\
\tilde{g}^{\mathcal{ M}_3} \;\;& = &\;\; \frac{1}{2C_1}\left[\frac{64\pi^2m_B^3m_{K^*}\sqrt{q^2}(m_B + m_{K^*})}{\lambda(q^2) A_2(q^2)q^2}h_0(q^2)\right.\nonumber\\
&&\;\;\left. - \frac{16m_B^3\pi^2(m_B + m_{K^*})(m_B^2 - q^2 -
   m_{K^*}^2)}{\lambda(q^2) A_2(q^2)q^2}
   \left(h_-(q^2) + h_+(q^2)\right)\right]\,, \nonumber
\end{eqnarray}
where the form factor definition is given in Appendix
\ref{sec:FF}. Notice that the nonfactorizable contribution to
$\Delta C_9^i(q^2)$ is given by $2 C_1 \tilde{g}^{\mathcal{M}_i}$.
For the reader's convenience, we also give the expression of
$\Delta C_7^i(0)$ in terms of $h_\lambda(0)$:
\begin{eqnarray}
\Delta C_7^{1}(0)\;\; & = &\;\; -
                            \frac{8\pi^2m_B^3}{\lambda^{1/2}(0)m_bT_1(0)}
                            \left(h_-(0) - h_+(0)\right)\,, \nonumber\\
\Delta C_7^{2}(0)\;\; & = &\;\;-
                        \frac{8\pi^2m_B^3}{\lambda^{1/2}(0)m_bT_1(0)}
                            \left(h_-(0)
                        + h_+(0)\right)\,. \label{eq:DeltaC7tohlambda}
\end{eqnarray}

In our analysis we let the complex parameters $h_\lambda^{(0,1,2)}$
vary in the range $\vert h_\lambda^{(0,1,2)}\vert < 2 \times 10^{-3}$
with arbitrary phase using flat priors. To comply with the expected
power suppression of $h_+^{(0)}$ with respect to $h_-^{(0)}$, we
impose that $\vert h_+^{(0)}/h_-^{(0)}\vert \le 0.2$. We use the
results in table 1 of ref.~\cite{Khodjamirian:2010vf} at 1 GeV$^2$ as a
constraint on $\vert h_\lambda \vert$ via
eq.~(\ref{eq:gtildetohlambda}). We also use the results in
eqs. (6.2)-(6.3) in the same paper at $q^2=0$ to further constrain
$\vert h_\lambda \vert$ via eq.~(\ref{eq:DeltaC7tohlambda}). As useful
cross-checks, we also present in Appendix \ref{sec:other} the results
of the analysis using as a constraint the phenomenological model of
ref.~\cite{Khodjamirian:2010vf} over the full $q^2$ range, obtaining
results in agreement with the recent analysis of
ref.~\cite{Descotes-Genon:2014uoa}, as well as the results of the
analysis without using the constraints from
ref.~\cite{Khodjamirian:2010vf} at all.

\section{Main results}
\label{sec:fit}

We present the results for the Branching Ratios (BRs) and angular
observables obtained performing a Bayesian analysis. We use the tool
\HEP\ \cite{HEPfit} to compute all relevant observables and to
estimate the p.d.f.\ performing a Markov Chain Monte Carlo
(MCMC).\footnote{\HEP\ uses a parallelized version of the Bayesian
  Analysis Toolkit (BAT) library \cite{Caldwell:2008fw} to perform
  MCMC runs.} The main input parameters are collected in
table~\ref{Tab:SM}. They are the strong coupling, quark masses, meson
decay constants, CKM parameters, the matching scale $\mu_W$ for the
effective Hamiltonian, and the parameters $\lambda_B$ and
$a_{1,2}(\bar{K}^{*})_{\perp, \, ||}$ describing properties of meson
distribution functions entering the QCD factorization leading power
expressions. The LHCb results from
refs.~\cite{Aaij:2013iag,Aaij:2013hha,LHCb:2015dla,Aaij:2015dea,Aaij:2015oid}
are reported in tables~\ref{tab:mainmumu} and \ref{tab:mainee} for the
reader's convenience (we do not report here the correlation matrices
for LHCb results, which are used in our analysis).\footnote{In
  ref.~\cite{Aaij:2015oid} the data are analysed using three different
  methods. We use the unbinned maximum likelihood fit, which is the
  most accurate one.} We use the form factors
from~\cite{Straub:2015ica} (details can be found in
Appendix~\ref{sec:FF}). All Wilson coefficients are computed at NNLO
at $4.8$
GeV~\cite{Bobeth:1999mk,Bobeth:2003at,Gambino:2003zm,Misiak:2004ew}.

\begin{table}[htb!]
\centering
\begin{tabular}{|c|c|c|c|}
\hline
\textbf{Parameters} & \textbf{Mean Value} & \textbf{Uncertainty} & \textbf{Reference} \\[1mm]
\hline
$\alpha_{s}(M_{Z})$ & $0.1185$ & $0.0005$ & \cite{Agashe:2014kda}\\
$m_{t}$ (GeV) & $173.34$ & $0.76$ & \cite{ATLAS:2014wva}\\
$m_{c}(m_c)$ (GeV) & $1.28$ & $0.02$ & \cite{Lubicz}\\
$m_{b}(m_b)$ (GeV) & $4.17$ &  $0.05$ & \cite{Sanfilippo:2015era}\\
$f_{B_{s}}$ (MeV) & $226$ & $5$ & \cite{Aoki:2013ldr}\\
$f_{B_{s}}/ f_{B_{d}}$ & $1.204$ & $0.016$ & \cite{Aoki:2013ldr}\\
$f_{K^*,\vert\vert}$ (MeV) & $225$ & $30$ & \cite{Ball:1998kk}\\
$f_{K^*,\perp}(1\mathrm{GeV})$ (MeV) & $185$ & $10$ & \cite{Ball:1998kk}\\
$\lambda $ & $0.2250$ & $0.0006$ & \cite{Bona:2006ah,UTfit}\\
$A $ & $0.829$ & $0.012$ & \cite{Bona:2006ah,UTfit}\\
$\bar{\rho} $ & $0.132$ & $0.018$ & \cite{Bona:2006ah,UTfit}\\
$\bar{\eta} $ & $0.348$ & $0.012$ & \cite{Bona:2006ah,UTfit}\\
\hline
$\mu_{W}$ (GeV) & $100$ &  60 & \\
$\lambda_{B}$ (MeV) & $350$  & $ 150$ & \cite{Bosch:2001gv}\\ 
$a_{1}(\bar{K}^{*})_{\perp, \, ||}$ & $0.2$ &  $ 0.1$ & \cite{Ball:1998kk,Beneke:2001at}\\ 
$a_{2}(\bar{K}^{*})_{\perp, \, ||}$ & $0.05$ &  $ 0.1$ & \cite{Ball:1998kk,Beneke:2001at}\\ 
\hline
\end{tabular}
\caption{\textit{Parameters varied in the
    analysis. The
    last four parameters have flat priors with half width reported in
    the third column. The remaining ones 
    have Gaussian prior. Meson masses, lepton masses, $s$-quark mass and electroweak couplings are
    fixed at the PDG value \cite{Agashe:2014kda}.}}
 \label{Tab:SM}
\end{table}

\begin{figure}[!htbp]
\centering

\subfigure{\includegraphics[width=.4\textwidth]{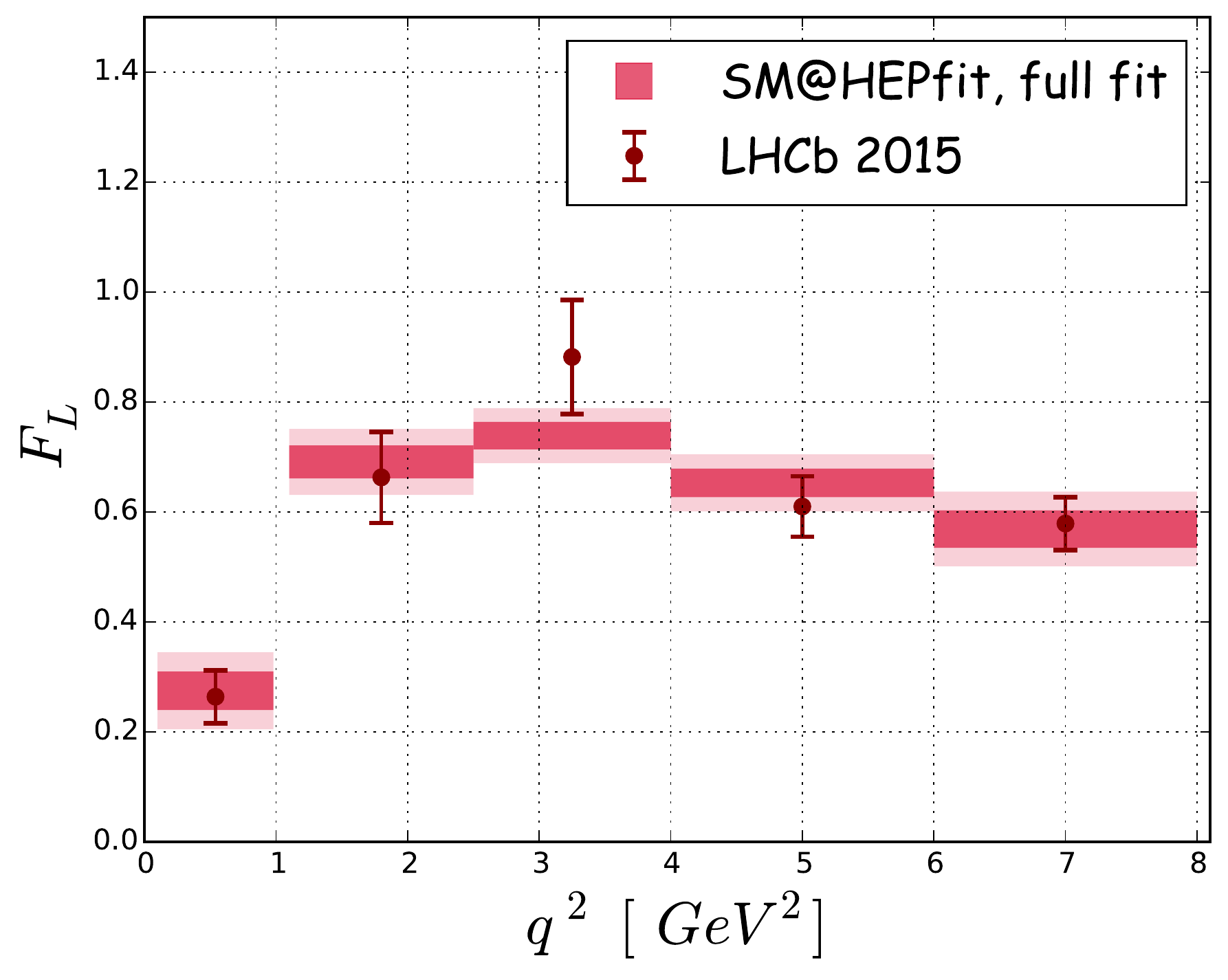}}
\subfigure{\includegraphics[width=.4\textwidth]{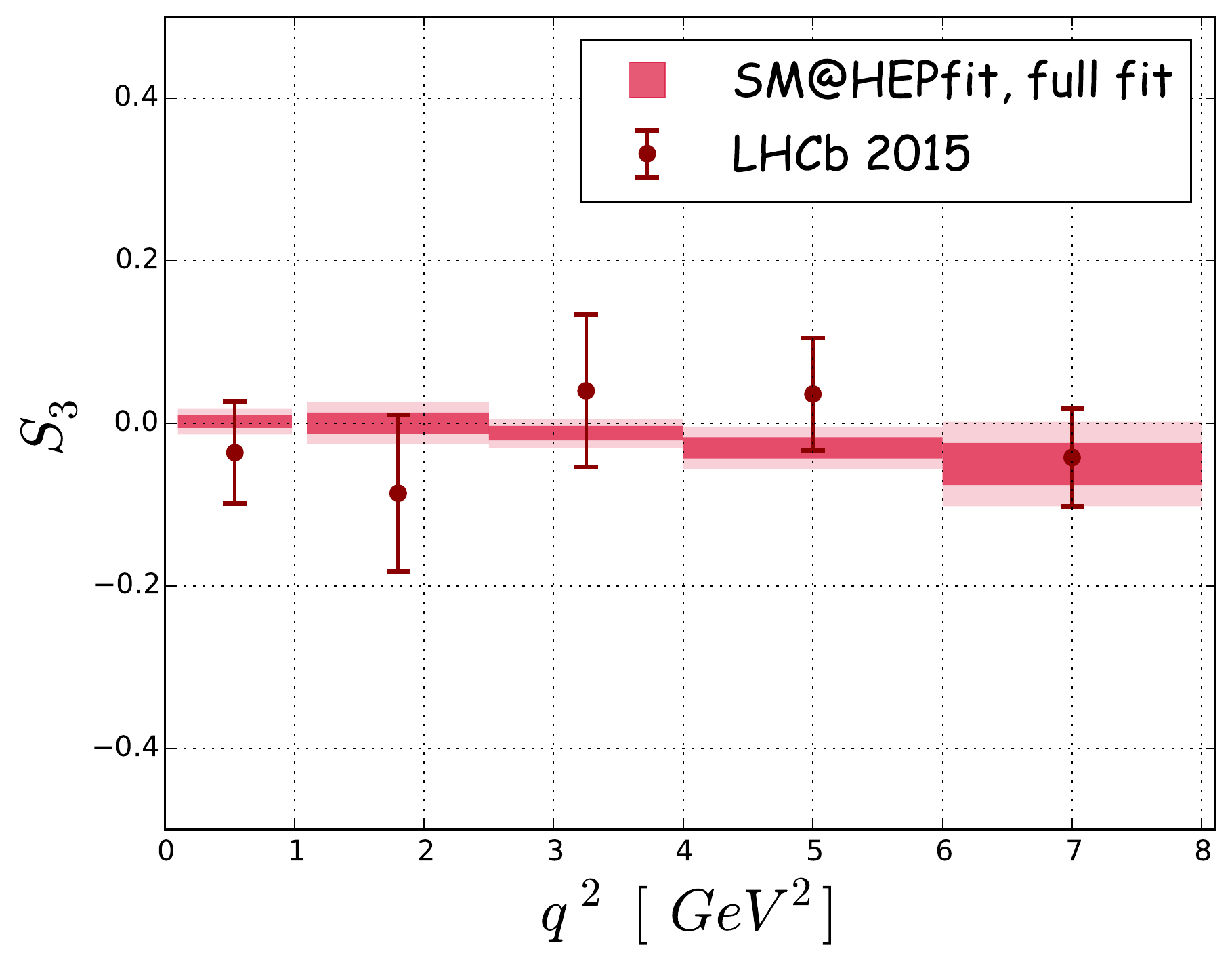}}
\subfigure{\includegraphics[width=.4\textwidth]{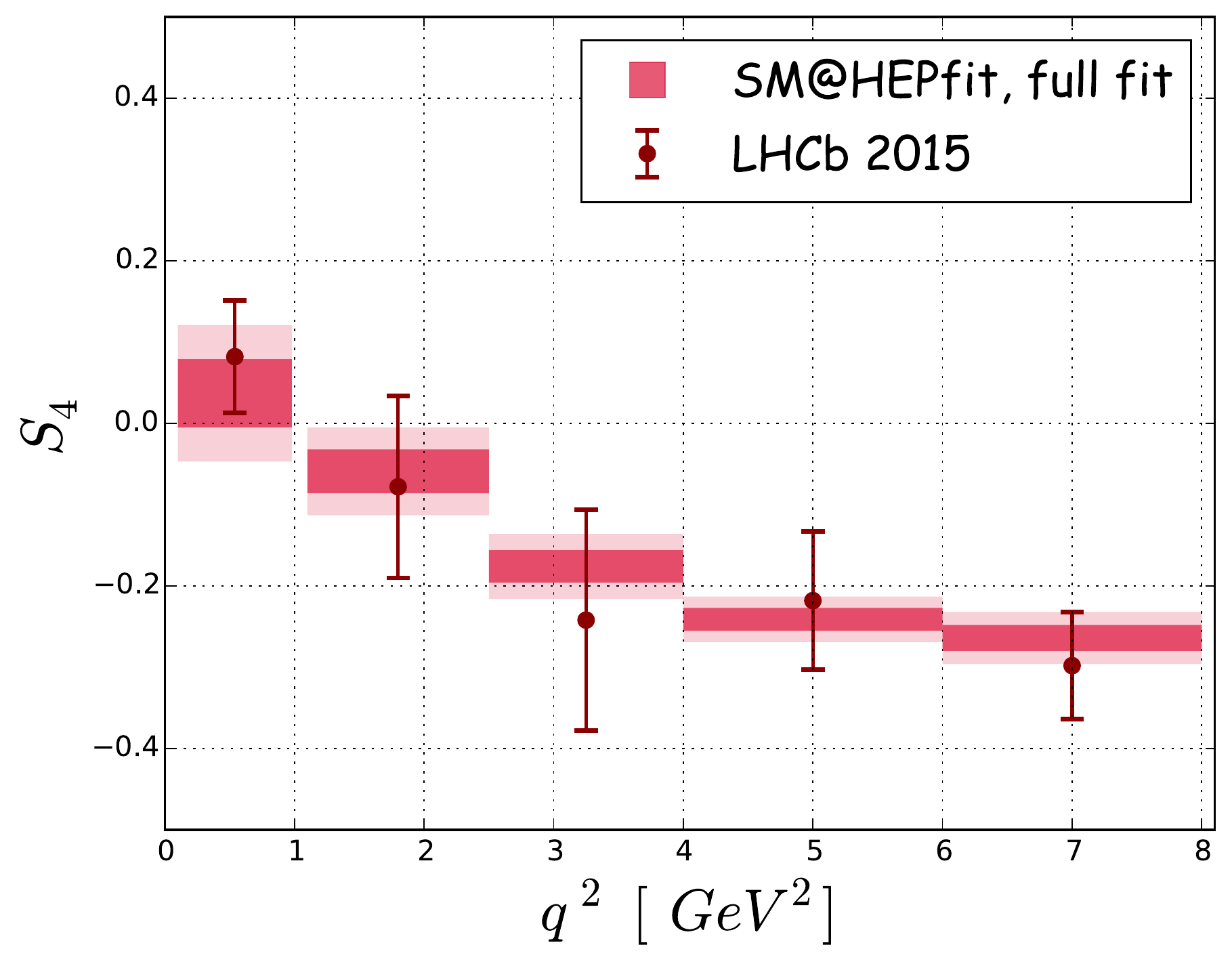}}
\subfigure{\includegraphics[width=.4\textwidth]{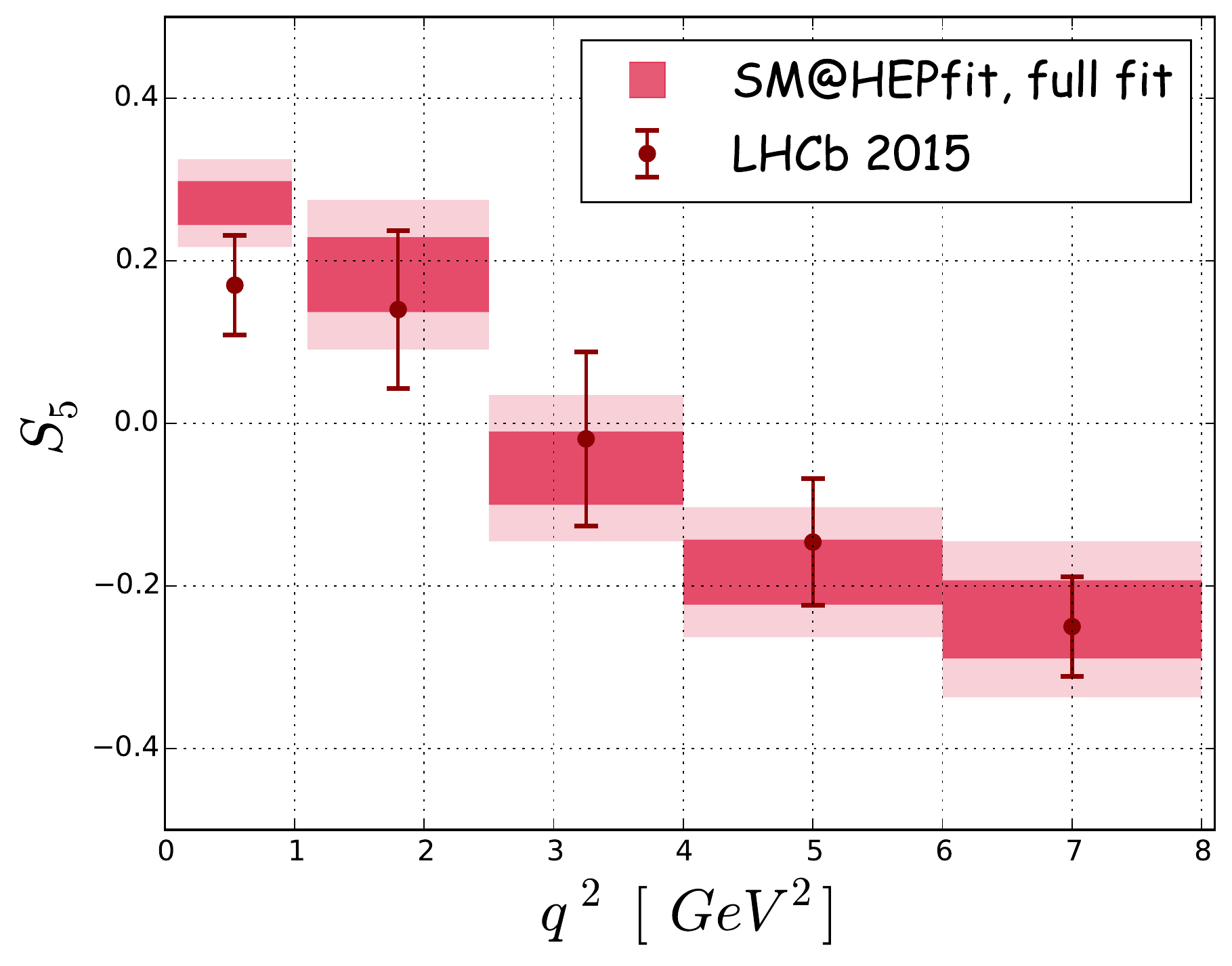}}
\subfigure{\includegraphics[width=.4\textwidth]{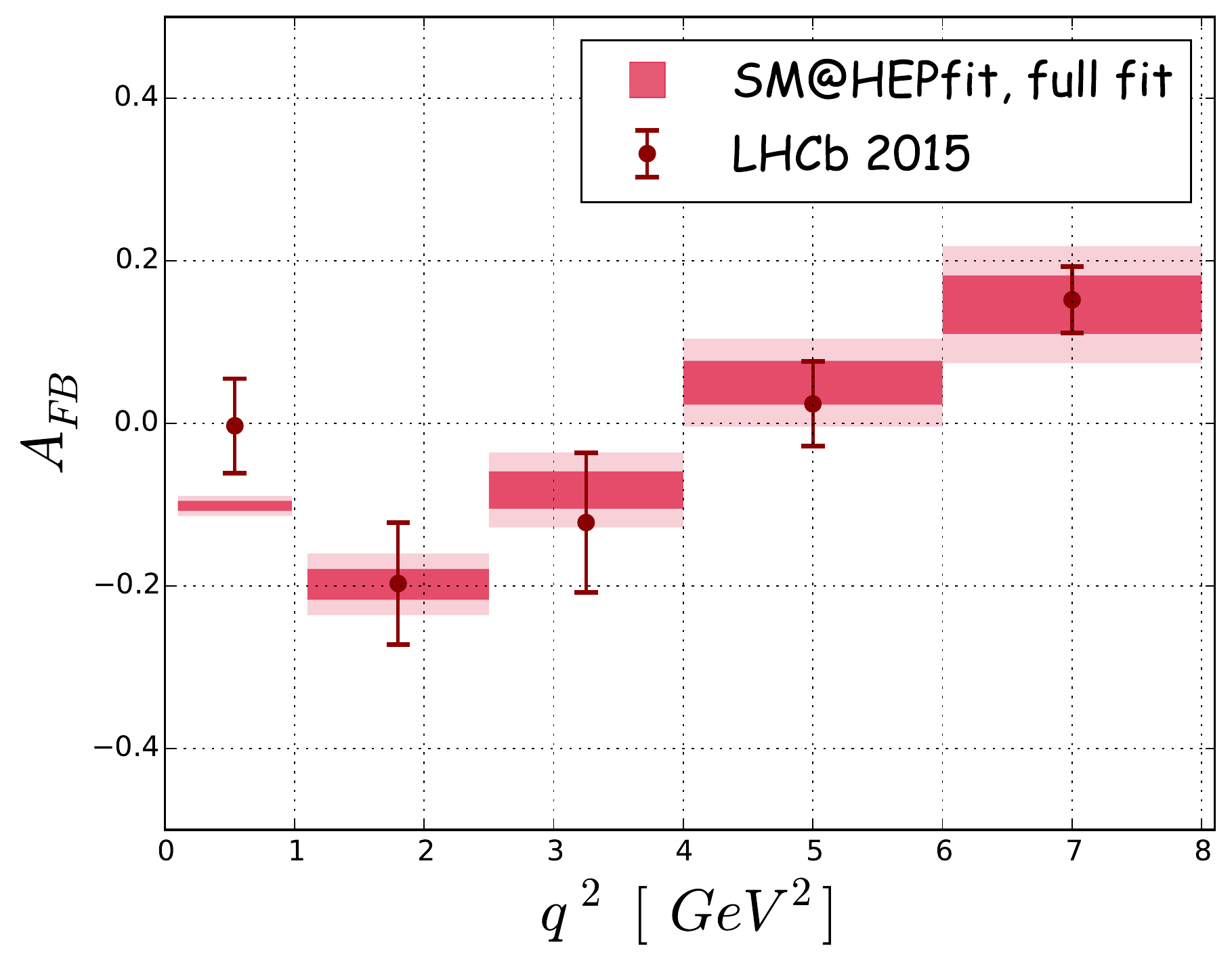}}
\subfigure{\includegraphics[width=.4\textwidth]{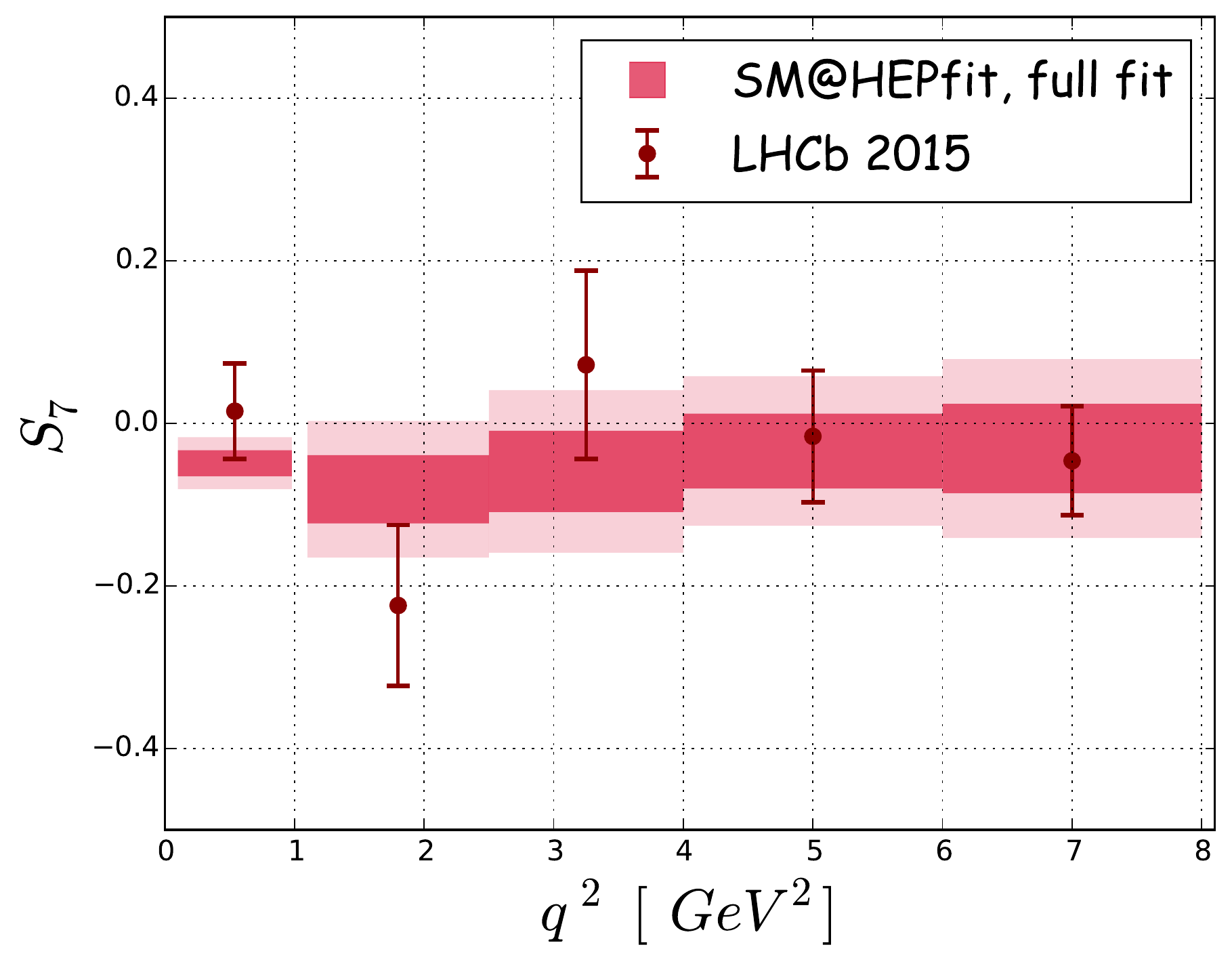}}
\subfigure{\includegraphics[width=.4\textwidth]{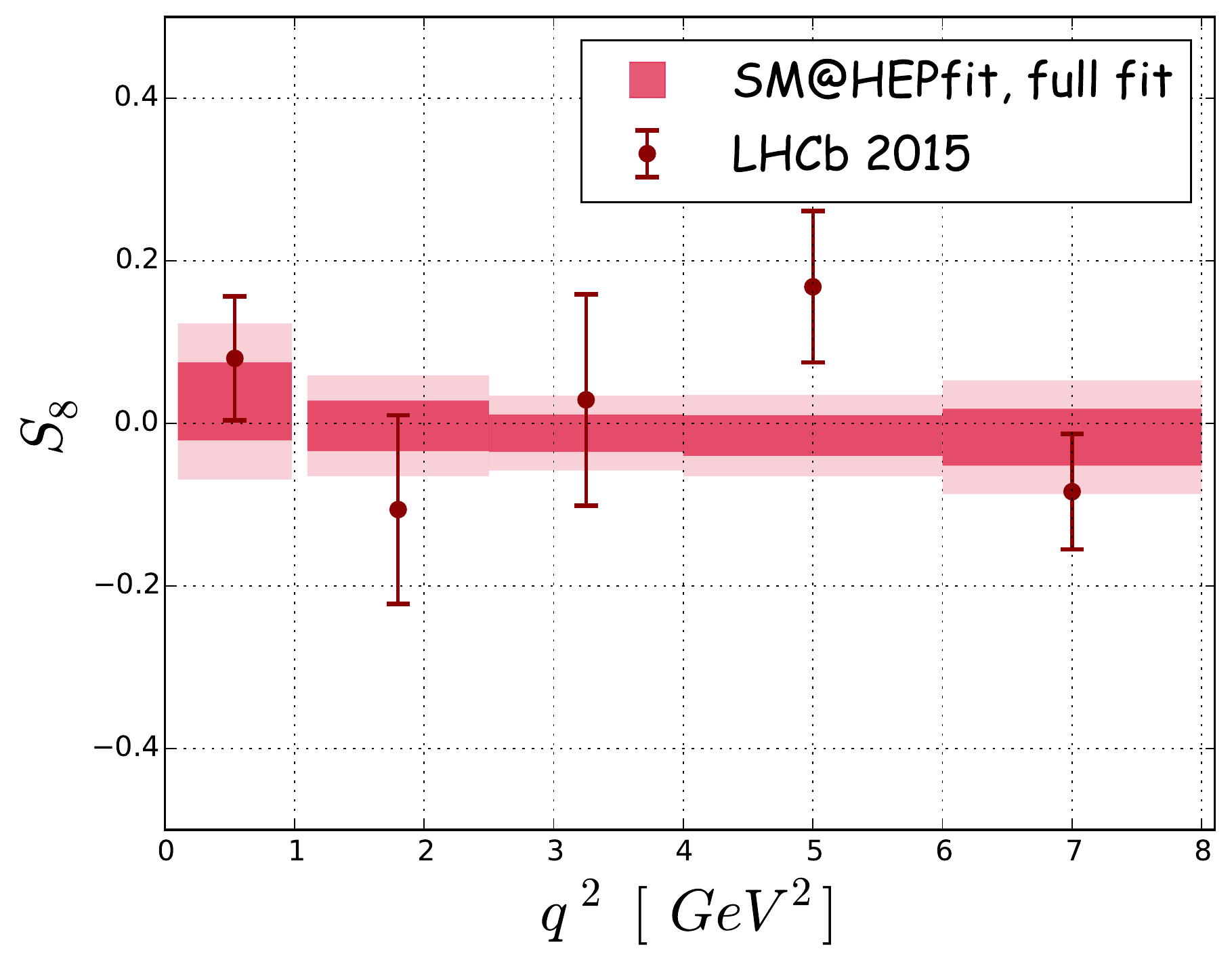}}
\subfigure{\includegraphics[width=.4\textwidth]{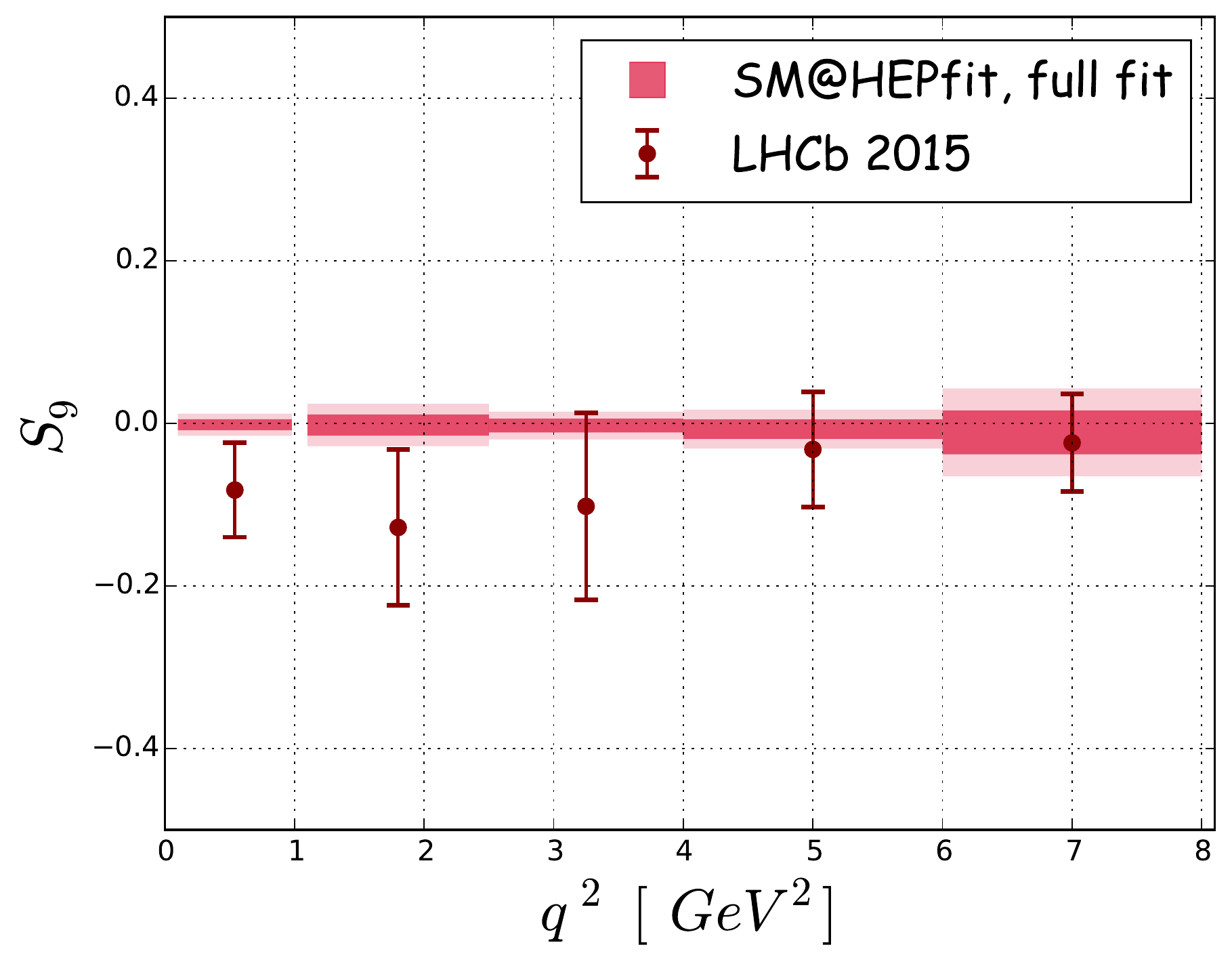}}

\caption{\textit{Results of the full fit and experimental results for
    the $B \to K^* \mu^+ \mu^-$ angular observables. Here and in the
    following, we use darker (lighter) colours for the $68\%$ ($95\%$)
    probability regions.}}
\label{fig:fullfit}
\end{figure}

In figure~\ref{fig:fullfit} we present the results for the
$B \to K^* \mu^+ \mu^-$ angular observables of the full fit to all the
LHCb measurements reported in tables~\ref{tab:mainmumu} and
\ref{tab:mainee}. The corresponding numerical results are reported in
the ``full fit'' column of table~\ref{tab:mainmumu}, while in
table~\ref{tab:mainee} we report the numerical results for the
$B \to K^* e^+ e^-$ observables. 

Let us now discuss the compatibility of the SM with experimental data,
taking theoretical and experimental correlations into account. For
uncorrelated observables, such as BR's and $B \to K^* e^+ e^-$ angular
observables, one can simply remove the experimental information on a
particular observable from the fit to obtain a ``prediction'' for that
observable, and then compute the $p$-value (see
table~\ref{tab:mainmumu} and~\ref{tab:mainee}). In the case of
correlated observables, one can generalize this procedure to take all
correlations into account. Since the angular observables in each bin
are correlated, we proceed as follows: we remove the experimental
information in one bin at a time from the fit to obtain the
``predictions'' reported in the corresponding column in
table~\ref{tab:mainmumu}, as well as their correlation matrix. Adding
the experimental covariance matrix to the one obtained from the fit,
we compute the log likelihood and report in table~\ref{tab:mainmumu}
the corresponding $p$-value. For completeness, we also give in
table~\ref{tab:mainmumu} our results and predictions for the
$B \to K^* \mu^+ \mu^-$ optimized observable $P_5^\prime$, which is
however not independent from the other observables in
table~\ref{tab:mainmumu}.\footnote{In this case, we quote a ``naive''
  $p$-value obtained neglecting the correlation with other observables.}

\begin{figure}[!htbp]
  \centering
  \includegraphics[width=.6\textwidth]{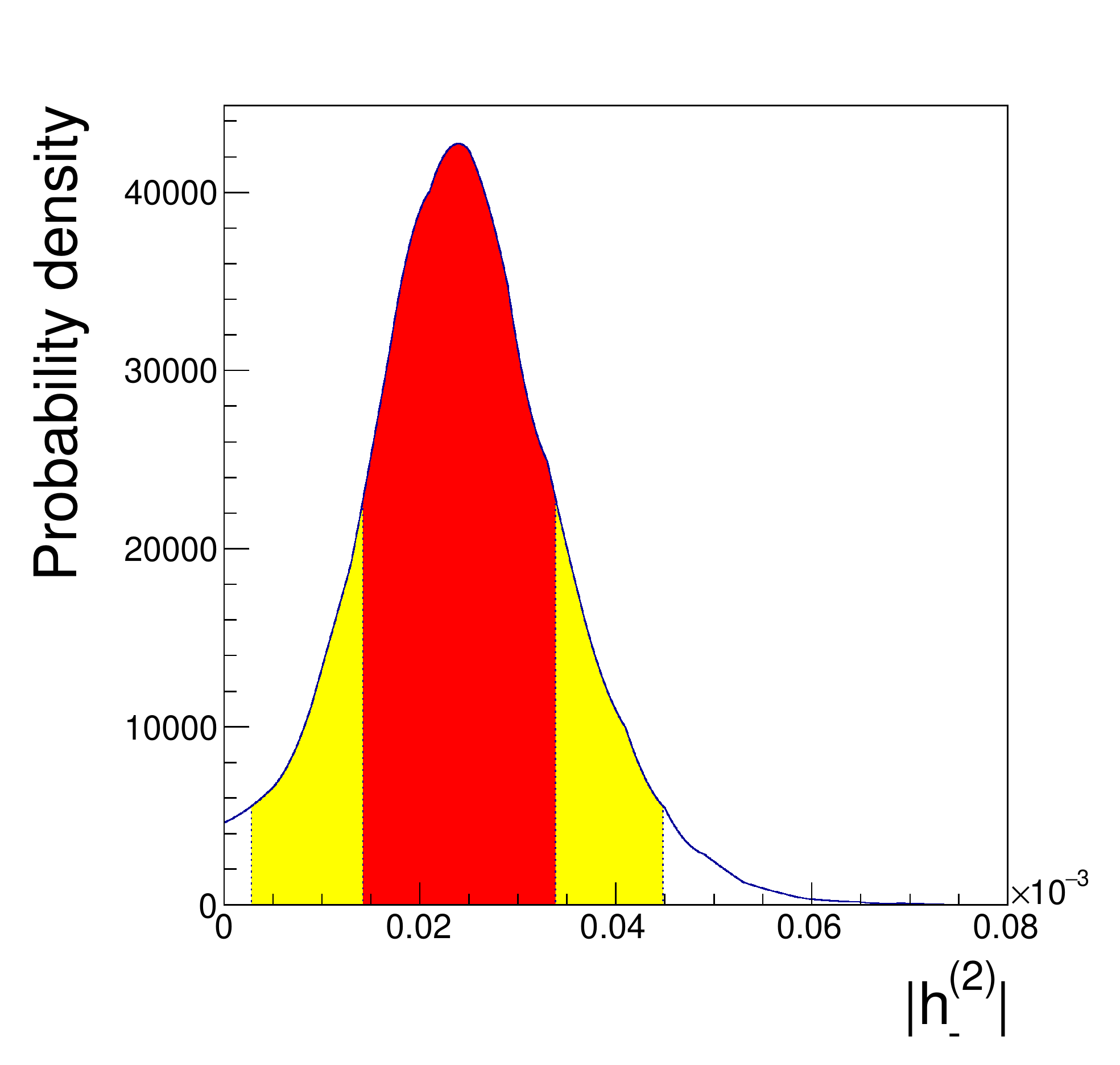}
  \caption{\textit{P.d.f.\ for the hadronic parameter $\vert h_{-}^{(2)}\vert$
    obtained using the numerical information from
    ref.~\cite{Khodjamirian:2010vf} for $q^2 \leq 1$ GeV$^2$.}}
  \label{fig:hm2}
\end{figure}

The results for the parameters defining the nonfactorizable power
corrections $h_\lambda$ are reported in table~\ref{tab:hfit} (in this
case, the distributions are not Gaussian). It is interesting to notice
that $\vert h_{-}^{(2)}\vert$ is different from zero at more than
$95.45\%$ probability (see figure~\ref{fig:hm2}), thus disfavouring the
interpretation of the hadronic correction as a modified Wilson
coefficient for operators $Q_{7,9}$, possibly generated by NP
contributions.

\begin{figure}[!htbp]
\centering

\subfigure{\includegraphics[width=.4\textwidth]{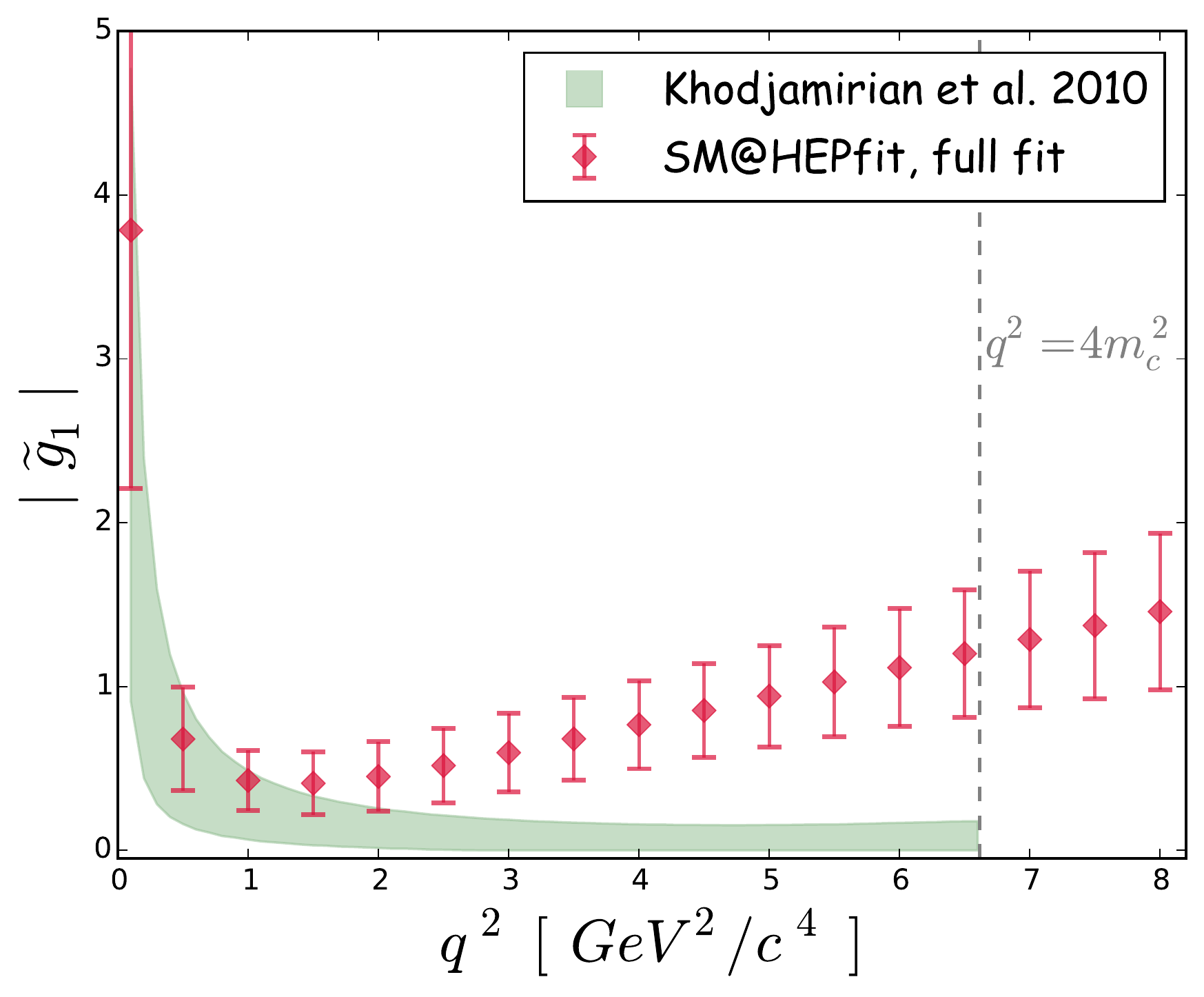}}
\subfigure{\includegraphics[width=.4\textwidth]{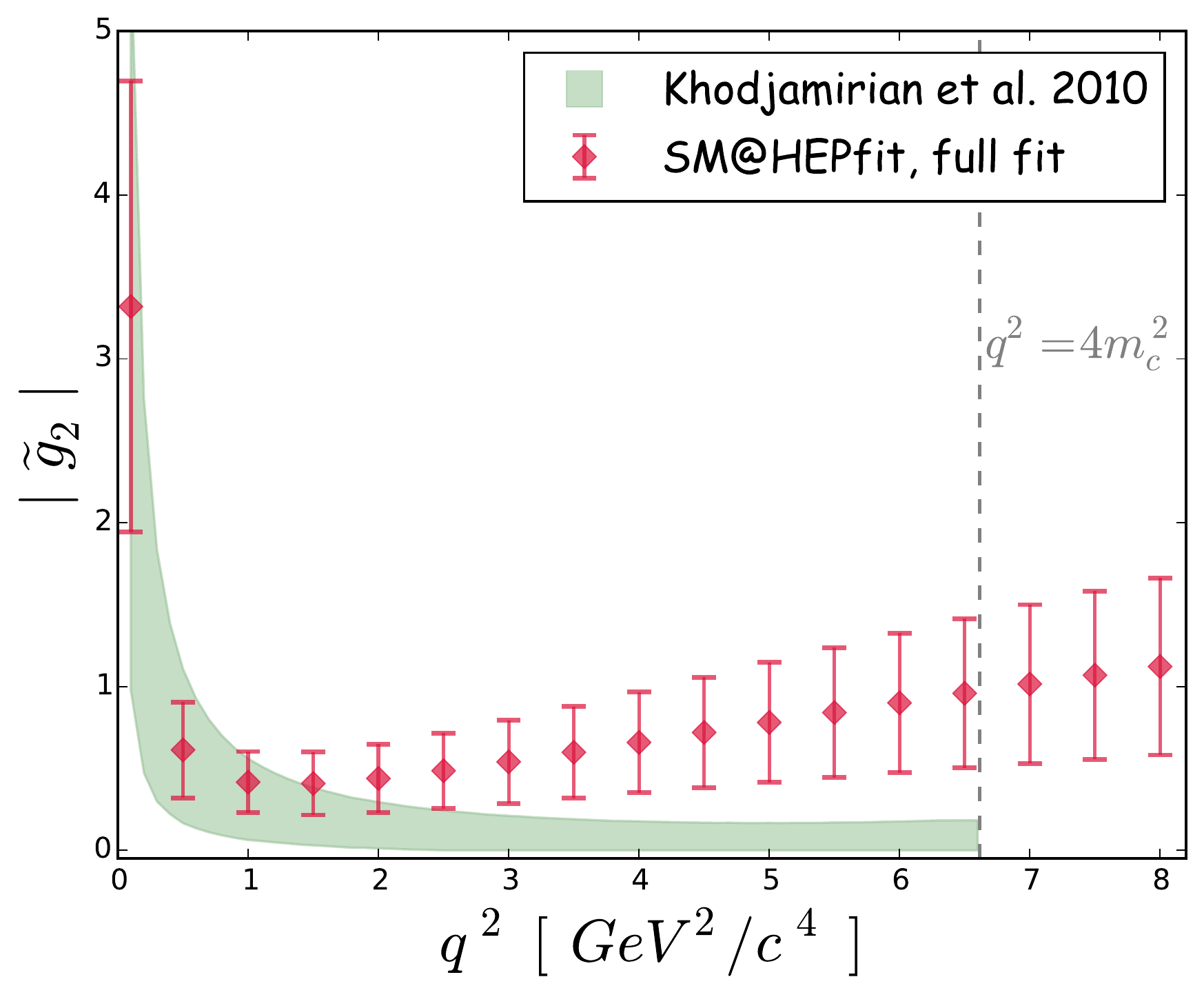}}
\subfigure{\includegraphics[width=.4\textwidth]{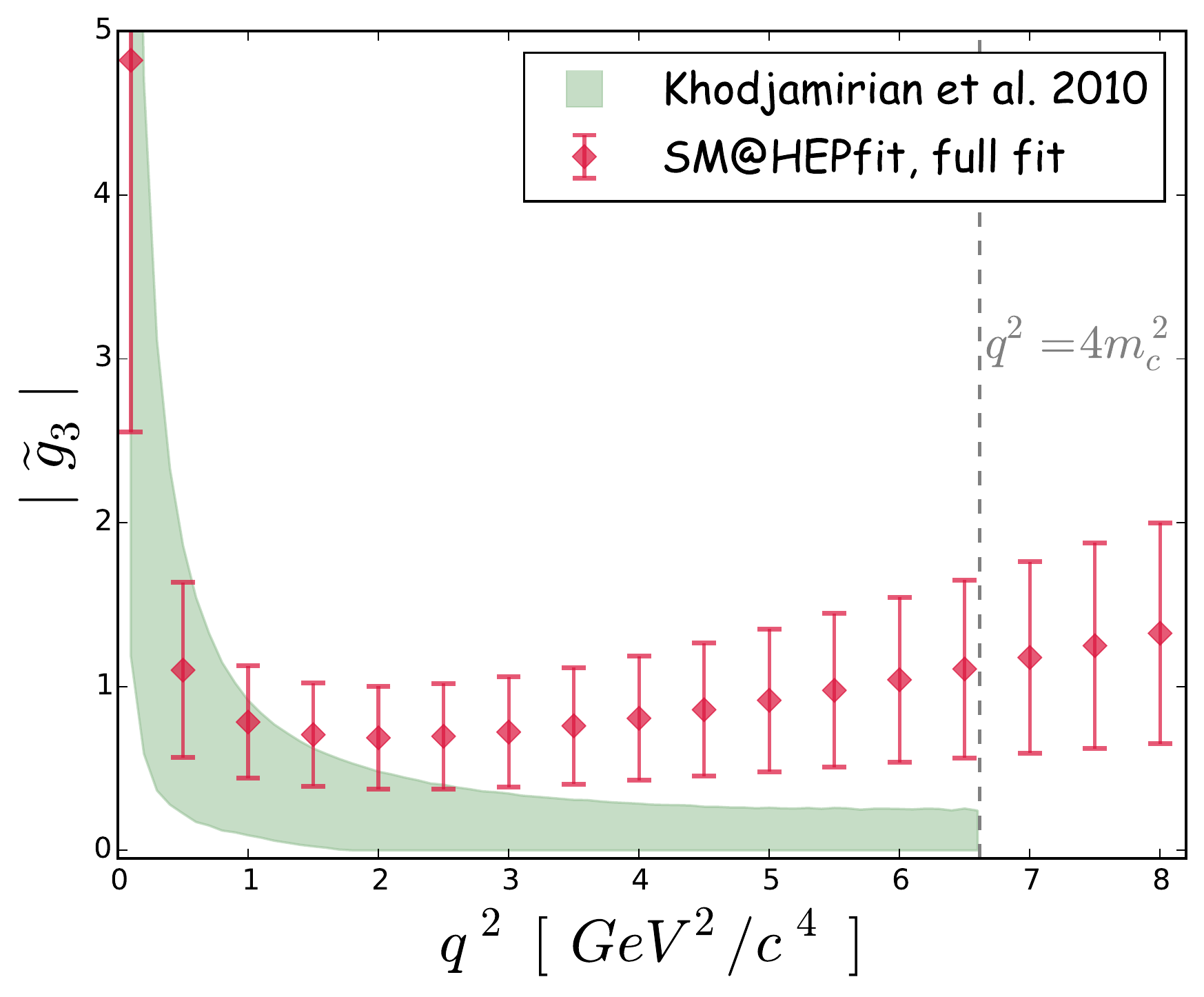}}

\caption{\textit{Results of the fit for $\vert\widetilde{g}_{1,2,3}\vert$ defined in ref.~\cite{Khodjamirian:2010vf} as a function of $q^2$ together with the phenomenological parametrization suggested in the same
paper.}}
\label{fig:gtildes}
\end{figure}

For an easy comparison with ref.~\cite{Khodjamirian:2010vf}, we also
report in figure~\ref{fig:gtildes} the results of the fit for the
absolute value of the $\tilde{g_i}$ functions, together with the
phenomenological model proposed in the same work. The sizable $q^2$
dependence of hadronic corrections is visible by eye in this plot.

\begin{table}[!htbp]
\fontsize{8}{8}\selectfont
\centering
\begin{tabular}{|c|c|c|c|c|c|}
\hline
&&&&&\\[-1mm]
\textbf{$q^2$ bin [GeV$^2$]} & \textbf{Observable} & \textbf{measurement} & \textbf{full fit} & \textbf{prediction} & $\mathbf{p-value}$ \\[1mm]
\hline
&&&&&\\[-2mm]
\multirow{9}{*}{\normalsize $ [0.1,0.98] $}
& $F_L
$ & $
\phantom{-}
0.264 \pm 0.048$ & $
\phantom{-}
0.275 \pm 0.035$ & $
\phantom{-}
0.257 \pm 0.035$ & \multirow{8}{*}{0.13}
\\
& $S_3
$ & $
-0.036 \pm 0.063$ & $
\phantom{-}
0.002 \pm 0.008$ & $
\phantom{-}
0.002 \pm 0.008$ & 
\\
& $S_4
$ & $
\phantom{-}
0.082 \pm 0.069$ & $
\phantom{-}
0.037 \pm 0.042$ & $
-0.025 \pm 0.047$ & 
\\
& $S_5
$ & $
\phantom{-}
0.170 \pm 0.061$ & $
\phantom{-}
0.271 \pm 0.027$ & $
\phantom{-}
0.301 \pm 0.024$ & 
\\
& $ A_{FB}
$ & $
-0.003 \pm 0.058$ & $
-0.102 \pm 0.006$ & $
-0.104 \pm 0.006$ & 
\\
& $S_7
$ & $
\phantom{-}
0.015 \pm 0.059$ & $
-0.049 \pm 0.016$ & $
-0.043 \pm 0.017$ & 
\\
& $S_8
$ & $
\phantom{-}
0.080 \pm 0.076$ & $
\phantom{-}
0.027 \pm 0.048$ & $
-0.004 \pm 0.046$ & 
\\
& $S_9
$ & $
-0.082 \pm 0.058$ & $
-0.002 \pm 0.007$ & $
-0.002 \pm 0.007$ & 
\\
\cline{2-6}
&&&&&\\[-2mm]
& $ P_5'
$ & $
\phantom{-}
0.387 \pm 0.142$ & $
\phantom{-}
0.774 \pm 0.094$ & $
\phantom{-}
0.881 \pm 0.082$ & 0.0026
\\[.5mm]
\hline
&&&&&\\[-2mm]
 \multirow{9}{*}{\normalsize $ [1.1,2.5] $}
& $F_L
$ & $
\phantom{-}
0.663 \pm 0.083$ & $
\phantom{-}
0.691 \pm 0.030$ & $
\phantom{-}
0.688 \pm 0.034$ & \multirow{8}{*}{0.63}
\\
& $S_3
$ & $
-0.086 \pm 0.096$ & $
\phantom{-}
0.000 \pm 0.013$ & $
\phantom{-}
0.001 \pm 0.013$ & 
\\
& $S_4
$ & $
-0.078 \pm 0.112$ & $
-0.059 \pm 0.027$ & $
-0.070 \pm 0.032$ & 
\\
& $S_5
$ & $
\phantom{-}
0.140 \pm 0.097$ & $
\phantom{-}
0.183 \pm 0.046$ & $
\phantom{-}
0.208 \pm 0.057$ & 
\\
& $ A_{FB}
$ & $
-0.197 \pm 0.075$ & $
-0.198 \pm 0.019$ & $
-0.200 \pm 0.022$ & 
\\
& $S_7
$ & $
-0.224 \pm 0.099$ & $
-0.081 \pm 0.042$ & $
-0.056 \pm 0.049$ & 
\\
& $S_8
$ & $
-0.106 \pm 0.116$ & $
-0.003 \pm 0.031$ & $
-0.004 \pm 0.033$ & 
\\
& $S_9
$ & $
-0.128 \pm 0.096$ & $
-0.002 \pm 0.013$ & $
\phantom{-}
0.002 \pm 0.013$ & 
\\
\cline{2-6}
&&&&&\\[-2mm]
& $ P_5'
$ & $
\phantom{-}
0.298 \pm 0.212$ & $
\phantom{-}
0.410 \pm 0.099$ & $
\phantom{-}
0.460 \pm 0.120$ & 0.51
\\[.5mm]
\hline
&&&&&\\[-2mm]
 \multirow{9}{*}{\normalsize $ [2.5,4] $}
& $F_L
$ & $
\phantom{-}
0.882 \pm 0.104$ & $
\phantom{-}
0.739 \pm 0.025$ & $
\phantom{-}
0.729 \pm 0.028$ & \multirow{8}{*}{0.80}
\\
& $S_3
$ & $
\phantom{-}
0.040 \pm 0.094$ & $
-0.012 \pm 0.009$ & $
-0.014 \pm 0.010$ & 
\\
& $S_4
$ & $
-0.242 \pm 0.136$ & $
-0.176 \pm 0.020$ & $
-0.179 \pm 0.021$ & 
\\
& $S_5
$ & $
-0.019 \pm 0.107$ & $
-0.055 \pm 0.045$ & $
-0.055 \pm 0.052$ & 
\\
& $ A_{FB}
$ & $
-0.122 \pm 0.086$ & $
-0.082 \pm 0.023$ & $
-0.082 \pm 0.025$ & 
\\
& $S_7
$ & $
\phantom{-}
0.072 \pm 0.116$ & $
-0.059 \pm 0.050$ & $
-0.080 \pm 0.055$ & 
\\
& $S_8
$ & $
\phantom{-}
0.029 \pm 0.130$ & $
-0.012 \pm 0.023$ & $
-0.012 \pm 0.023$ & 
\\
& $S_9
$ & $
-0.102 \pm 0.115$ & $
-0.003 \pm 0.009$ & $
-0.003 \pm 0.009$ & 
\\
\cline{2-6}
&&&&&\\[-2mm]
& $ P_5'
$ & $
-0.077 \pm 0.354$ & $
-0.130 \pm 0.100$ & $
-0.130 \pm 0.120$ & 0.89
\\[.5mm]
\hline
&&&&&\\[-2mm]
 \multirow{9}{*}{\normalsize $ [4,6] $}
& $F_L
$ & $
\phantom{-}
0.610 \pm 0.055$ & $
\phantom{-}
0.653 \pm 0.026$ & $
\phantom{-}
0.661 \pm 0.030$ & \multirow{8}{*}{0.50}
\\
& $S_3
$ & $
\phantom{-}
0.036 \pm 0.069$ & $
-0.030 \pm 0.013$ & $
-0.030 \pm 0.015$ & 
\\
& $S_4
$ & $
-0.218 \pm 0.085$ & $
-0.241 \pm 0.014$ & $
-0.239 \pm 0.016$ & 
\\
& $S_5
$ & $
-0.146 \pm 0.078$ & $
-0.183 \pm 0.040$ & $
-0.205 \pm 0.046$ & 
\\
& $ A_{FB}
$ & $
\phantom{-}
0.024 \pm 0.052$ & $
\phantom{-}
0.050 \pm 0.027$ & $
\phantom{-}
0.067 \pm 0.032$ & 
\\
& $S_7
$ & $
-0.016 \pm 0.081$ & $
-0.034 \pm 0.046$ & $
-0.037 \pm 0.055$ & 
\\
& $S_8
$ & $
\phantom{-}
0.168 \pm 0.093$ & $
-0.015 \pm 0.025$ & $
-0.026 \pm 0.026$ & 
\\
& $S_9
$ & $
-0.032 \pm 0.071$ & $
-0.007 \pm 0.012$ & $
-0.012 \pm 0.014$ & 
\\
\cline{2-6}
&&&&&\\[-2mm]
& $ P_5'
$ & $
-0.301 \pm 0.160$ & $
-0.388 \pm 0.087$ & $
-0.440 \pm 0.100$ & 0.46
\\[.5mm]
\hline
&&&&&\\[-2mm]
 \multirow{9}{*}{\normalsize $ [6,8] $}
& $F_L
$ & $
\phantom{-}
0.579 \pm 0.048$ & $
\phantom{-}
0.569 \pm 0.034$ & $
\phantom{-}
0.517 \pm 0.070$ & \multirow{8}{*}{0.82}
\\
& $S_3
$ & $
-0.042 \pm 0.060$ & $
-0.050 \pm 0.026$ & $
-0.006 \pm 0.054$ & 
\\
& $S_4
$ & $
-0.298 \pm 0.066$ & $
-0.264 \pm 0.016$ & $
-0.224 \pm 0.037$ & 
\\
& $S_5
$ & $
-0.250 \pm 0.061$ & $
-0.241 \pm 0.048$ & $
-0.164 \pm 0.100$ & 
\\
& $ A_{FB}
$ & $
\phantom{-}
0.152 \pm 0.041$ & $
\phantom{-}
0.146 \pm 0.036$ & $
\phantom{-}
0.099 \pm 0.077$ & 
\\
& $S_7
$ & $
-0.046 \pm 0.067$ & $
-0.031 \pm 0.055$ & $
\phantom{-}
0.010 \pm 0.110$ & 
\\
& $S_8
$ & $
-0.084 \pm 0.071$ & $
-0.017 \pm 0.035$ & $
\phantom{-}
0.039 \pm 0.055$ & 
\\
& $S_9
$ & $
-0.024 \pm 0.060$ & $
-0.011 \pm 0.027$ & $
\phantom{-}
0.018 \pm 0.047$ & 
\\
\cline{2-6}
&&&&&\\[-2mm]
& $ P_5'
$ & $
-0.505 \pm 0.124$ & $
-0.491 \pm 0.098$ & $
-0.330 \pm 0.200$ & 0.46
\\[.5mm]
\hline
\hline&&&&&\\[-2mm]
 $ [0.1,2] $ & 
\multirow{3}{*}{\normalsize $ {\rm BR } \cdot 10^7 $}
 & $ 0.58 \pm 0.09$ & $
0.65 \pm 0.04$ & $
0.67 \pm 0.04$ & 0.36
\\
 $ [2,4.3] $ & 
 & $ 0.29 \pm 0.05$ & $
0.33 \pm 0.03$ & $
0.35 \pm 0.04$ & 0.35
\\
 $ [4.3,8.68] $ & 
 & $ 0.47 \pm 0.07$ & $
0.45 \pm 0.05$ & $
0.47 \pm 0.11$ & 1.0
\\
\hline

&&&&&\\[-2mm]
& { \normalsize $ {\rm BR }_{B \to K^* \gamma} \cdot 10^5 $}
& $ 4.33 \pm 0.15$ & $
4.35 \pm 0.14$ & $
4.61 \pm 0.56$ & 0.63
\\[1mm]
\hline
\end{tabular}
\caption{\textit{Experimental results (with symmetrized errors), results from the full fit,
    predictions and $p$-values for $B \to K^* \mu^+ \mu^-$ BR's and angular
    observables. The predictions are obtained removing the 
    corresponding observable from the fit. For the angular observables,
    since their measurements are correlated in each bin, we remove from
    the fit the
    experimental information on all angular observables in one bin at a
    time to obtain the predictions. See the text for details. 
    We also report the results for BR$(B \to K^* \gamma)$
    (including the experimental value from refs.~\cite{Coan:1999kh,Nakao:2004th,Aubert:2009ak,Agashe:2014kda})
    and for the optimized observable $P_5^\prime$. The latter is however
    not explicitly used in the fit as a constraint, since it is not
    independent of $F_L$ and $S_5$. }}
\label{tab:mainmumu}
\end{table}

\begin{table}[!htbp]
\centering
\begin{tabular}{|c|c|c|c|c|}
\hline
&&&&\\[-4mm]
\textbf{Observable} & \textbf{measurement} & \textbf{full fit} & \textbf{prediction} & \textbf{p-value} \\[1mm]
\hline
$ {P_1}$
 & $
-0.23 \pm 0.24$ & $
\phantom{-}0.00 \pm 0.01$ & $
\phantom{-}0.00 \pm 0.01$ & 0.34
\\
$ {P_2}$
 & $
\phantom{-}
0.05 \pm 0.09$ & $
-0.040 \pm 0.00$ & $
-0.040 \pm 0.00$ & 0.32
\\
$ {P_3}$
 & $
-0.07 \pm 0.11$ & $
\phantom{-}0.00 \pm 0.01$ & $
\phantom{-}0.00 \pm 0.01$ & 0.53
\\
$ {F_L}$
 & $
\phantom{-}
0.16 \pm 0.08$ & $
\phantom{-}0.170 \pm 0.04$ & $
\phantom{-}0.18 \pm 0.05$ & 0.82
\\
$ {\rm BR }\cdot 10^7 $
 & $ \phantom{-}3.1 \pm 1.0$ & $
\phantom{-}1.4 \pm 0.1$ & $
\phantom{-}1.4 \pm 0.1$ & 0.06
\\
\hline
\end{tabular}
\caption{\textit{Experimental results (with symmetrized errors), results from the full fit,
    predictions and $p$-values for $B \to K^* e^+ e^-$ BR and angular
    observables. The predictions are obtained removing the corresponding
    observable from the fit.}}
\label{tab:mainee}
\end{table}

\begin{table}
\centering
\begin{tabular}{|c|c|c|}
\hline
&&\\[-3mm]
\textbf{Parameter} & \textbf{Absolute value} & \textbf{Phase} (rad) \\[1mm]
\hline
&&\\[-3mm]
$ h_0^{(0)} $ & 
$ (5.7 \pm 2.0) \cdot10^{-4} $ 
&
$\phantom{-} 3.57 \pm 0.55 $ 
\\
$ h_0^{(1)} $ &
$ (2.3 \pm 1.6) \cdot10^{-4} $
&
$\phantom{-} 0.1 \pm 1.1$
\\
$ h_0^{(2)} $ &
$ (2.8 \pm 2.1)\cdot10^{-5}$
&
$ -0.2 \pm 1.7$
\\
[1mm]
\hline
&&\\[-3mm]
$ h_+^{(0)} $ & 
$ (7.9 \pm 6.9)\cdot10^{-6} $ 
&
$\phantom{-} 0.1 \pm 1.7 $ 
\\
$ h_+^{(1)} $ &
$ (3.8 \pm 2.8)\cdot10^{-5}$
&
$ -0.7 \pm 1.9$
\\
$ h_+^{(2)} $ &
$ (1.4 \pm 1.0)\cdot10^{-5}$
&
$ \phantom{-}3.5 \pm 1.6$
\\
[1mm]
\hline
&&\\[-3mm]
$ h_-^{(0)} $ & 
$ (5.4 \pm 2.2)\cdot10^{-5} $ 
&
$ \phantom{-}3.2 \pm 1.4 $ 
\\[1mm]
$ h_-^{(1)} $ &
$ (5.2\pm 3.8)\cdot10^{-5}$
&
$ \phantom{-}0.0 \pm 1.7$
\\[1mm]
$ h_-^{(2)} $ &
$ (2.5 \pm 1.0)\cdot10^{-5}$
&
$\phantom{-} 0.09 \pm 0.77$
\\[1mm]
\hline
\end{tabular}
\caption{\it Results for the parameters defining the nonfactorizable power corrections $h_{\lambda}$ obtained 
using the numerical information from
    ref.~\cite{Khodjamirian:2010vf} for $q^2 \leq 1$ GeV$^2$.}
\label{tab:hfit}
\end{table}

\begin{figure}[!htbp]
\centering

\subfigure{\includegraphics[width=.4\textwidth]{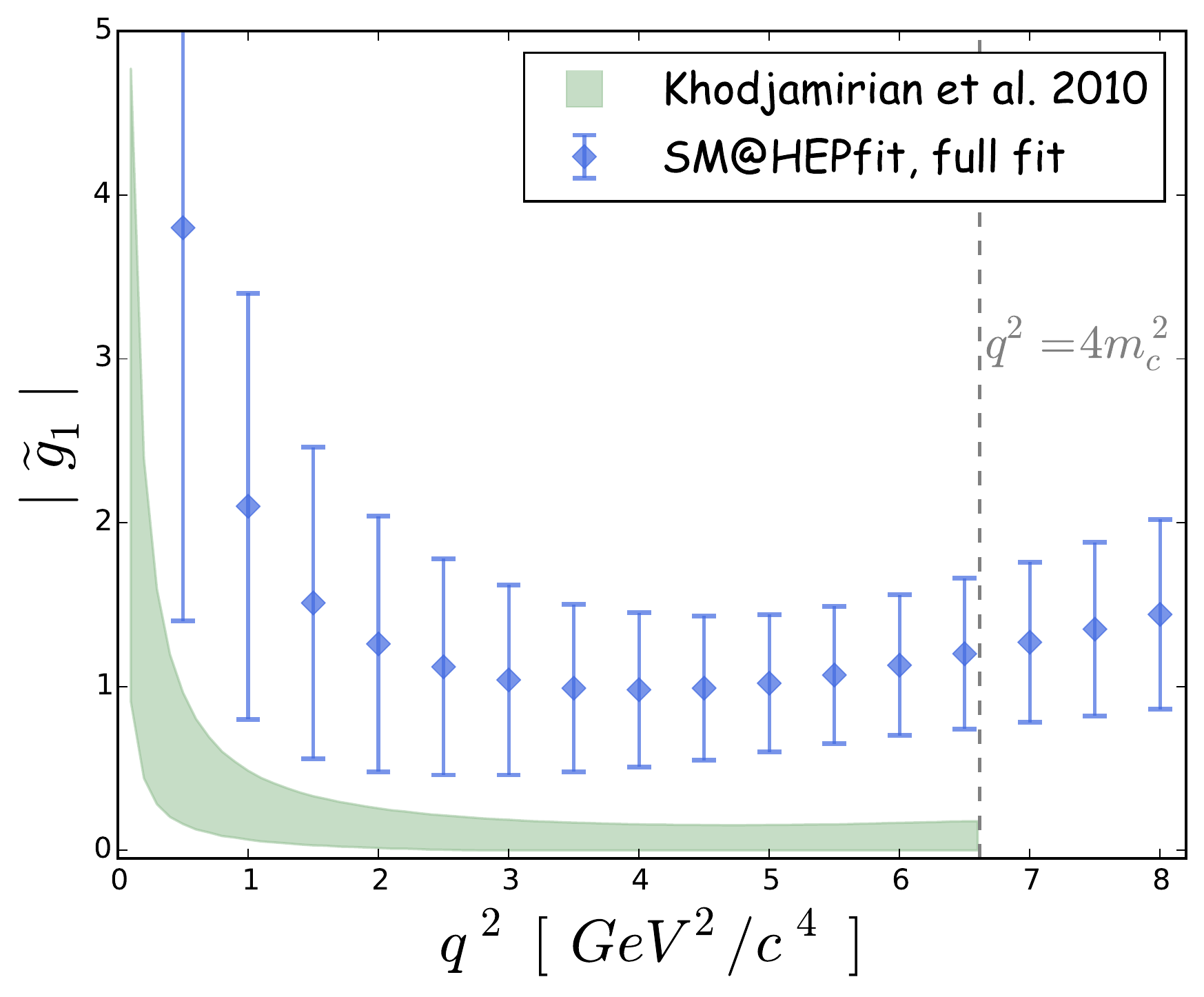}}
\subfigure{\includegraphics[width=.4\textwidth]{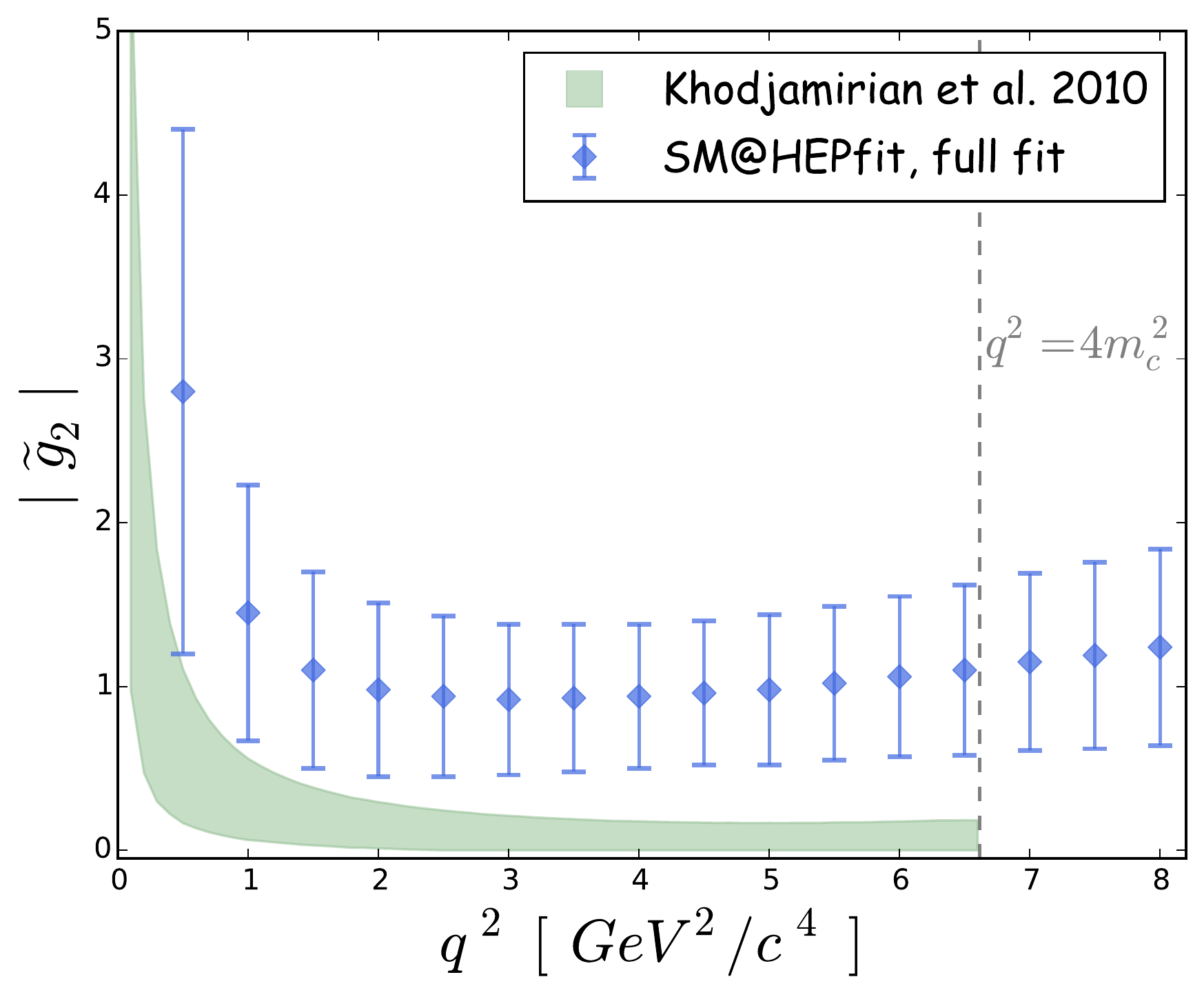}}
\subfigure{\includegraphics[width=.4\textwidth]{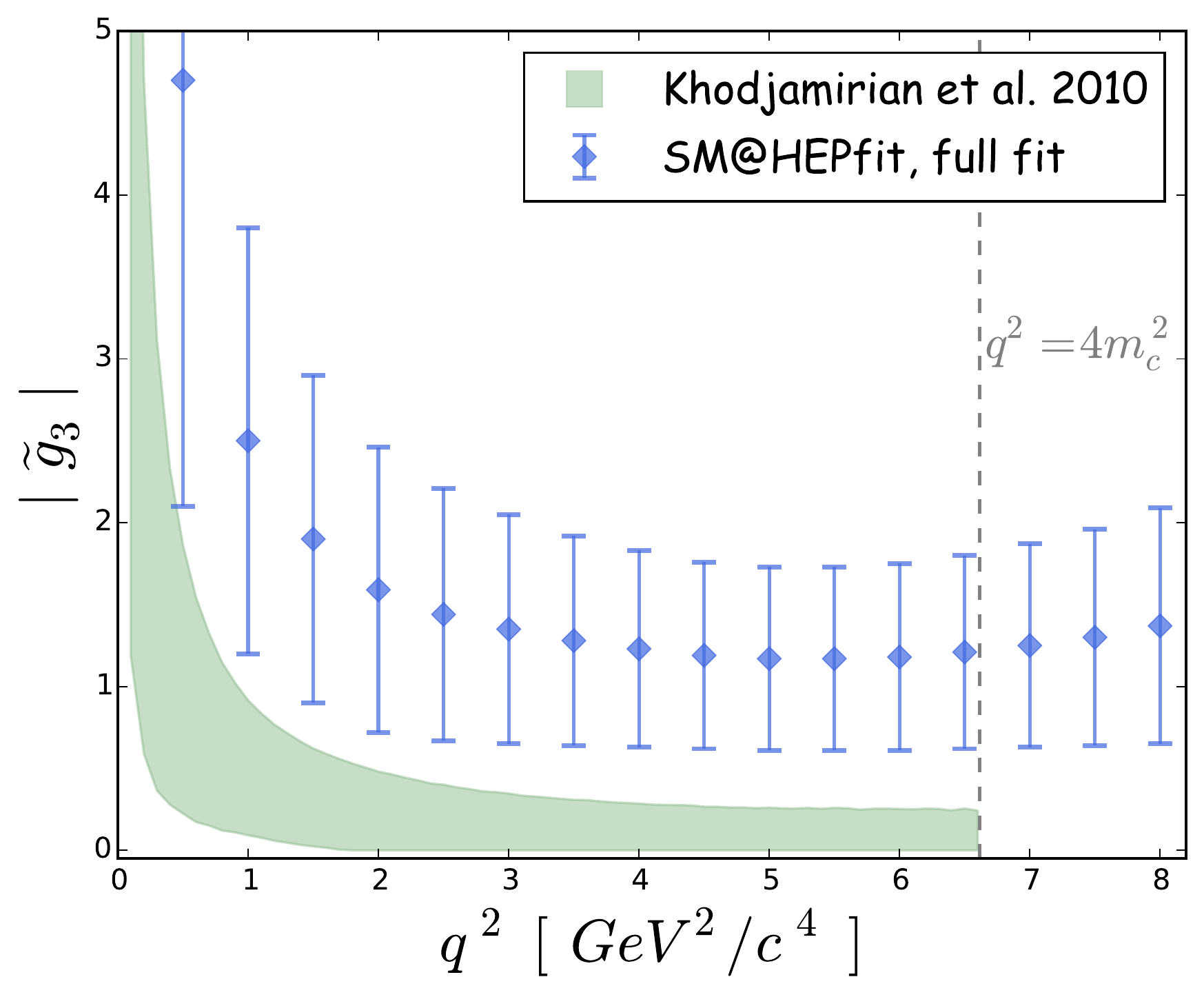}}

\caption{\textit{Same plots as in figure~\ref{fig:gtildes} obtained without using the
results of ref.~\cite{Khodjamirian:2010vf} for $q^2\leq 1$ GeV$^2$ in the fit.}}
\label{fig:gtildesnoK}
\end{figure}

The reader may wonder how the results presented so far depend on our
assumptions on the size and shape of nonfactorizable power
corrections. To elucidate this interesting point, we performed a
number of tests and cross-checks. Let us summarize our findings here
and relegate detailed numerical results to Appendix
\ref{sec:other}. If we do not use the numerical information from
ref.~\cite{Khodjamirian:2010vf}, we obtain (as expected) an even
better fit of experimental data (see tables~\ref{tab:mumunoK} and \ref{tab:eenoK}, and figure~\ref{fig:fullfitnoK}) with a completely
reasonable posterior for the power corrections, reported in
table~\ref{tab:hnoK} and in figure~\ref{fig:gtildesnoK}. It is evident
that the SM calculation supplemented with purely data-driven
nonfactorizable power corrections of the expected order of magnitude
is fully compatible with the data. In this case, however, the
determination from data of the $\tilde{g_i}$ functions is less
precise, and no firm conclusion can be drawn on the size of the
$h_\lambda^{(2)}$ term.

Finally, for the sake of comparison, we also present in Appendix
\ref{sec:other} the results obtained adopting the phenomenological
model of ref.~\cite{Khodjamirian:2010vf} for the $q^2$ dependence of
the power corrections, although we consider this model to be
inadequate for $q^2 \sim 4 m_c^2$ as discussed in
section~\ref{sec:pow}. In this case, we reproduce the results in the
literature, with large deviations in several angular observables (see
tables~\ref{tab:mumuallK} and \ref{tab:eeallK}, and
figure~\ref{fig:fullfitallK}). For
completeness, we also report in the same Appendix the results of a fit
assuming vanishing $h_\lambda^{(2)}$, i.e.\ hadronic corrections fully
equivalent to a shift in $C_{7,9}$ (tables~\ref{tab:mumunoq4}
and \ref{tab:eenoq4}, and figure~\ref{fig:fullfitnoq4}).

We close this section by comparing the above scenarios using the
Information Criterion \cite{IC,MR2027492}, defined as
\begin{equation}
  \label{eq:IC}
  \mathit{IC} = -2 \overline{\log L} + 4 \sigma^2_{\log L}\,,
\end{equation}
where $\overline{\log L}$ is the average of the log-likelihood and
$\sigma^2_{\log L}$ is its variance. Preferred models are expected
to give smaller IC values.
If we ignore the constraints from the
calculation in ref.~\cite{Khodjamirian:2010vf}, we obtain
$\mathit{IC} \sim 72$; using the calculation of
ref.~\cite{Khodjamirian:2010vf} at $q^2 \leq 1$ GeV$^2$ yields
$\mathit{IC} \sim 78$; doing the same but dropping the
$h_\lambda^{(2)}$ terms gives $\mathit{IC} \sim 81$, while using the
model of ref.~\cite{Khodjamirian:2010vf} over the full $q^2$ range
yields $\mathit{IC} \sim 111$. This confirms that the phenomenological
model proposed in ref.~\cite{Khodjamirian:2010vf} does not give a
satisfactory description of experimental data, while the Standard
Model supplemented with the hadronic corrections in
eq.~(\ref{eq:hlambda}) provides a much better fit, even when the
results of ref.~\cite{Khodjamirian:2010vf} at $q^2 \leq 1$ GeV$^2$ are
used. In this case, a nonvanishing $q^{4}$ term is preferred.  

\FloatBarrier

\begin{table}[!htbp]
\centering
\begin{tabular}{|c|c|c|}
\hline
&&\\[-3mm]
\textbf{Parameter} & \textbf{Absolute value} & \textbf{Phase} (rad) \\[1mm]
\hline
&&\\[-3mm]
$ h_0^{(0)} $ &
$ (5.8\pm 2.1)\cdot10^{-4} $
&
$ \phantom{-}3.54 \pm 0.56 $
\\
$ h_0^{(1)} $ &
$ (2.9\pm 2.1)\cdot10^{-4}$
&
$ \phantom{-}0.2\pm 1.1$
\\
$ h_0^{(2)} $ &
$ (3.4 \pm 2.8)\cdot10^{-5}$
&
$ -0.4 \pm 1.7$
\\
[1mm]
\hline
&&\\[-3mm]
$ h_+^{(0)} $ &
$ (4.0 \pm 4.0)\cdot10^{-5} $
&
$ \phantom{-}0.2 \pm 1.5 $
\\
$ h_+^{(1)} $ &
$ (1.4 \pm 1.1)\cdot10^{-4}$
&
$ \phantom{-}0.1 \pm 1.7$
\\
$ h_+^{(2)} $ &
$ (2.6 \pm 2.0)\cdot10^{-5}$
&
$ \phantom{-}3.8 \pm 1.3$
\\
[1mm]
\hline
&&\\[-3mm]
$ h_-^{(0)} $ &
$ (2.5 \pm 1.5)\cdot10^{-4} $
&
$ \phantom{-}1.85 \pm 0.45 \cup 4.75 \pm 0.75 $
\\[1mm]
$ h_-^{(1)} $ &
$ (1.2\pm 0.9)\cdot10^{-4}$
&
$ -0.90 \pm 0.70 \cup 0.80 \pm 0.80$
\\[1mm]
$ h_-^{(2)} $ &
$ (2.2 \pm 1.4)\cdot10^{-5}$
&
$\phantom{-} 0.0 \pm 1.2$
\\[1mm]
\hline
\end{tabular}
\caption{\textit{Results for the parameters defining the
nonfactorizable power corrections $h_{\lambda}$ obtained without
using the numerical information from
ref.~\cite{Khodjamirian:2010vf}.}}
\label{tab:hnoK}
\end{table}

\FloatBarrier

\section{Impact of improved measurements}
\label{sec:future}
In this section, we study how our determination of $h_\lambda(q^2)$ would improve if all
experimental errors in Table~\ref{tab:mainmumu} were improved by an order of magnitude, keeping
fixed the central values of the hadronic parameters. 
\begin{table}[!htbp]
\centering
\begin{tabular}{|c|c|c|c|c|}
\hline
&\multicolumn{2}{c|}{using ref.~\cite{Khodjamirian:2010vf} at $q^2<1$ GeV$^2$}
&\multicolumn{2}{c|}{not using ref.~\cite{Khodjamirian:2010vf}}\\
\hline
&&&&\\[-3mm]
\textbf{Parameter} & \boldmath$\frac{\delta\,\mathrm{abs}}{\mathrm{abs}}$ &
 \boldmath$\delta\,\mathrm{arg\; (rad)}$ &\boldmath$\frac{\delta\,\mathrm{abs}}{\mathrm{abs}}$ &
 \boldmath$\delta\,\mathrm{arg\; (rad)}$\\[1mm]
\hline
&&&&\\[-3mm]
$ h_0^{(0)} $ &
$ \pm10\% $
&
$ \pm 0.07 $
&
$ \pm10\% $
&
$ \pm 0.09 $

\\
$ h_0^{(1)} $ &
$ \pm 20\%$
&
$ \pm 0.2$
&
$ \pm 20\%$
&
$ \pm 0.3$
\\
$ h_0^{(2)} $ &
$ \pm 30\%$
&
$ \pm 0.3$
&
$ \pm 30\%$
&
$ \pm 0.4$
\\
[1mm]
\hline
&&&&\\[-3mm]
$ h_+^{(0)} $ &
$ \pm 80\% $
&
$ \pm 1.4 $
&
$ \pm 90\% $
&
$ \pm 1.4 $
\\
$ h_+^{(1)} $ &
$\pm 70\%$
&
$ \pm 1.6$
&
$\pm 60\%$
&
$ \pm 1.4$
\\
$ h_+^{(2)} $ &
$ \pm 30\%$
&
$ \pm 0.4$
&
$ \pm 30\%$
&
$ \pm 0.3$
\\
[1mm]
\hline
&&&&\\[-3mm]
$ h_-^{(0)} $ &
$\pm 40\% $
&
$ \pm 0.8 $
&
$\pm 50\% $
&
$ \pm 1.0 $
\\[1mm]
$ h_-^{(1)} $ &
$\pm 30\%$
&
$ \pm 0.5$
&
$\pm 30\%$
&
$ \pm 0.5$
\\[1mm]
$ h_-^{(2)} $ &
$ \pm 14\%$
&
$\pm 0.1$
&
$ \pm 14\%$
&
$\pm 0.2$
\\[1mm]
\hline
\end{tabular}
\caption{\textit{Results for the parameters defining the
nonfactorizable power corrections $h_{\lambda}$ obtained using experimental
errors reduced by one order of magnitude. }}
\label{tab:future}
\end{table}

We show the results for the coefficients $h_\lambda^{(0,1,2)}$ in Table~\ref{tab:future}.
There is a significant reduction of the uncertainty on the coefficients $h_0^{(0,1,2)}$ and on
$h_\pm^{(2)}$. Furthermore, the dependence of the fit on the theoretical estimate of
ref.~\cite{Khodjamirian:2010vf} is removed to a large extent. This exercise shows that future
measurements, depending of course on their central values, could allow for an unambiguous determination of the $q^4$ terms in $h_\lambda$, even without theoretical input.

\section{Conclusions}
\label{sec:conclusions}

In this work, we critically examined the theoretical uncertainty in
the SM analysis of $B \to K^* \ell^+ \ell^-$ decays, with particular
emphasis on the nonfactorizable corrections in the region of
$q^2 \lesssim 4 m_c^2$. Using all available theoretical information
within its domain of validity we performed a fit to the experimental
data and found no significant discrepancy with the SM. This requires
the presence of sizable, yet perfectly acceptable, nonfactorizable
power corrections. Assuming the validity of the QCD sum rules estimate
of these power corrections at $q^2 \leq 1$ GeV$^2$, we observe a $q^2$
dependence of the nonfactorizable contributions (in particular a
nonvanishing $h_-^{(2)}$), which disfavours their interpretation as a
shift of the SM Wilson coefficients at more than $95.45\%$
probability. A fit performed without using any theoretical estimate of
the nonfactorizable corrections yields a range for these contributions
larger than, but in the same ballpark of, the QCD sum rule
calculation. In this case, unfortunately, no conclusion on the
presence of $q^4$ terms in $h_\lambda$ can be drawn. We conclude that
no evidence of CP-conserving NP contributions to the Wilson
coefficients $C_{7,9}$ can be inferred from these decays unless a
theoretical breakthrough allows us to obtain an accurate estimate of
nonfactorizable power corrections and to disentangle possible NP
contributions from hadronic uncertainties. Nevertheless, an improved
set of measurements could possibly clarify the issue of the $q^2$ dependence
of $h_\lambda$.

Of course, there might be
other measurements, such as $R_K$~\cite{Aaij:2014ora}, hinting at
possible NP contributions which may well play a role also in
$B\to K^* \ell^+ \ell^-$. In this case, a global fit could benefit
also from the information provided by $B\to K^* \ell^+ \ell^-$
decays~\cite{Beaujean:2013soa,Hurth:2013ssa,Altmannshofer:2014rta,Mandal:2014kma,Jager:2014rwa,Altmannshofer:2015sma,Straub:2015ica,Mandal:2015bsa,Descotes-Genon:2015uva}.

\acknowledgments The research leading to these results has received
funding from the European Research Council under the European Union's
Seventh Framework Programme (FP/2007-2013) / ERC Grant Agreements
n. 279972 ``NPFlavour'' and n. 267985 ``DaMeSyFla''. M.C. is
associated to the Dipartimento di Matematica e Fisica, Universit{\`a}
di Roma Tre, and E.F. and L.S. are associated to the Dipartimento di
Fisica, Universit{\`a} di Roma ``La Sapienza''. We thank J. Matias for
pointing out a numerical error in our evaluation of $S_4$ in the first
version of this manuscript. It is a pleasure to thank T. Blake,
C. Bobeth, J. Camalich, G. D'Agostini, S. J{\"a}ger, A. Khodjamirian,
N. Serra, V. Vagnoni, J. Virto, M. Wingate, and R. Zwicky for
interesting discussions.

\appendix

\section{Form factors}
\label{sec:FF}
There have been some recent developments in the computation of the
form factor in both the large and small recoil regions. In the low
$q^2$ regime the form factors derived using LCSR
\cite{Ball:2004rg,Khodjamirian:2006st} have been recomputed with more
precise hadronic inputs for $q^2 = 0$. The extrapolation of the form
factors into the finite $q^2$ region below 10 GeV$^2$ is now done with
a new parametrization~\cite{Straub:2015ica}, as opposed to the old one
found in \cite{Ball:2004rg}. This parametrization is akin to what has
been used by the lattice group \cite{Horgan:2013hoa,Horgan:2015vla}
for their computations of the form factors in the high $q^2$ region
and is adopted to follow the explicit symmetry relations that need to
be imposed on the form factors at the lower kinematic endpoint. The
other new development is that the parametrization now comes with a
full correlation matrix that we use in our fits. In this section we
shall briefly outline these developments so as to make the
presentation comprehensive.

In the helicity basis the seven $B\to V$ form factors, with $V$ being
a vector meson, can be written in terms of those in the transversality
basis
\begin{eqnarray}
V_0\left( q^{2}\right) &=& \frac{1}{2m_V\lambda^{1/2}(m_B + m_V)} [(m_B + m_V)^2(m_B^2 - q^2 - m_V^2)A_1\left( q^{2}\right) -\lambda A_2\left( q^{2}\right)]\,, \nonumber\\
V_{\pm}\left( q^{2}\right) &\:\:=\:\:& \frac{1}{2} \bigg[ \Big( 1 + \frac{m_V}{m_B} \Big) A_1\left( q^{2}\right) \mp \frac{\lambda^{1/2}}{m_B(m_B + m_V)} V\left( q^{2}\right) \bigg]\,, \nonumber\\
T_0\left( q^{2}\right) &=& \frac{m_B}{2m_V\lambda^{1/2}} \bigg[ (m_B^2 + 3m_V^2 - q^2)T_2\left( q^{2}\right) - \frac{\lambda}{m_B^2 - m_V^2}T_3\left( q^{2}\right) \bigg]\,, \nonumber\\
T_{\pm}\left( q^{2}\right) &=& \frac{m_B^2 - m_V^2}{2m_B^2}T_2\left( q^{2}\right) \mp \frac{\lambda^{1/2}}{2m_B^2}T_1\left( q^{2}\right)\,, \nonumber\\
S\left( q^{2}\right) &=& A_0\left( q^{2}\right)\,.
\label{eq:formfactors}
\end{eqnarray}
Adopting the notation of \cite{Horgan:2013hoa} one can redefine
\begin{eqnarray}
A_{12}(q^2) \;\;&=&\;\;  \frac{ \left(m_B+m_{V}\right){}^2 \left(m_B^2-m_{V}^2-q^2\right)A_1(q^2) -\lambda(q^2) A_2(q^2)  }{16 m_B m_{V}^2 \left(m_B+m_{V}\right)}\,,
\nonumber \\ 
T_{23}(q^2)\;\;&=&\;\;\frac{ \left(m_B^2-m_{V}^2\right) \left(m_B^2+3 m_{V}^2-q^2\right)T_2(q^2)-\lambda(q^2)  T_3(q^2)}{8 m_B m_{V}^2 \left(m_B-m_{V}\right)}\,.
\end{eqnarray}
The form factors $\tilde{V}^0$ and $\tilde{T}^0$ that appear in the helicity amplitudes are defined as 
\begin{equation}
\tilde{V}^0(q^2) = \frac{4m_V}{\sqrt{q^2}}A_{12}(q^2)\;\;\mathrm{and}\;\; \tilde{T}^0(q^2)=\frac{2\sqrt{q^2}m_V}{m_B(m_B + m_V)}T_{23}(q^2).
\end{equation}
The rest of the helicity form factors are defined as
\begin{eqnarray}
\tilde{V}_{L\pm}(q^2)\;\;&=&\;\;-\tilde{V}_{R\mp}(q^2)=V_\pm(q^2)\,,\nonumber\\
\tilde{T}_{L\pm}(q^2)\;\;&=&\;\;-\tilde{T}_{R\mp}(q^2)=T_\pm(q^2)\,,\nonumber\\
\tilde{S}_L(q^2)\;\;&=&\;\;-\tilde{S}_R(q^2)=S(q^2)\,.
\end{eqnarray}
There are some symmetry relations between the form factors at $q^2=0$. These relations are used in deriving the parametric fits in \cite{Straub:2015ica} resulting in a correlation between the different form factors which we use in our computation of the observables. These can be written as
\begin{equation}
A_{12}(0) =  \frac{m_B^2 - m_{K^*}^2}{8 m_B m_{K^*}} A_0(0)\;\;\mathrm{and}\;\;T_1(0) = T_2(0). 
\label{eq:symmetry}
\end{equation}
The form factors can now be parametrized in terms of $z(t)$ defined as
\begin{equation}
z(t) = \frac{\sqrt{t_+-t}-\sqrt{t_+-t_0}}{\sqrt{t_+-t}+\sqrt{t_+-t_0}}\,,
\end{equation}
with 
\begin{equation}
t_\pm=(m_B\pm m_V)^2,\;\;  t_0=t_+(1-\sqrt{1-t_-/t_+})\;\;\mathrm{and}\;\; t = q^2\,.
\end{equation}
The fit function used in \cite{Straub:2015ica} fits the form factors with the expansion
\begin{equation}
F_i(q^2) = P_i(q^2) \sum_k \alpha_k^i \,\left[z(q^2)-z(0)\right]^k\,,
\label{eq:SSE}
\end{equation}
where $P_i(q^2)=(1-q^2/m_{R,i}^2)^{-1}$. The central values of the
parameters $\alpha^i_k$ along with the errors and correlations can be
found in the ancillary files in the arXiv entry of
\cite{Straub:2015ica}.\footnote{We use the fit based on LCSR results
  only.} The vales of $m_{R,i}$ corresponding to the
first resonance in the spectrum can be found in table 3 of
\cite{Straub:2015ica}.

\section{Helicity amplitudes in the Standard Model}
\label{sec:helas}

Since our analysis primarily focuses on the SM, we shall present here
the helicity amplitudes that are relevant for this analysis. The
entire list of helicity amplitudes including the chirality flipped
contributions can be found in \cite{Jager:2012uw} from where we derive
our notation. The most significant amongst the helicity amplitudes are
the vector and axial ones. The psuedoscalar one gets contributions
from SM but is suppressed by the mass of the lepton and hence is
numerically significant only in the lowest $q^2$ bin. The scalar
helicity amplitude does not get any contribution from the SM. The
tensor helicity amplitudes will not be considered here since they are
missing in the literature and addressing that is out of the scope of
this work. However, their expected contribution to the observables is
not significant \cite{Jager:2012uw}. Stripping the relevant helicity
amplitudes to the bare minimum relevant for our SM computation we
have:\footnote{While we do not present the entire basis here for
  clarity, \HEP\ has all of those encoded in it.}

\begin{eqnarray}
H_V^{\lambda} &\:\:=\:\:& -i N \left\{C^{\mathrm{eff}}_9\tilde{V}_{L\lambda}  + \frac{m_B^2}{q^2} \left[\frac{2m_b}{m_B}C_7^{\mathrm{eff}}\tilde{T}_{L\lambda}  - 16\pi^2h_{\lambda} \right]\right\},\nonumber\\
H_A^{\lambda} &=& -i N C_{10}\tilde{V}_{L\lambda}, \nonumber\\
H_P &=& i N \frac{2m_{l}m_b}{q^2}  C_{10} \left( \tilde{S}_{L} - \frac{m_s}{m_B}\tilde{S}_{R} \right) \label{Hp},
\label{eq:helamp}
\end{eqnarray}
where 
\begin{equation}
\label{N}
N = -\frac{4G_Fm_B}{\sqrt{2}} \frac{e^2}{16\pi^2}\lambda_t
\end{equation}
is a normalisation factor, and $h_{\lambda}$ contains all the
non-factorizable hadronic contributions, as discussed in section~\ref{sec:pow}.\\
Observing now that the radiative decay $B \to V \gamma$ is described
by a subset of the helicity amplitudes involved in the
$B \to V \ell^+\ell^-$ decay, following \cite{Jager:2012uw} it is
possible to write
\begin{eqnarray}
A(\bar B \to V(\lambda) \gamma(\lambda)) \:\:=\;\;\frac{i N m_B^2}{e} \left[\frac{2 m_b}{m_B} C_7 \tilde T_\lambda(q^2=0) - 16 \pi^2 h_\lambda(q^2=0) \right]\,.
\end{eqnarray}
The definitions and values of all the parameters used in this analysis are given in table \ref{Tab:SM}.
\section{Kinematic distribution}
Considering the full decay of the $K^*$ channel
\begin{equation}
\bar{B}(p) \rightarrow \bar{K}^{*}(k)[\to \bar{K}(k_1)\pi(k_2)]\ell^+(q_1)\ell^-(q_2)
\end{equation}
where $\bar{K} = \bar{K}^0$ or $K^-$ and $\pi = \pi^+$ or $\pi^0$ it
is important to define the kinematic variables used since different
conventions can be found in the literature. We define $\phi$ as the
angle between the normals to the planes defined by $K^-\pi^+$ and
$\ell^+\ell^-$ in the B meson rest frame. The angle $\theta_\ell$ is
the angle between the direction of flight of the $\bar{B}$ and the
$\ell^-$ in the dilepton rest frame, and $\theta_K$ is the angle
between the direction of motion of the $\bar{B}$ and the $\bar{K}$ in
the dimeson rest frame (note that $\theta_\ell$ and $\theta_K$ are
defined in the interval $[0,\pi)$ ). Squaring the amplitude and
summing over lepton spins allow us to write the fully differential
decay rate as:
\begin{eqnarray}
\frac{d^{(4)}\Gamma}{dq^2d(\cos{\theta_\ell})d(\cos{\theta_K})d\phi} &\:\:=\:\:& \frac{9}{32\pi} \Big( I_1^s\sin^2{\theta_K} + I_1^c\cos^2{\theta_K} + (I_2^s\sin^2{\theta_K} + I_2^c\cos^2{\theta_K})\cos{2\theta_{\ell}} \nonumber \\
&&+ I_3\sin^2{\theta_K}\sin^2{\theta_{\ell}}\cos{2\phi} + I_4\sin{2\theta_K}\sin{2\theta_{\ell}}\cos{\phi} \notag\\
&& + I_5\sin{2\theta_K}\sin{\theta_{\ell}}\cos{\phi} \nonumber+ (I_6^s\sin^2{\theta_K} + I_6^c\cos^2{\theta_K})\cos{\theta_{\ell}}  \nonumber \\
&& + I_7\sin{2\theta_K}\sin{\theta_{\ell}}\sin{\phi} + I_8\sin{2\theta_K}\sin{2\theta_{\ell}}\sin{\phi} \notag \\
&&+ I_9\sin^2{\theta_K}\sin^2{\theta_{\ell}}\sin{2\phi} \Big) \label{degamma}.
\end{eqnarray}
The angular coefficients $I_i$, as functions of $q^2$, can be
expressed in terms of the helicity amplitudes as\footnote{Again,
  please note that the angular coefficients are only in terms of the
  helicity amplitudes that appear in the SM. For the complete
  expressions see ref.~\cite{Jager:2012uw}.}
\begin{eqnarray}
\label{Istart}
I_1^c &\:\:=\:\:& F \left( \frac{1}{2}\left(|H_V^0|^2 + |H_A^0|^2\right) + |H_P|^2 + \frac{2m_{\ell}^2}{q^2}\left(|H_V^0|^2 - |H_A^0|^2\right) \right), \nonumber\\
I_1^s &=& F \left(  \frac{\beta^2 + 2}{8}\left(|H_V^+|^2 + |H_V^-|^2 + |H_A^+|^2 + |H_A^-|^2\right) + \frac{m_{\ell}^2}{q^2}\left(|H_V^+|^2 - |H_V^-|^2 - |H_A^+|^2 + |H_A^-|^2 \right) \right),\:\:\: \nonumber\\
I_2^c &=& -F \frac{\beta^2}{2}\left(|H_V^0|^2 + |H_A^0|^2\right), \nonumber\\
I_2^s &=& F \frac{\beta^2}{8}\left(\left(|H_V^+|^2 + |H_V^-|^2\right) + \left(|H_A^+|^2 + |H_A^-|^2\right)\right), \nonumber\\
I_3 &=& -\frac{F}{2}\mathrm{Re}\left[H_V^+(H_V^-)^* + H_A^+(H_A^-)^*\right], \nonumber\\
I_4 &=& F\frac{\beta^2}{4}\mathrm{Re}\left[(H_V^+ + H_V^-)(H_V^0)^* + (H_A^+ + H_A^-)(H_A^0)^*\right], \nonumber\\
I_5 &=& F \frac{\beta}{4}\mathrm{Re}\left[(H_V^- - H_V^+)(H_A^0)^* + (H_A^- - H_A^+)(H_V^0)^*\right], \nonumber\\
I_6^s &=& F\beta\mathrm{Re}\big[H_V^-(H_A^-)^* - H_V^+(H_A^+)^*\big], \nonumber\\
I_6^c &=& 0, \nonumber\\
I_7 &=& F \frac{\beta}{2}\mathrm{Im}\left[(H_A^+ + H_A^-)(H_V^0)^* + (H_V^+ + H_V^-)(H_A^0)^*\right],\nonumber \\
I_8 &=& F \frac{\beta^2}{4}\mathrm{Im}\left[(H_V^- - H_V^+)(H_V^0)^* + (H_A^- - H_A^+)(H_A^0)^*\right], \nonumber\\
I_9 &=& F \frac{\beta^2}{4}\mathrm{Im}\left[H_V^+ (H_V^-)^* + H_A^+ (H_A^-)^*\right],
\label{eq:angcoeff}
\end{eqnarray}
where 
\begin{eqnarray}
F \;\;&=&\;\; \frac{\lambda^{1/2}\beta q^2}{3 \times 2^5\pi^3m_B^3}\mathrm{BR}(K^* \to K\pi), \quad \beta = \sqrt{1 - \frac{4m_{\ell}^2}{q^2}}, \nonumber\\
\lambda \;\;&=&\;\;  m_B^4 + m_V^4 + q^4 -2(m_B^2m_V^2 +m_B^2q^2 + m_V^2q^2).
\label{eq:beta}
\end{eqnarray}
For the CP-conjugate decay $B \to K^*\ell^+\ell^-$, the angular coefficients can be defined by
\begin{equation}
I_{1s(c),2s(c),3,4,7} \to \bar{I}_{1s(c),2s(c),3,4,7}, \qquad\qquad I_{5,6s(c),8,9} \to -\bar{I}_{5,6s(c),8,9},
\label{eq:CP}
\end{equation}
when one uses the angles defined as in the $\bar{B}$ decays with $K^- \to K^+$ and with conjugated CKM elements.

\section{Angular observables}
\label{sec:ang}

From the full angular distribution one can define angular observables
in multiple ways. Two different prescriptions have been advocated in
the past \cite{Altmannshofer:2008dz,Matias:2012xw}. While both sets of
definitions are equivalent in their physics content, the two different
sets have been used for experimental analyses
\cite{Aaij:2013iag,Aaij:2013qta,Aaij:2013hha,LHCb:2015dla,Aaij:2015dea,Aaij:2015oid}.
These two definitions can be easily related to each other. Since we
shall present our results cast into both sets it is best to define
both here.

Following \cite{Altmannshofer:2008dz}, one can define
\begin{equation}
S_i = \frac{I_i + \bar{I}_i}{2 \Gamma^\prime}, \qquad \qquad A_i =
\frac{I_i - \bar{I}_i}{2 \Gamma^\prime}.
\end{equation}

The twelve $q^2$-dependent observables $I_i$ derived in the previous
section are all accessible through a full angular analysis of the
$\bar{B} \to \bar{K}^*\ell^+\ell^-$ decay rate. The analysis of the
CP-conjugate decay $B \to K^*\ell^+\ell^-$ gives the same number of
independent observables, so that it is useful to define the following
combinations:
\begin{equation}
\Sigma_i = \frac{I_i + \bar{I}_i}{2}, \qquad \qquad \Delta_i = \frac{I_i - \bar{I}_i}{2}.
\end{equation}

In an attempt to reduce the uncertainties coming from form factors and
hadronic contributions one can define the ratios of the angular
coefficients. However, this comes with a caveat. These observables are
really ``clean'' of uncertainties in their analytic functional form
and when the form factors are assumed to come with small corrections
to the soft form factors in addition to negligible hadronic
contributions. In case the latter assumptions break down, which seems
to be the most likely case, these observables are no longer ``clean''
of uncertainties in the form factor and hadronic
contributions. Nevertheless, one defines the observables

\begin{eqnarray}
P_1 \;\;&=&\;\; \frac{\Sigma_3}{2\Sigma_{2s}}, \qquad P_2 = \frac{\Sigma_{6s}}{8\Sigma_{2s}}, \qquad P_3 = -\frac{\Sigma_9}{4\Sigma_{2s}}, \\
P_4^\prime \;\;&=&\;\; \frac{\Sigma_4}{\sqrt{-\Sigma_{2s}\Sigma_{2c}}}, \;\; P_5^\prime = \frac{\Sigma_5}{2\sqrt{-\Sigma_{2s}\Sigma_{2c}}}, \;\;
P_6^\prime =-\frac{\Sigma_7}{2\sqrt{-\Sigma_{2s}\Sigma_{2c}}}, \;\; P_8^\prime = -\frac{\Sigma_8}{\sqrt{-\Sigma_{2s}\Sigma_{2c}}}. \nonumber
\end{eqnarray}
In addition to these there are the traditional observables, the
branching fraction, the longitudinal component and the
forward-backward asymmetry which can be defined in terms of the
angular coefficients as:
\begin{eqnarray}
\Gamma^\prime \;\;&=&\;\; \frac{1}{2}\frac{d\Gamma + d\bar{\Gamma}}{dq^2} = \frac{1}{4}\left[(3\Sigma_{1c} - \Sigma_{2c}) + 2(3\Sigma_{1s} - \Sigma_{2s})\right], \nonumber\\
F_L \;\;&=&\;\; \frac{3\Sigma_{1c} - \Sigma_{2c}}{4\Gamma^\prime}, \qquad \qquad A_{FB} = -\frac{3\Sigma_{6s}}{4\Gamma^\prime}.
\end{eqnarray}
In the limit $q^2 \gg m_{\ell}^2$ the terms proportional to
$m_{\ell}^2/q^2$ can be dropped from the angular coefficients in
eq.~(\ref{eq:angcoeff}) and the helicity amplitude $H_P\to0$ since it
is proportional to $m_i/q^2$. In this limit there are further
relations connecting the angular coefficients effectively reducing the
number of independent observables. These relations can be written as:
\begin{eqnarray}
\Sigma_{1c} = - \Sigma_{2c}\qquad \mathrm{and} \qquad \Sigma_{1s} = 3\Sigma_{2s}.
\end{eqnarray}
This simplifies the expressions for $F_L$ and $\Gamma^\prime$ to
\begin{eqnarray}
F_L \;\;&=&\;\; \frac{\Sigma_{1c}}{\Gamma^\prime}\qquad \mathrm{and} \qquad \Gamma^\prime = \Sigma_{1c} + 4\Sigma_{2s} \,.
\end{eqnarray}
Experimentally the observables are measured in binned data cut in
regions of $q^2$, the dilepton invariant mass. The translation of the
analytic expressions to the experimentally binned observables is as
follows:
\begin{eqnarray}
\av{P_1}\;\;& = &\;\;\frac{\av{\Sigma_3}}{2\av{\Sigma_{2s}}}, \qquad \av{P_2} = \frac{\av{\Sigma_{6s}}}{8\av{\Sigma_{2s}}},\qquad \av{P_3} = -\frac{\av{\Sigma_9}}{4\av{\Sigma_{2s}}}, \nonumber\\
\av{P_4^\prime}\;\;& =&\;\; \frac{\av{\Sigma_4}}{\sqrt{-\av{\Sigma_{2s}\Sigma_{2c}}}}, \quad \av{P_5^\prime} = \frac{\av{\Sigma_5}}{2\sqrt{-\av{\Sigma_{2s}\Sigma_{2c}}}}, \nonumber\\
\av{P_6^\prime}\;\;& = &\;\;-\frac{\av{\Sigma_7}}{2\sqrt{-\av{\Sigma_{2s}\Sigma_{2c}}}},\quad \av{P_8^\prime}=-\frac{\av{\Sigma_8}}{2\sqrt{-\av{\Sigma_{2s}\Sigma_{2c}}}},
\label{eq:binob}
\end{eqnarray}
where it should be noted that the ratio of the binned angular
coefficients are the relevant rather than the binned ratios since:
\begin{equation}
\av{\Sigma_i} = \int_{q^2_{min}}^{q^2_{max}}\Sigma(q^2)dq^2\,.
\end{equation}
Furthermore, the binned branching fraction, $F_L$ and $A_{FB}$ are defined as:
\begin{equation}
\av{\Gamma^\prime} = \av{ \Sigma_{1c} + 4\Sigma_{2s}}\,,  \qquad \av{F_L} = \frac{\av{3\Sigma_{1c} - \Sigma_{2c}}}{4\av{\Gamma^\prime}}\,,\qquad \av{A_{FB}} = -\frac{3\av{\Sigma_{6s}}}{4\av{\Gamma^\prime}}\,.
\end{equation}

Even though the angular observables built out of the angular
coefficients are measured over bins as we have described, in effect
defeating some of the purpose of being clean that they were originally
advocated for, it is informative to take a look at their analytic form
assuming only SM contributions being present. The extension to the
full expressions will not be presented here as the expressions become
quite lengthy. The simplified forms are given by:

\allowdisplaybreaks{
\footnotesize{
\begin{eqnarray}
P_1=-\frac{2}{1-4\frac{m_{\ell}^2}{q^2}}\frac{\mathrm{Re}\left[\left(C_{10}V_{+}\right)\left(C_{10}V_{-}\right)^*\right]+\mathrm{Re}\left[D_+D_-^*\right]}{\left| C_{10} V_- \right|^2 + \left|C_{10} V_+\right|^2 + \left|D_+\right|^2 + \left|D_-\right|^2}\,,
\end{eqnarray}

\begin{eqnarray}
P_2=\frac{1}{\sqrt{1-4\frac{m_{\ell}^2}{q^2}}}\frac{\mathrm{Re}\left[D_+\left(C_{10} V_+ \right)^* +D_- \left(
   C_{10} V_- \right)^*\right]}{\left| C_{10} V_-\right|^2 + \left| C_{10} V_+\right|^2 + \left|D_+\right|^2 + \left| D_- \right|^2}\,,
\end{eqnarray}

\begin{eqnarray}
P_3=-\frac{\mathrm{Im}\left[\left( C_{10}V_{+}\right)\left(C_{10}V_{-} \right)^*\right]+\mathrm{Im}\left[D_+D_-^*\right]}{\left| C_{10} V_- \right|^2 + \left| C_{10} V_+\right|^2 + \left|D_+\right|^2 + \left|D_-\right|^2}\,,
\end{eqnarray}

\begin{eqnarray}
P^\prime_4=\frac{\mathrm{Re}\left[C_{10}(V_- + V_+) (C_{10} \tilde{V}_0)^*\right] + \mathrm{Re}\left[\left(D_-+D_+\right) D_0^*\right]}{\sqrt{\left(\left|C_{10}\tilde{V}_0\right|^2 +\left|D_0\right|^2\right)\left( \left|C_{10} V_- \right|^2 + \left|C_{10} V_+\right|^2 + \left|D_+\right|^2 + \left|D_-\right|^2\right)}}\,,
\end{eqnarray}

\begin{eqnarray}
P^\prime_5=
-\frac{\mathrm{Re}\left[\left(D_--D_+\right) \left(C_{10} \tilde{V}_0\right)^*\right] +  \mathrm{Re}\left[C_{10}(V_- - V_+) \left(D_0\right)^*\right]}{\sqrt{\left(1 -  \frac{4m_{\ell}^2}{q^2}\right) \left(\left|C_{10}\tilde{V}_0\right|^2 +\left|D_0\right|^2\right)\left( \left|C_{10} V_- \right|^2 + \left|C_{10} V_+\right|^2 + \left|D_+\right|^2 + \left|D_-\right|^2\right)}}\,,
\end{eqnarray}

\begin{eqnarray}
P^\prime_6=
-\frac{\mathrm{Im}\left[\left( D_--D_+\right) \left(C_{10} \tilde{V}_0\right)^*\right] +  \mathrm{Im}\left[C_{10}(V_- - V_+) D_0^*\right]}{\sqrt{\left(1 -  \frac{4m_{\ell}^2}{q^2}\right) \left(\left|C_{10}\tilde{V}_0\right|^2 +\left|D_0\right|^2\right)\left( \left|C_{10} V_- \right|^2 + \left|C_{10} V_+\right|^2 + \left|D_+\right|^2 + \left|D_-\right|^2\right)}}\,,
\end{eqnarray}
}

\begin{eqnarray}
P^\prime_8=\frac{\mathrm{Im}\left[C_{10}(V_- - V_+) (C_{10} \tilde{V}_0)^*\right] + \mathrm{Im}\left[\left(D_--D_+\right) D_0^*\right]}{\sqrt{\left(\left|C_{10}\tilde{V}_0\right|^2 +\left|D_0\right|^2\right)\left( \left|C_{10} V_- \right|^2 + \left|C_{10} V_+\right|^2 + \left|D_+\right|^2 + \left|D_-\right|^2\right)}}\,,
\end{eqnarray}
}
where

\begin{eqnarray}
D_0 &\;\;=\;\;& \frac{m_B^2}{q^2}\left(16\pi^2 h_0(q^2)  - 2\frac{m_b}{m_B} C^{\mathrm{eff}}_7\tilde{T}_0\right) - C^{\mathrm{eff}}_9(q^2) \tilde{V}_0\,,\nonumber\\
D_+ &\;\;=\;\;& \frac{m_B^2}{q^2}\left(16  \pi^2 h_+(q^2) - 2 \frac{m_b}{m_B}  C_7^{\mathrm{eff}}T_+\right)  - C_9^{\mathrm{eff}}(q^2) V_+\,,\nonumber\\
D_- &\;\;=\;\;& \frac{m_B^2}{q^2}\left(16 \pi^2 h_-(q^2) - 2 \frac{m_b}{m_B} C_7^{\mathrm{eff}}T_-\right) - C_9^{\mathrm{eff}}(q^2) V_-\,,
\end{eqnarray}
which are proportional to $H_V^{\lambda}$ given in
eq.~(\ref{eq:helamp}). In the SM $C^{\mathrm{eff}}_7$ and
$C_{10}$ do not pick up any $q^2$ dependence at low energy and remain
purely real. $C^{\mathrm{eff}}_9(q^2)$ is defined as
\begin{equation}
C^{\mathrm{eff}}_9(q^2) = C^{\mathrm{eff}}_9 + Y(q^2)\,,
\end{equation}
where $Y(q^2)$ comes from the perturbative part of the charm loop
contribution \cite{Grinstein:1988me,Misiak:1992bc,Buras:1994dj}. We
emphasize that we do not include the latter contribution in our
definition for $h_{\lambda}$ since it contains non-factorizable
contributions only.

It is instrumental at this point to underline the connection between
the two different sets of observables that are generally advocated in
the literature. There are some simple relations between them in the
$q^2 \gg m_{\ell}^2$ limit. While this limit does not strictly hold in
the lower $q^2$ region it does provide some insight into the way these
sets are connected so we shall collect the formula here
\cite{Matias:2012xw,DescotesGenon:2012zf}.

\begin{eqnarray}
P_1=A_T^{(2)}=\frac{2S_3}{1-F_L},\qquad &&P_2 = -\frac{2}{3}\frac{A_{FB}}{1-F_L},\;\;\qquad\quad P_3=-\frac{S_9}{1-F_L}\nonumber\\
P^\prime_4=\frac{2S_4}{\sqrt{F_L(1-F_L)}},\;\qquad &&P^\prime_5=\frac{S_5}{\sqrt{F_L(1-F_L)}},\qquad P^\prime_6=-\frac{S_7}{\sqrt{F_L(1-F_L)}},\nonumber\\
&&P^\prime_8=-\frac{2S_8}{\sqrt{F_L(1-F_L)}}.
\label{eq:Ps}
\end{eqnarray}
In all the above relations, both the left and the right hand sides
pertain to the definitions of the kinematic variables used in theory
computations.  It should also be noted that due to the difference in
the definitions of the kinematic variable between the convention used
for theory calculation and for experimental measurements at the LHCb,
the numerical results between the two are connected by
\cite{Lyon:2014hpa,Descotes-Genon:2015uva}:
\begin{equation}
P_2^{\mathrm{LHCb}} = -P_2^{\mathrm{T}},\;\;P_3^{\prime\mathrm{LHCb}} = -P_3^{\prime\mathrm{T}} , \;\;P_4^{\prime\mathrm{LHCb}} = -\frac{1}{2}P_4^{\prime\mathrm{T}}\;\;\mathrm{and}\;\;P_8^{\prime\mathrm{LHCb}} = -\frac{1}{2}P_8^{\mathrm{T}}\,,
\label{eq:thex}
\end{equation}
where the superscript T implies theory definitions. While the sign
difference stems from the change in the definition of the kinematic
variables the factors of two come from the difference in the
definitions of the variables themselves.

\section{Tests and Cross-checks}
\label{sec:other}
As explained in section~\ref{sec:fit}, we performed several tests and
cross-checks to assess the dependence of our results on our
assumptions on the size and shape of nonfactorizable power
corrections.

\begin{figure}[!htbp]
\centering

\subfigure{\includegraphics[width=.4\textwidth]{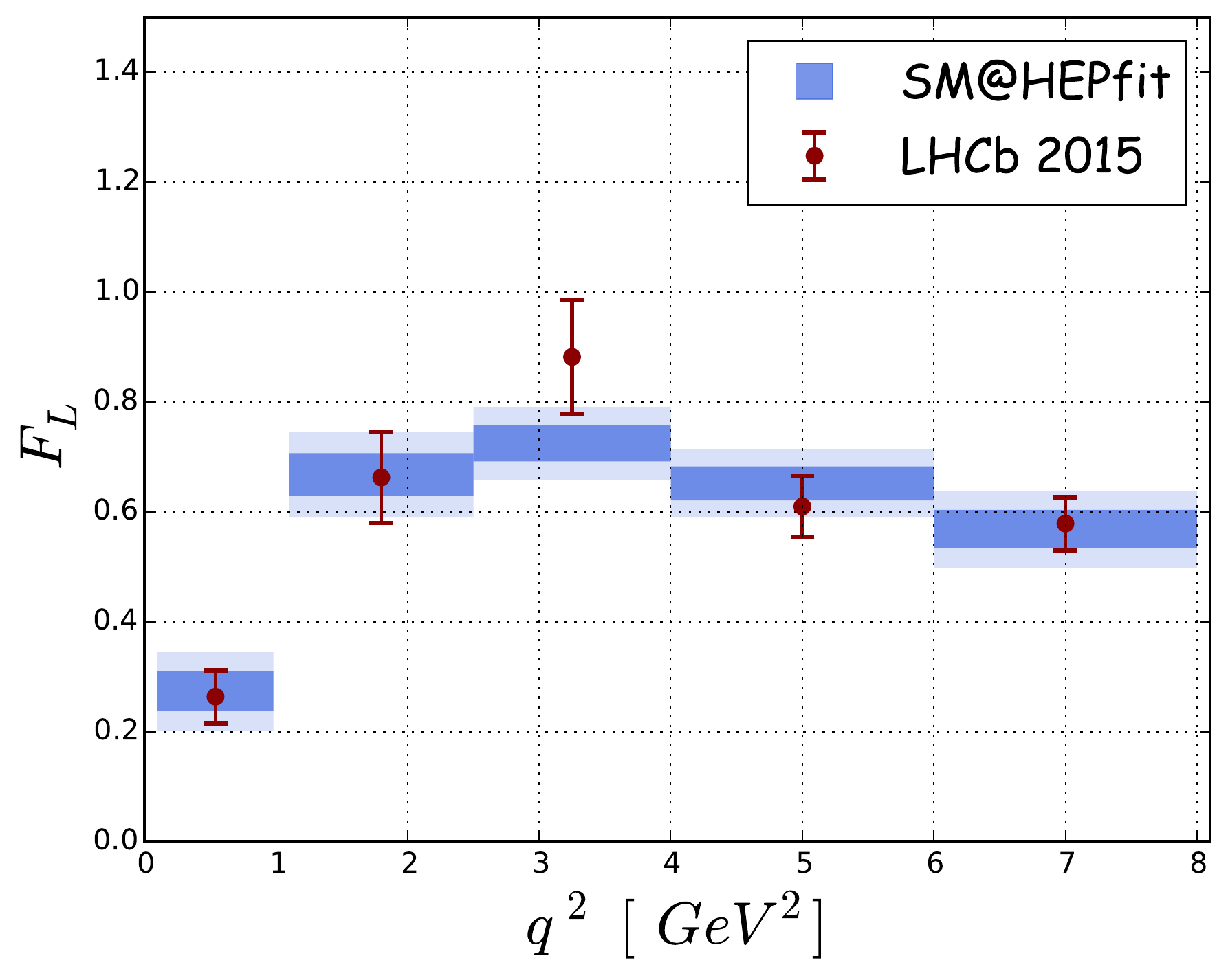}}
\subfigure{\includegraphics[width=.4\textwidth]{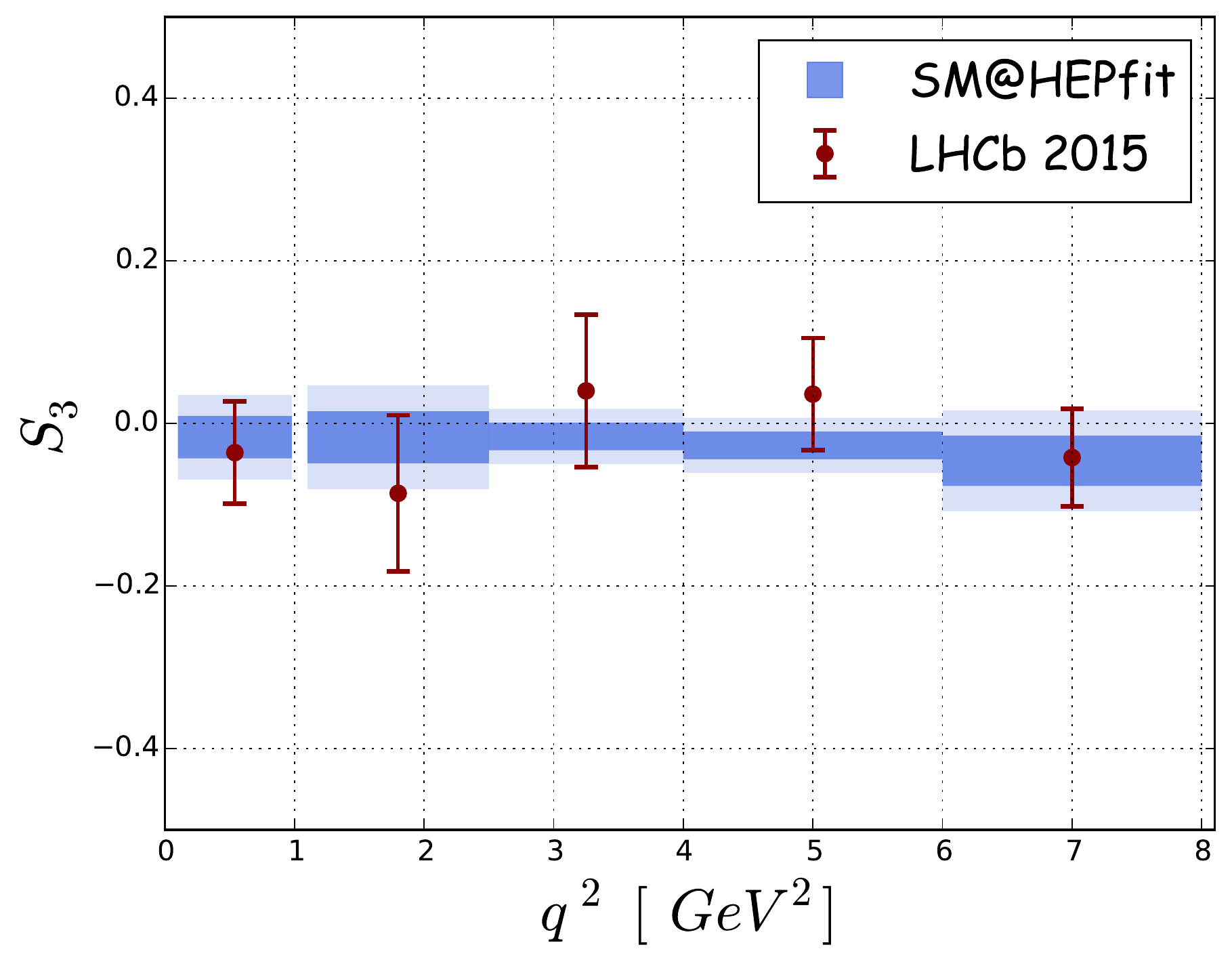}}
\subfigure{\includegraphics[width=.4\textwidth]{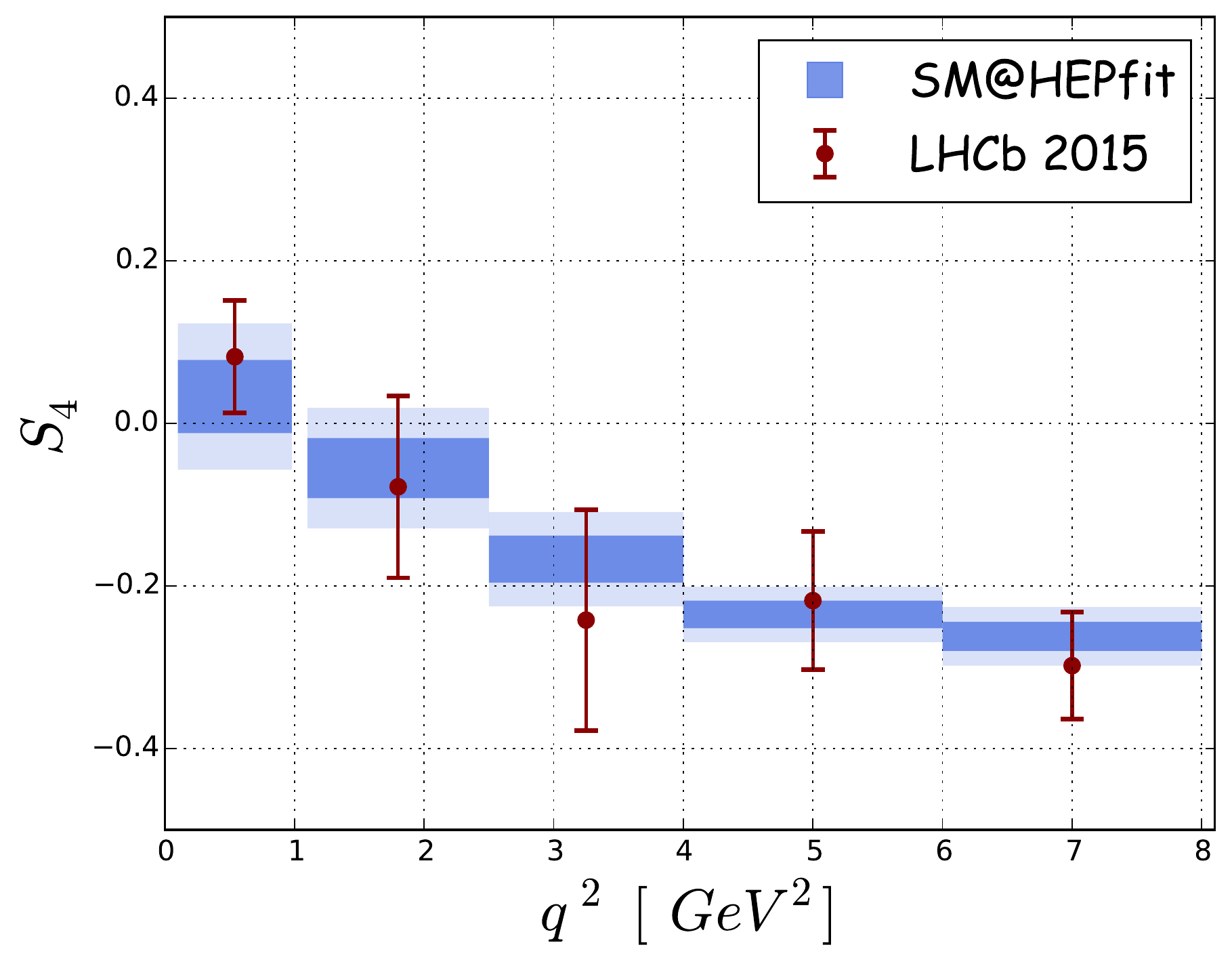}}
\subfigure{\includegraphics[width=.4\textwidth]{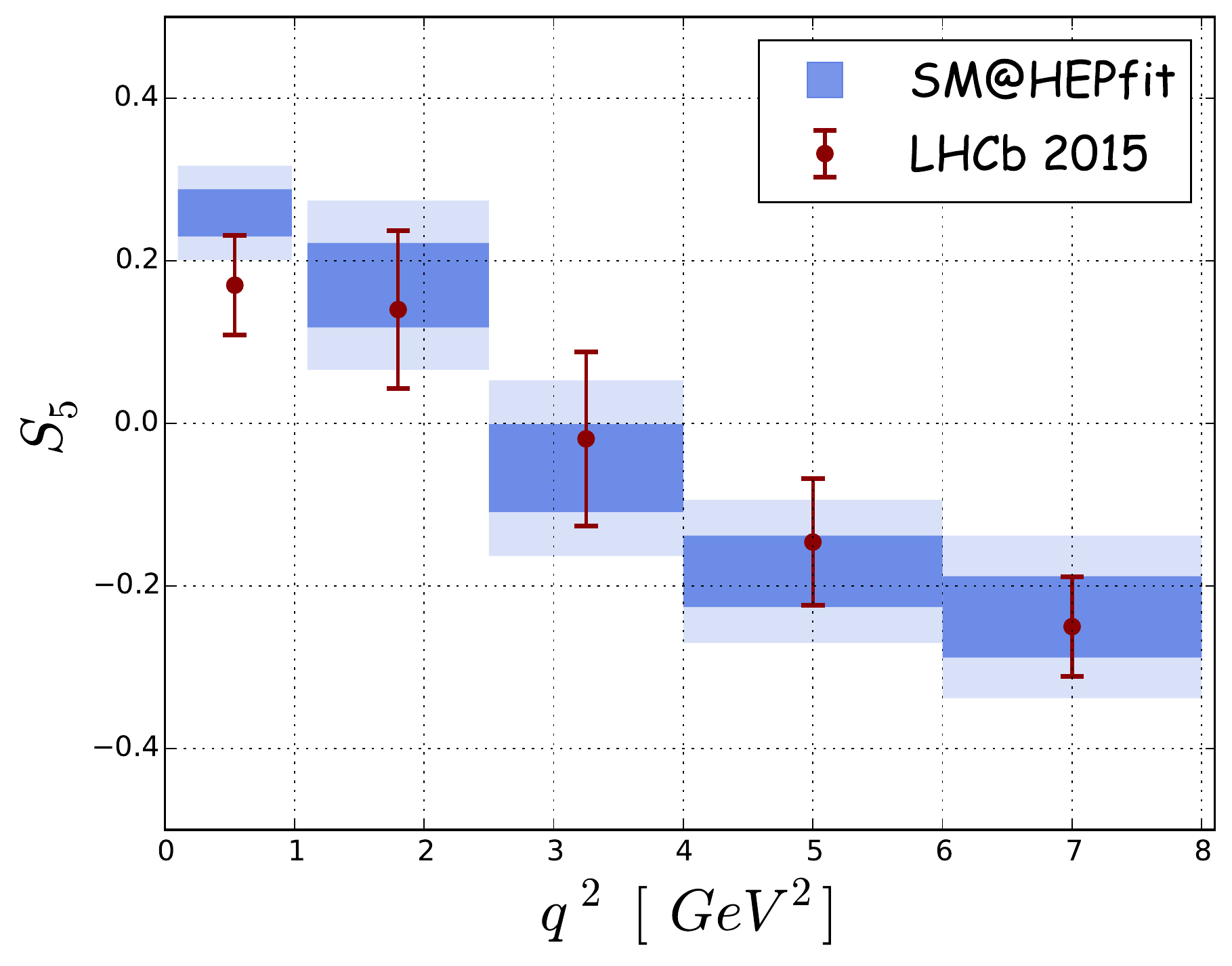}}
\subfigure{\includegraphics[width=.4\textwidth]{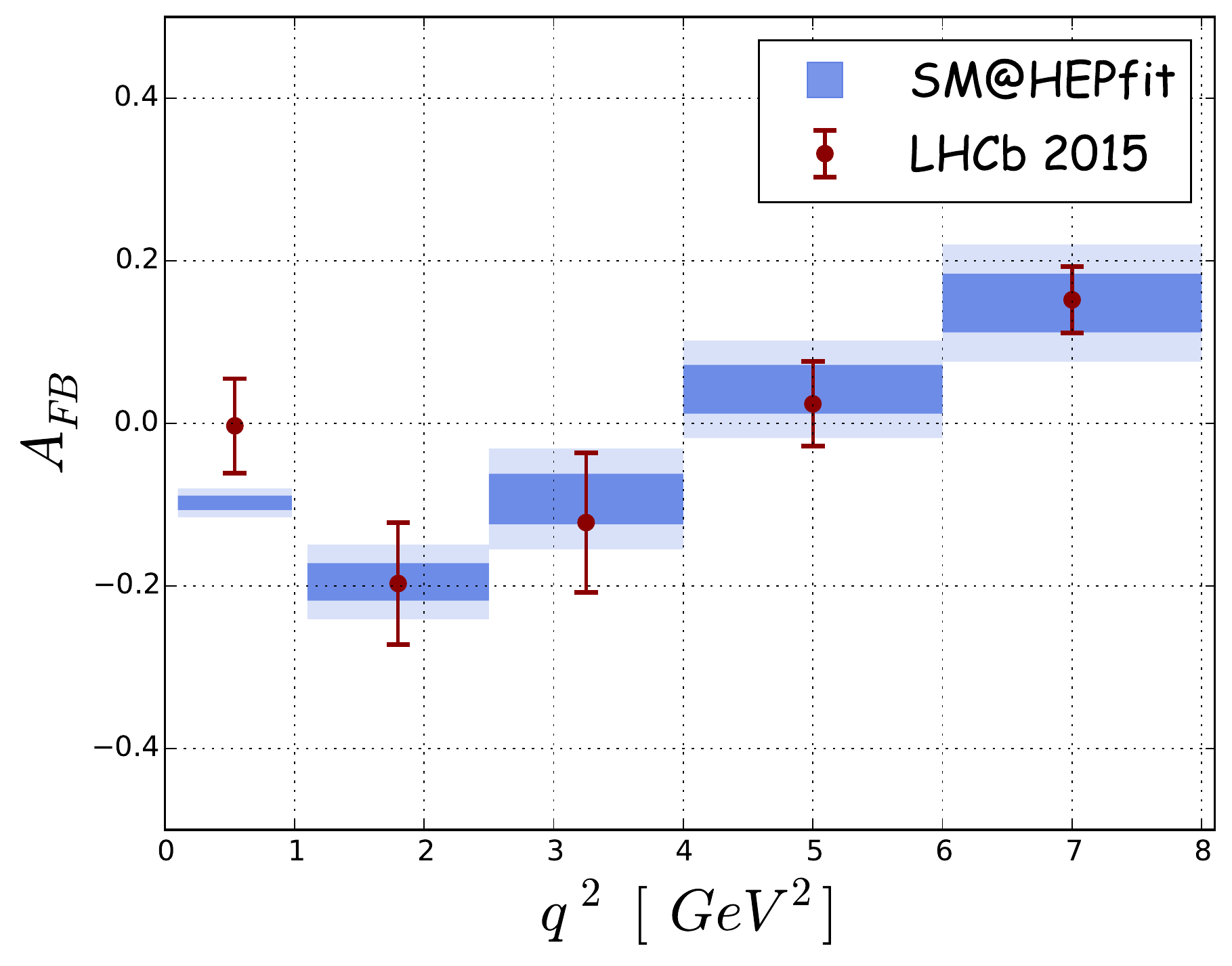}}
\subfigure{\includegraphics[width=.4\textwidth]{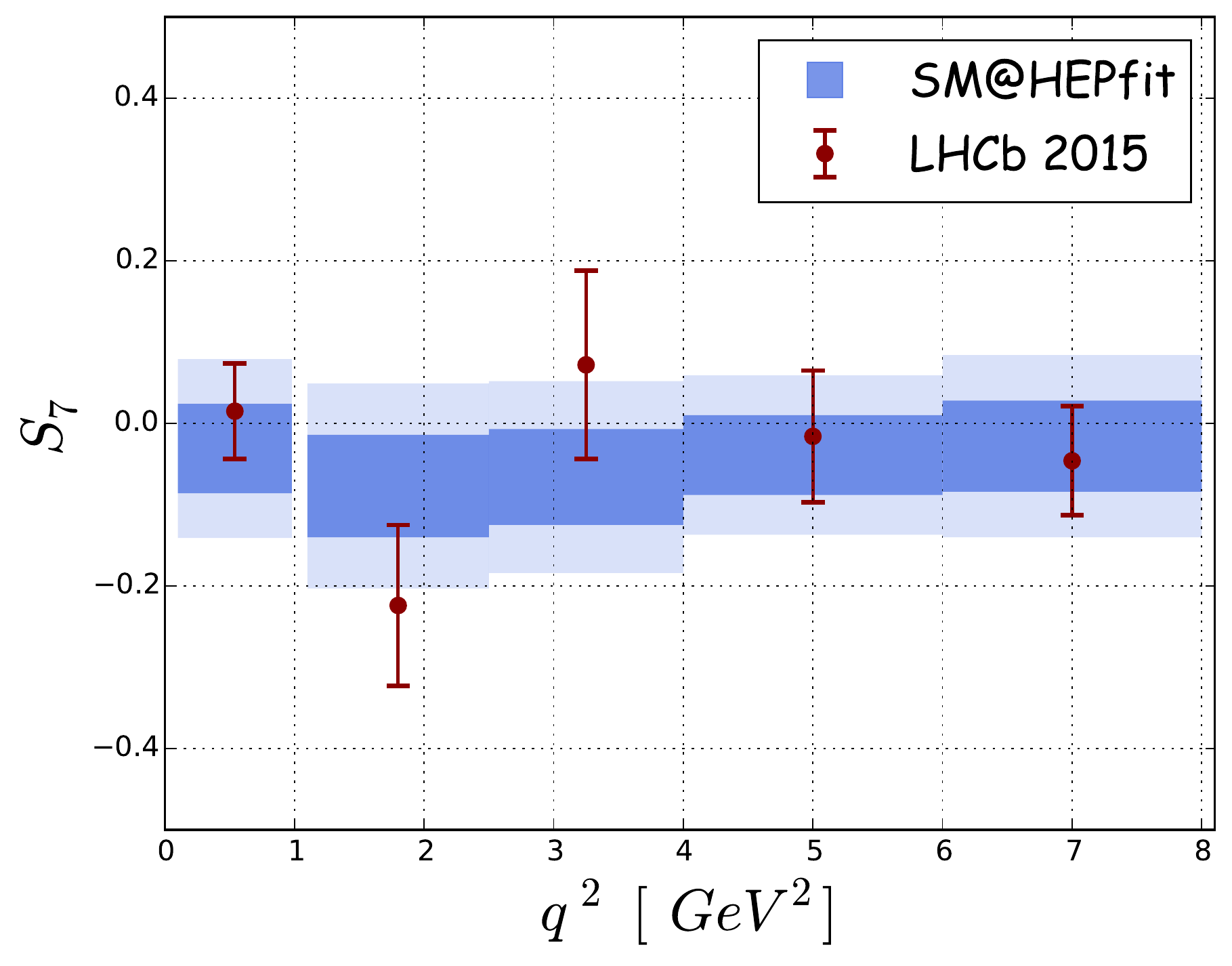}}
\subfigure{\includegraphics[width=.4\textwidth]{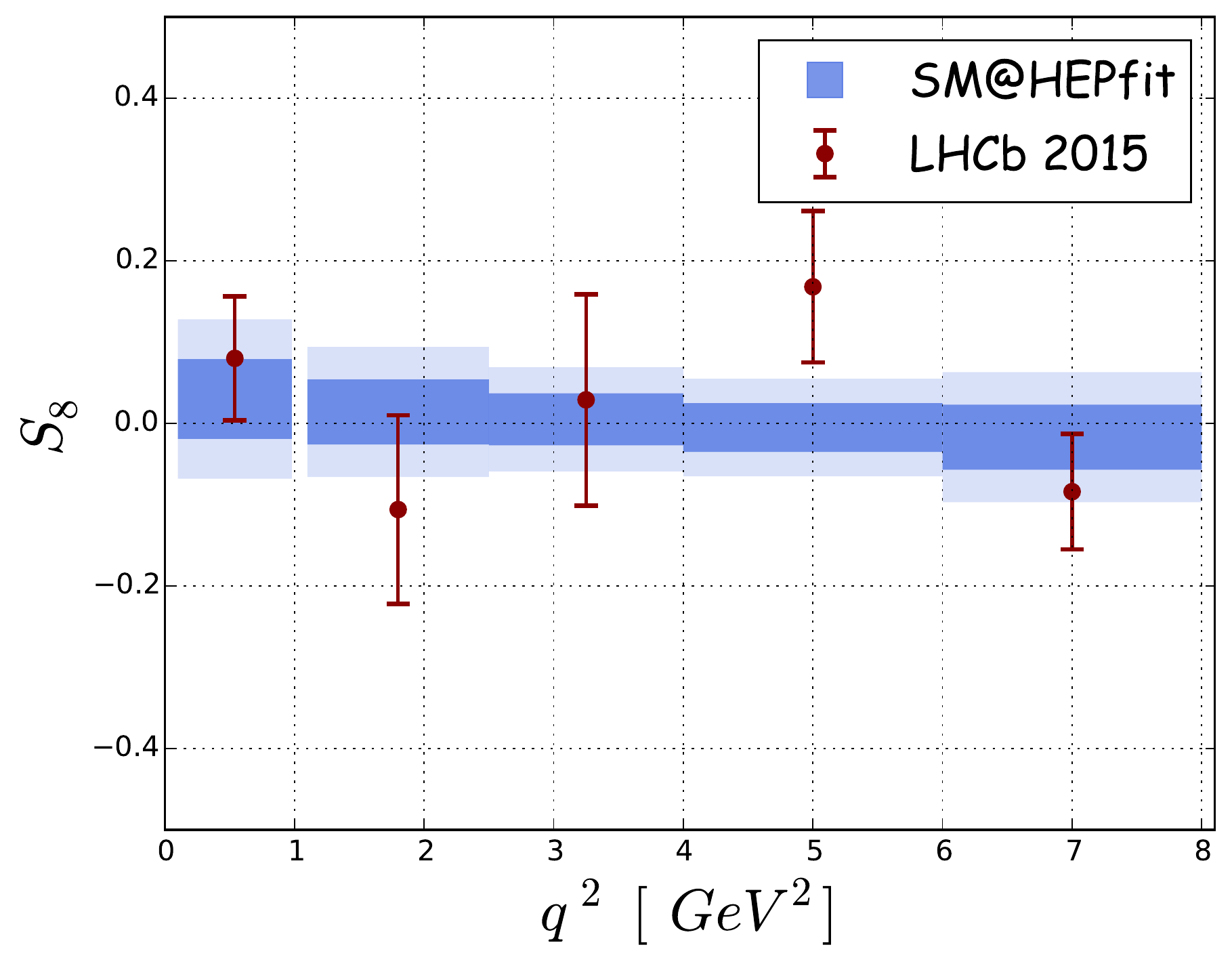}}
\subfigure{\includegraphics[width=.4\textwidth]{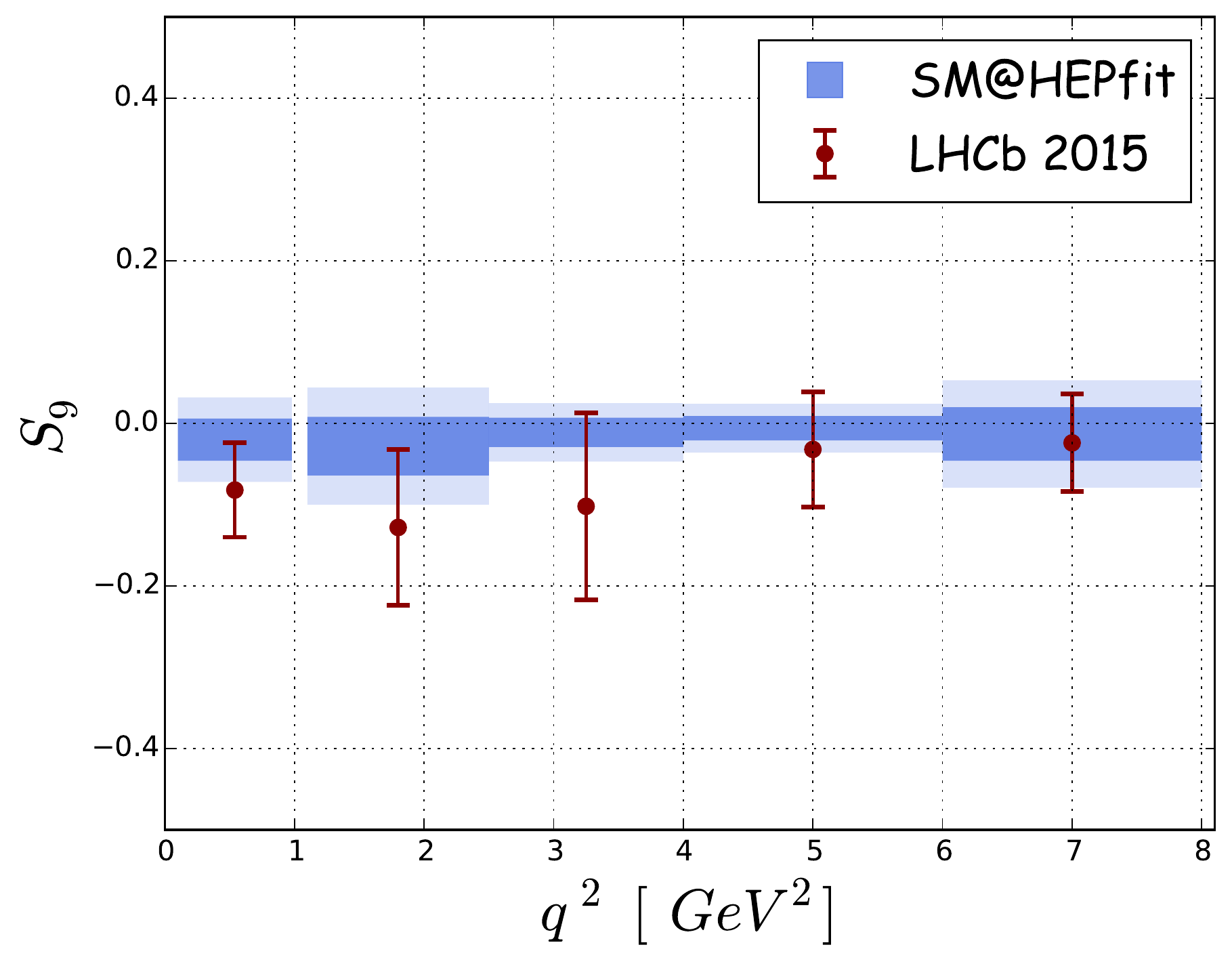}}

\caption{\textit{Results of the full fit and experimental results for
    the $B \to K^* \mu^+ \mu^-$ angular observables obtained without
    using the numerical information from
    ref.~\cite{Khodjamirian:2010vf}.}}
\label{fig:fullfitnoK}
\end{figure}

As a first test, we performed our fit without using the numerical
information from ref.~\cite{Khodjamirian:2010vf}. The results of the
fit for the $B \to K^* \mu^+ \mu^-$ observables are reported in
table~\ref{tab:mumunoK}, while the ones for the $B \to K^* e^+ e^-$
observables are in table~\ref{tab:eenoK}. Plots for the
$B \to K^* \mu^+ \mu^-$ angular observables are shown in
figure~\ref{fig:fullfitnoK}.

\begin{figure}[!htbp]
\centering

\subfigure{\includegraphics[width=.4\textwidth]{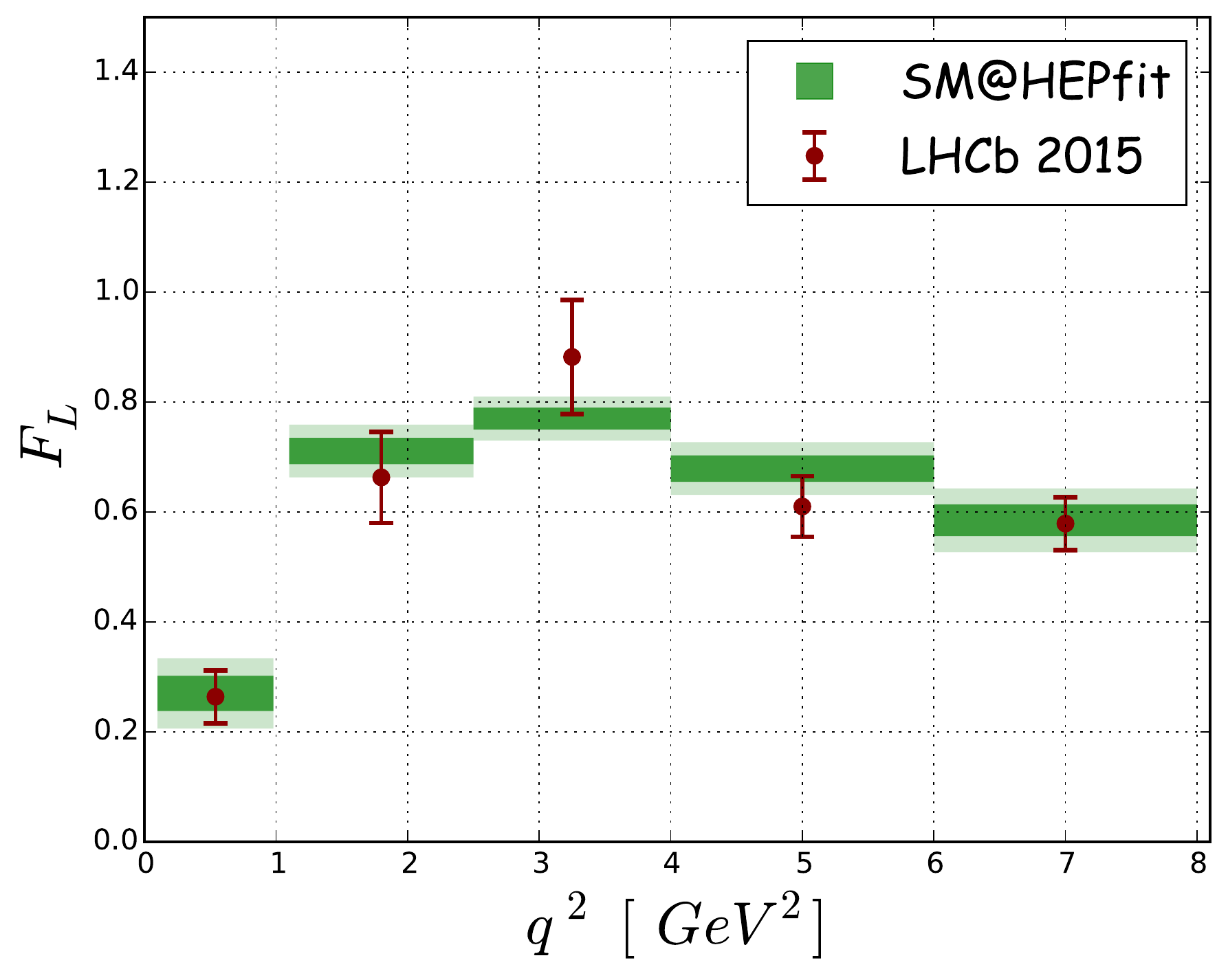}}
\subfigure{\includegraphics[width=.4\textwidth]{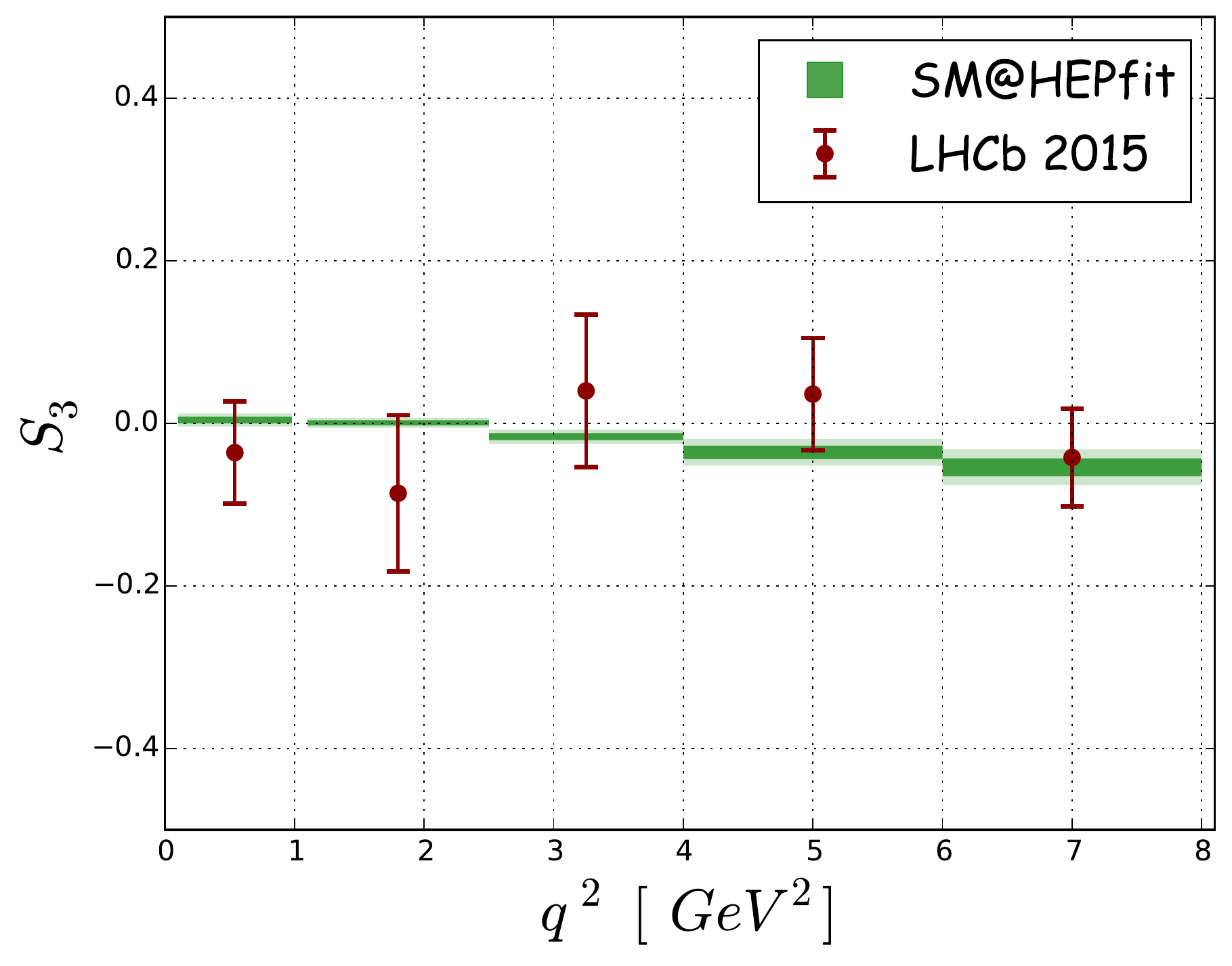}}
\subfigure{\includegraphics[width=.4\textwidth]{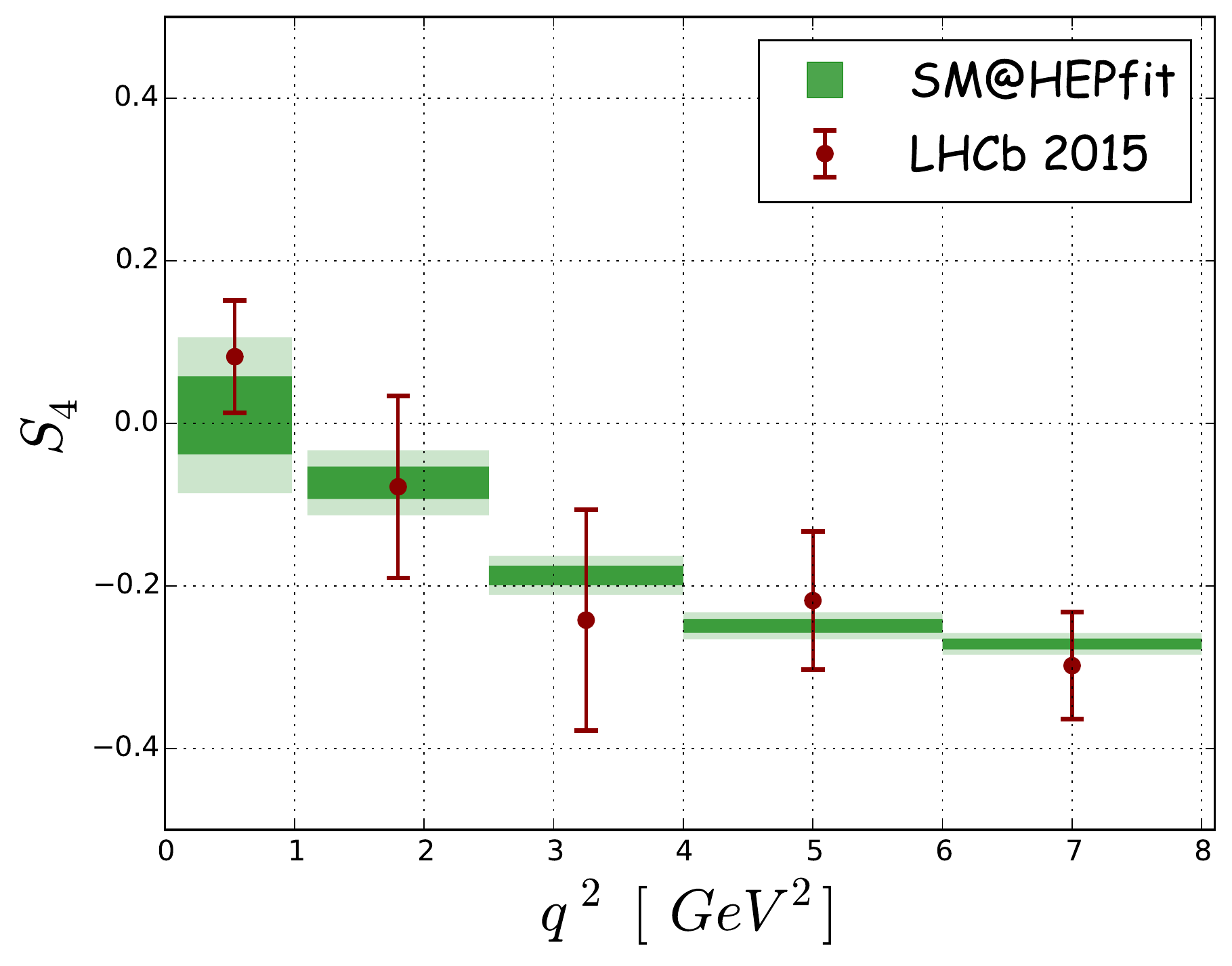}}
\subfigure{\includegraphics[width=.4\textwidth]{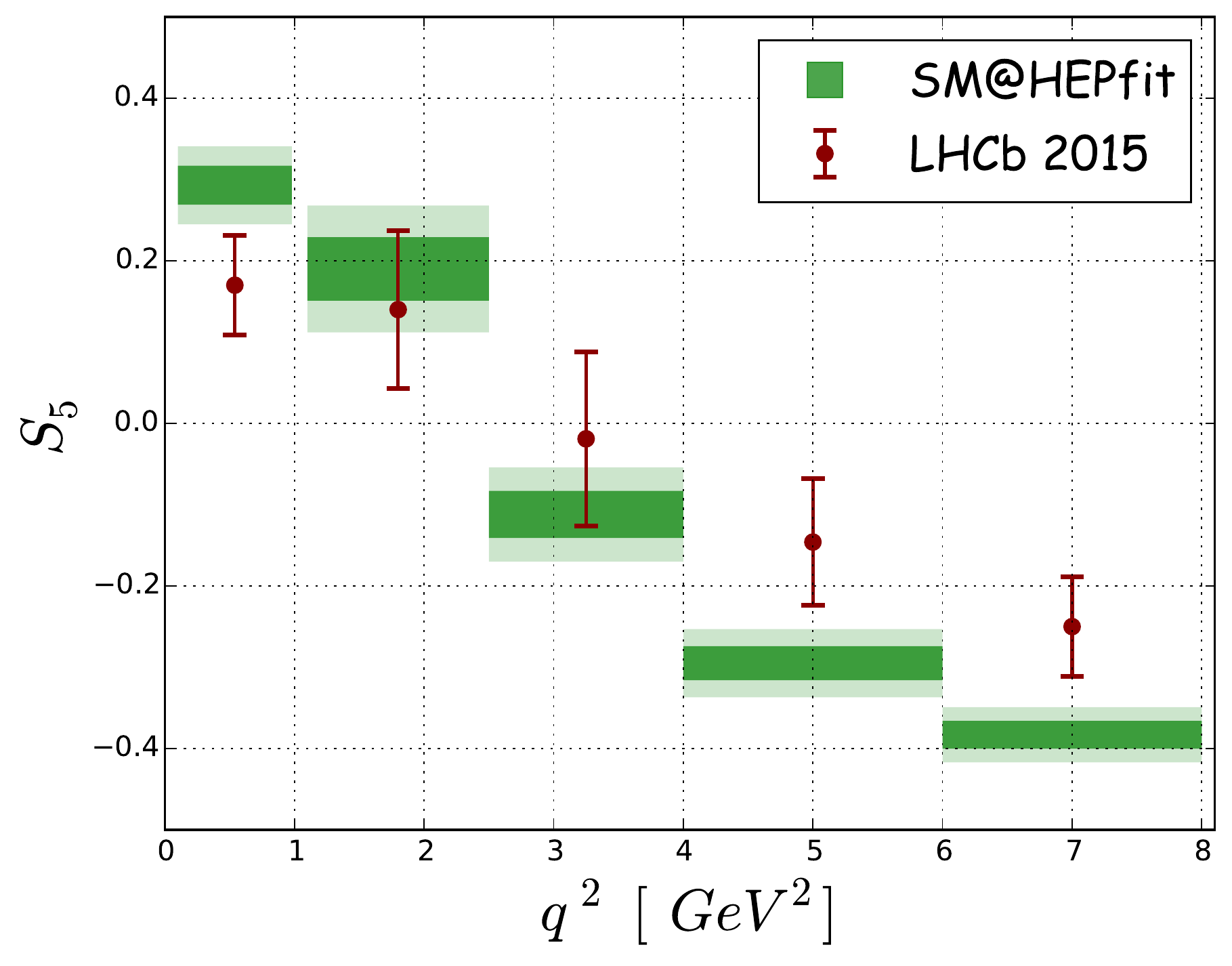}}
\subfigure{\includegraphics[width=.4\textwidth]{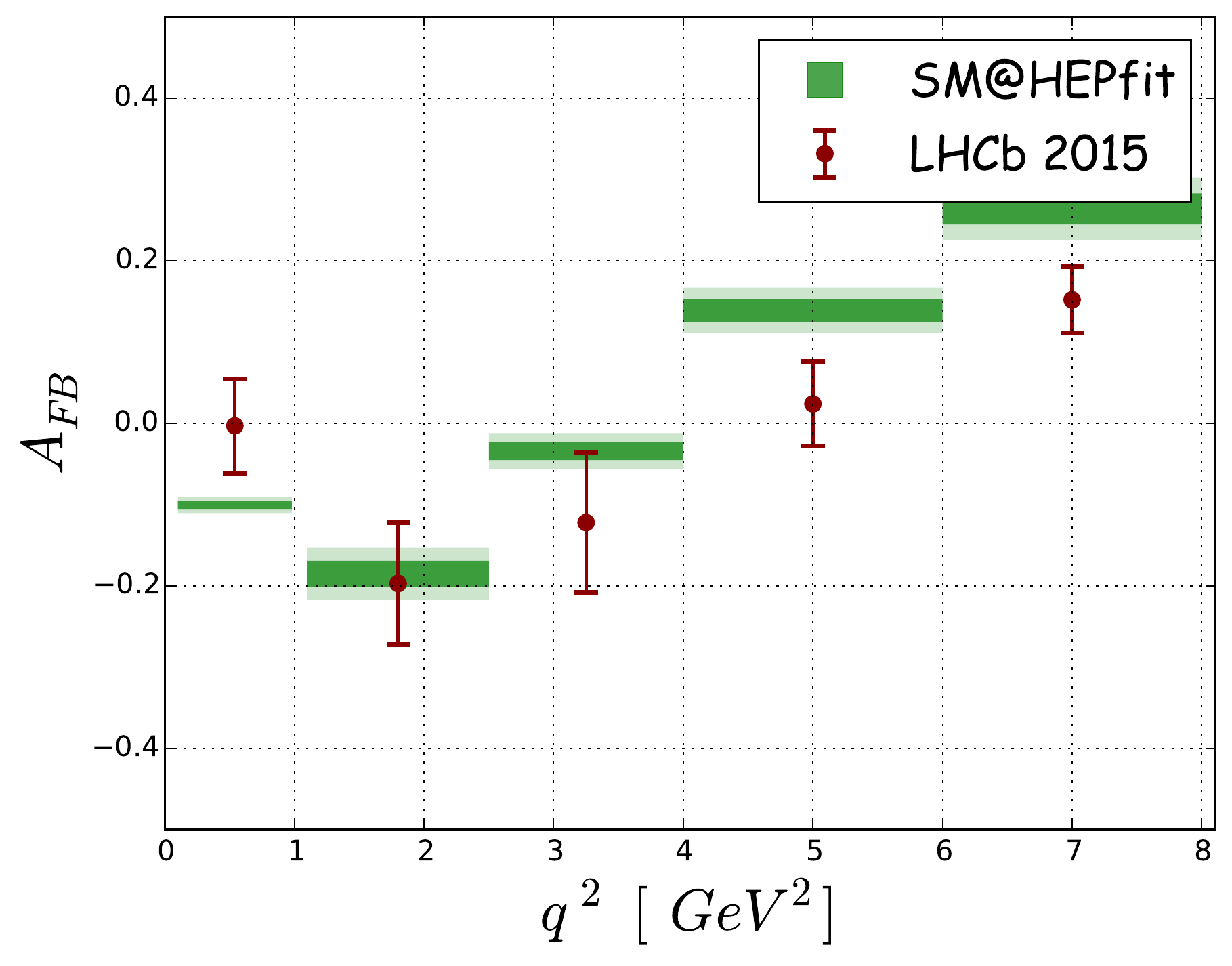}}
\subfigure{\includegraphics[width=.4\textwidth]{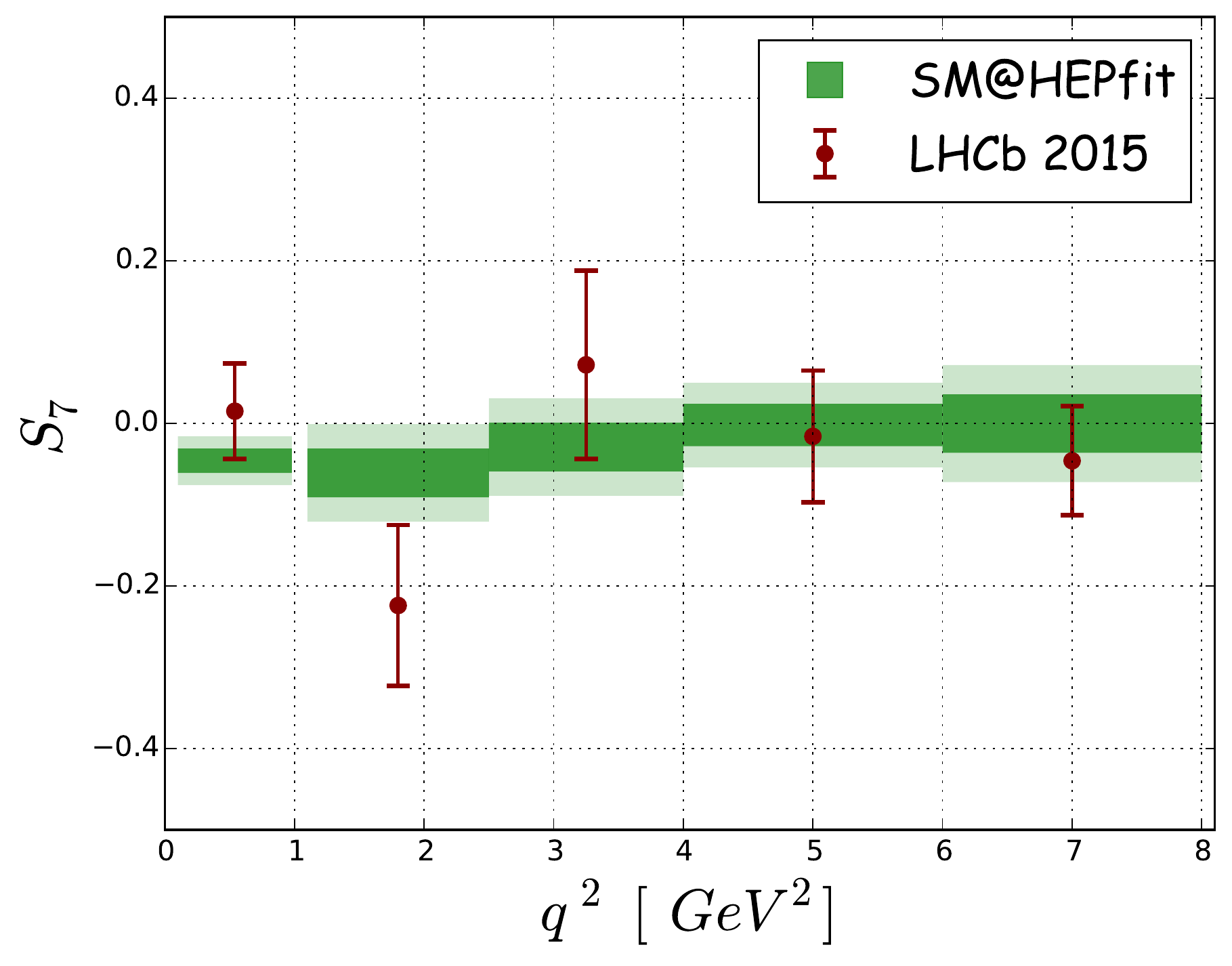}}
\subfigure{\includegraphics[width=.4\textwidth]{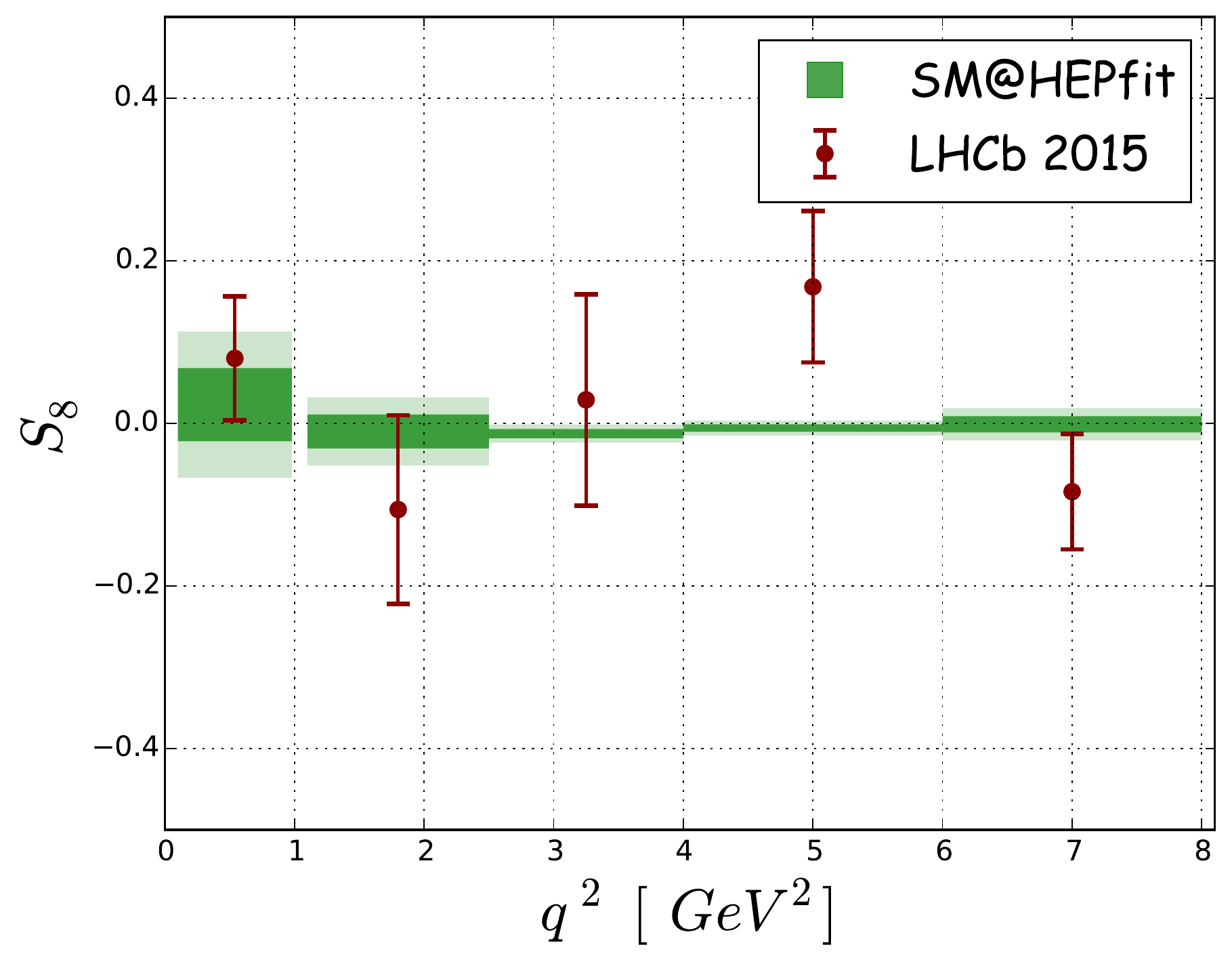}}
\subfigure{\includegraphics[width=.4\textwidth]{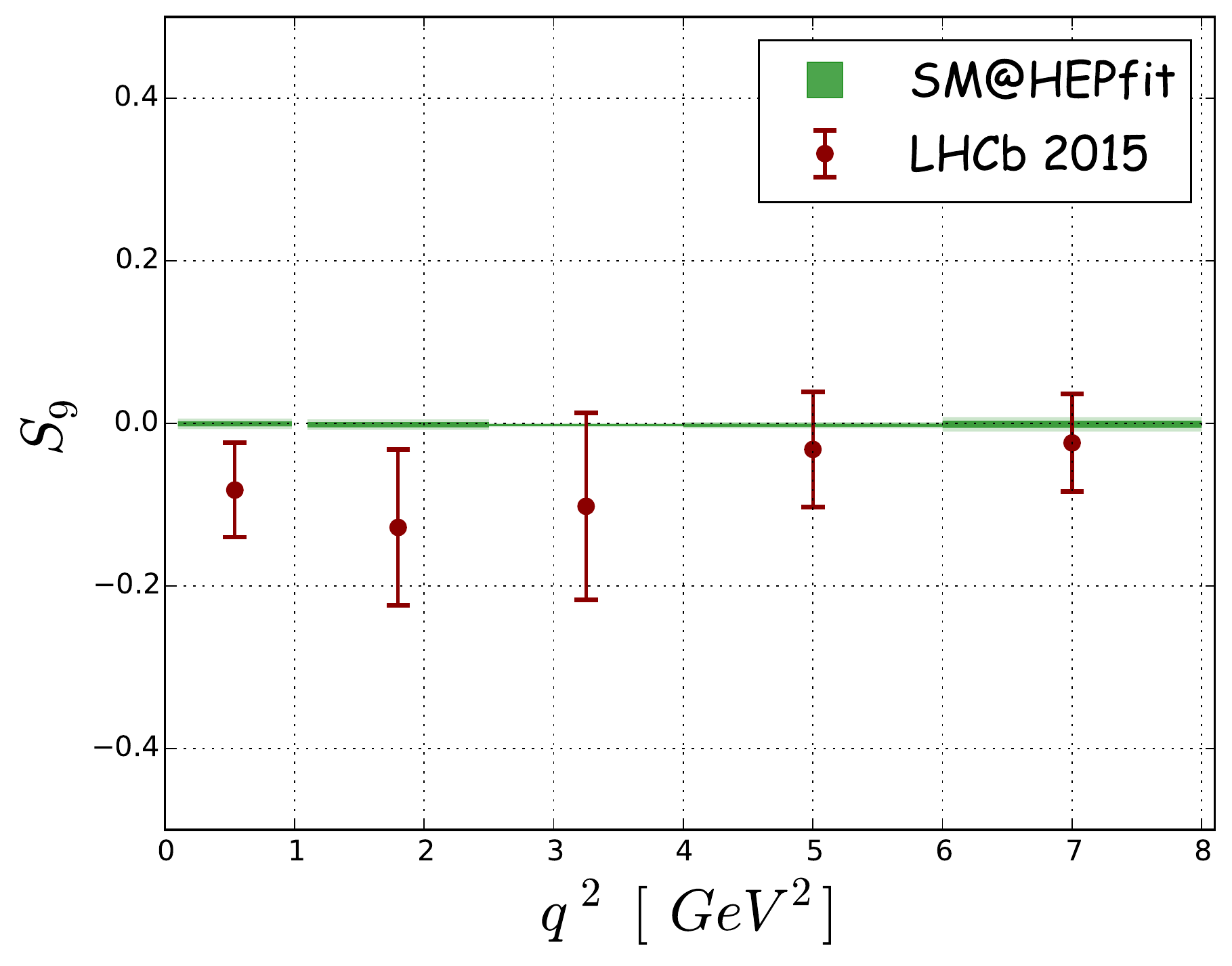}}

\caption{\textit{Results of the full fit and experimental results for
    the $B \to K^* \mu^+ \mu^-$ angular observables obtained using the
    phenomenological model from ref.~\cite{Khodjamirian:2010vf}.}}
\label{fig:fullfitallK}
\end{figure}

As a further test, we performed our fit adopting the phenomenological
model of ref.~\cite{Khodjamirian:2010vf} for the $q^2$ dependence of
the power corrections, although we consider this model to be
inadequate for $q^2 \sim 4 m_c^2$ as discussed in
section~\ref{sec:pow}. The results for the $B \to K^* \mu^+ \mu^-$
observables are reported in table~\ref{tab:mumuallK}, while the ones
for the $B \to K^* e^+ e^-$ observables are in
table~\ref{tab:eeallK}. Plots for the $B \to K^* \mu^+ \mu^-$ angular
observables are shown in figure~\ref{fig:fullfitallK}.

\begin{figure}[!htbp]
\centering

\subfigure{\includegraphics[width=.4\textwidth]{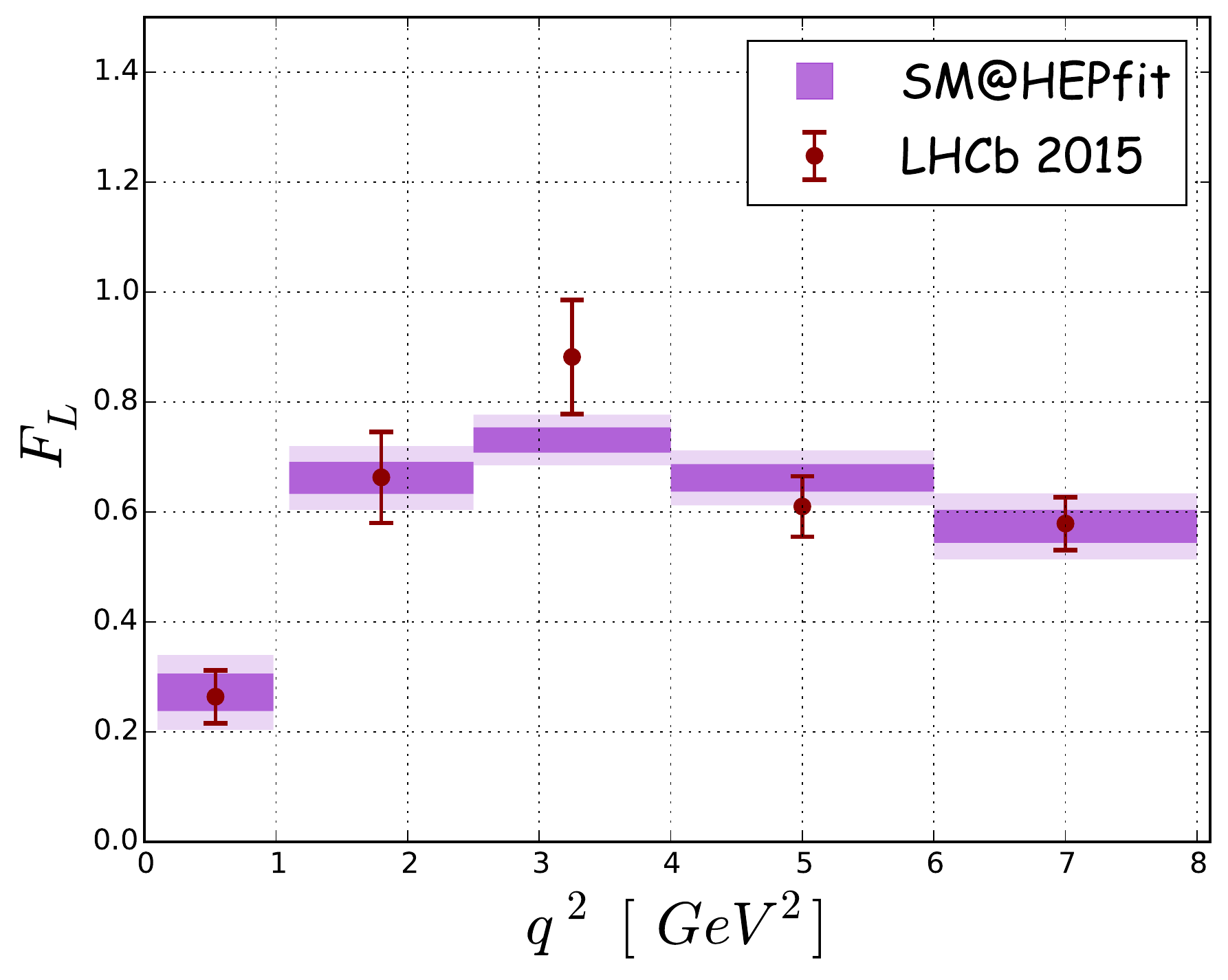}}
\subfigure{\includegraphics[width=.4\textwidth]{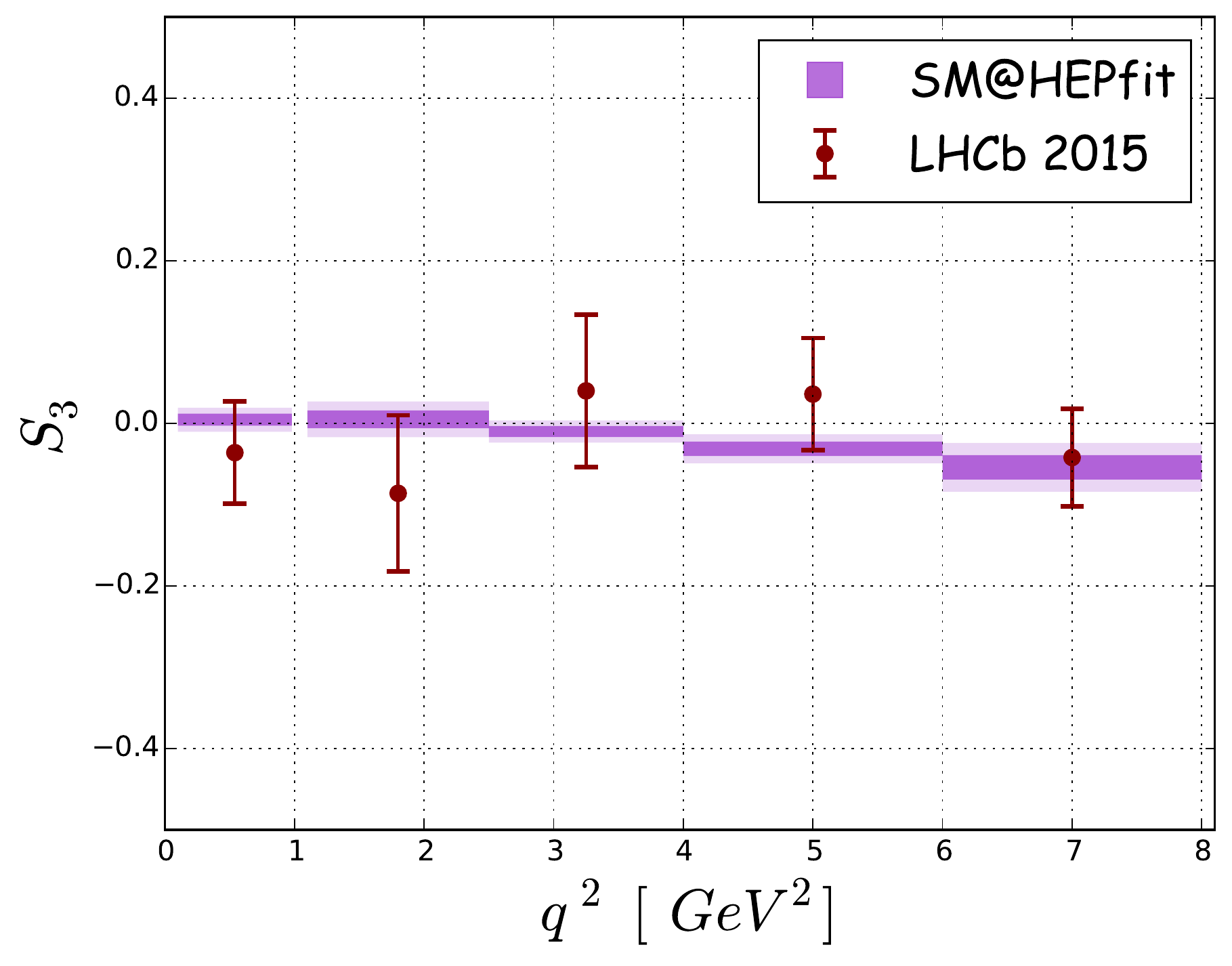}}
\subfigure{\includegraphics[width=.4\textwidth]{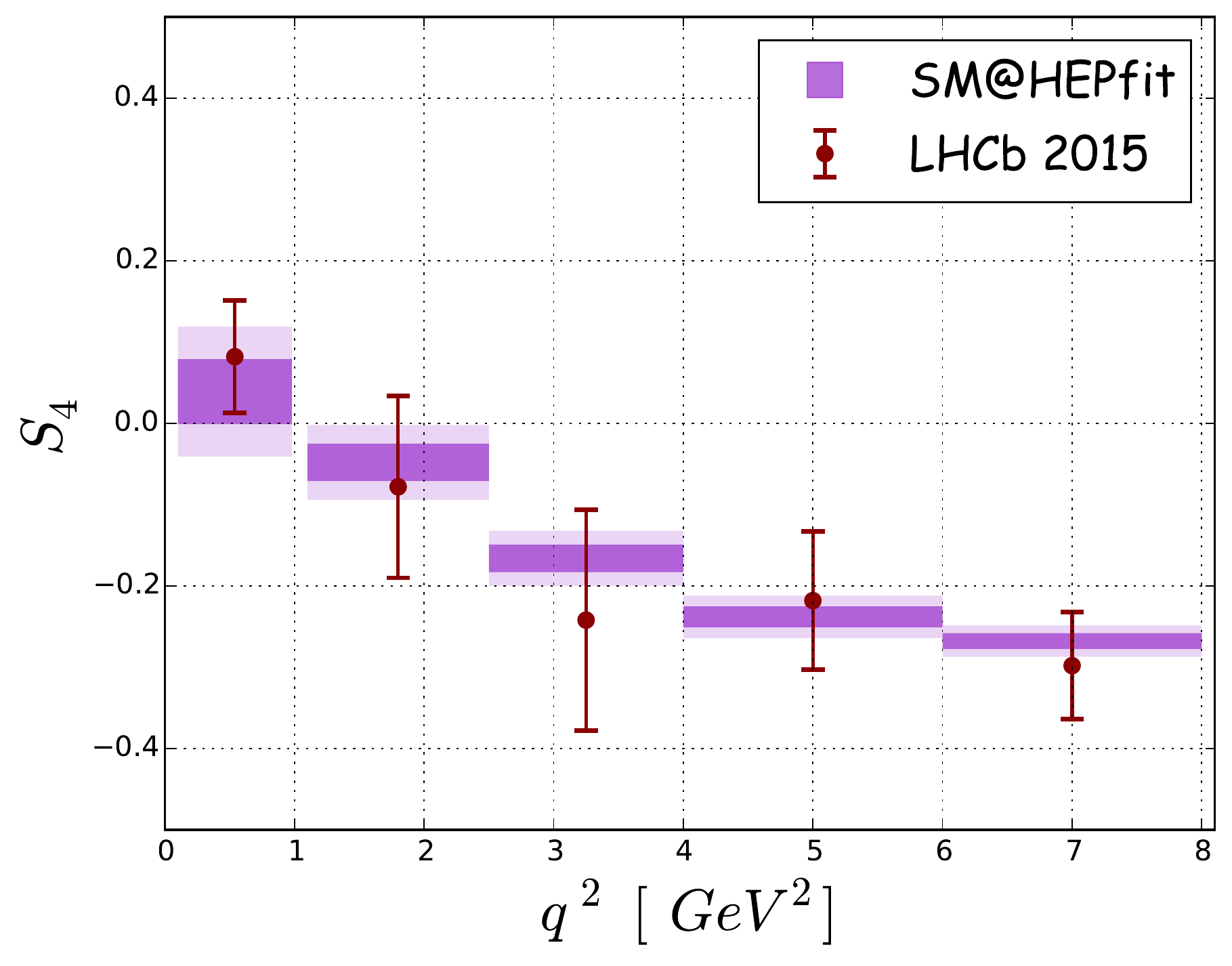}}
\subfigure{\includegraphics[width=.4\textwidth]{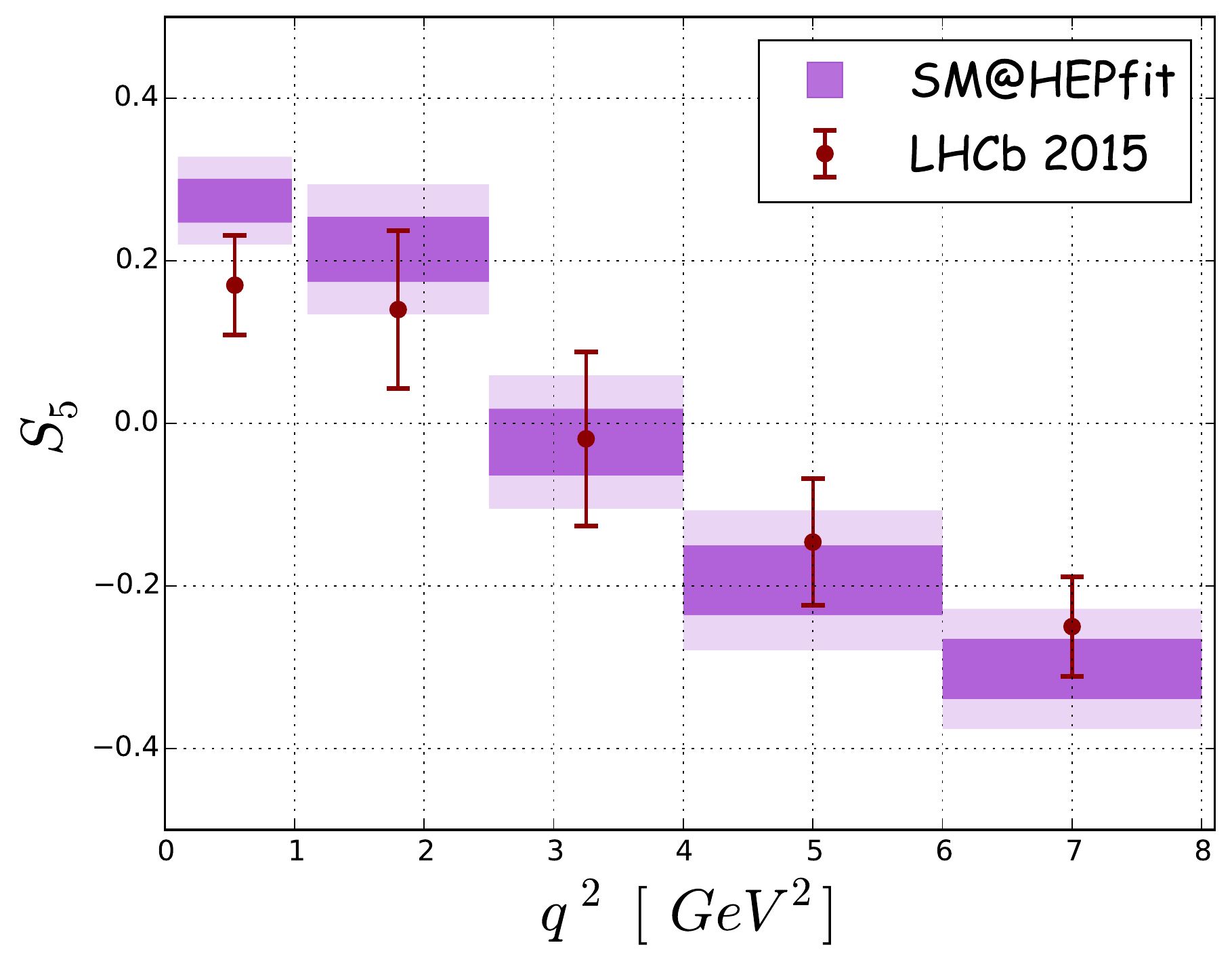}}
\subfigure{\includegraphics[width=.4\textwidth]{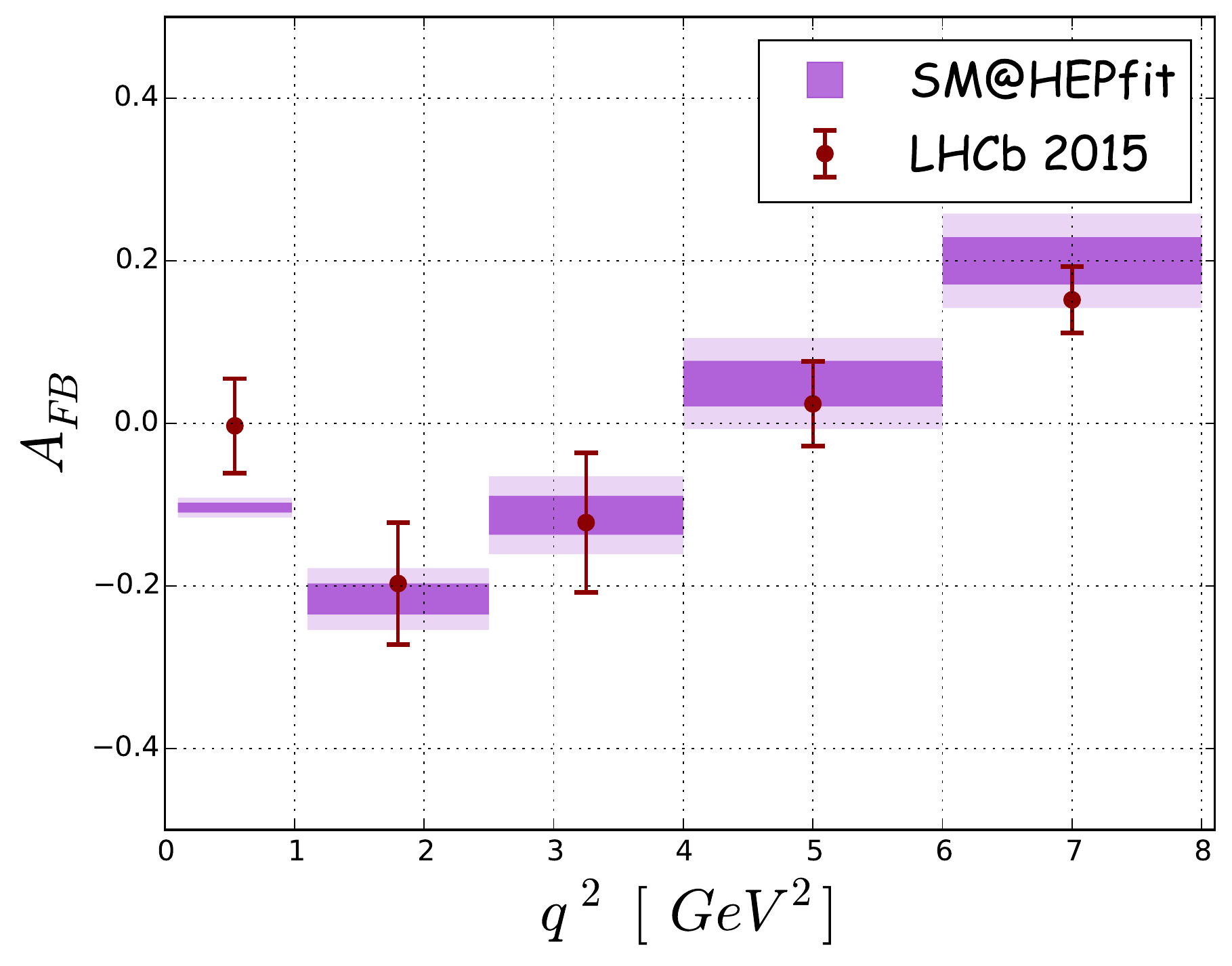}}
\subfigure{\includegraphics[width=.4\textwidth]{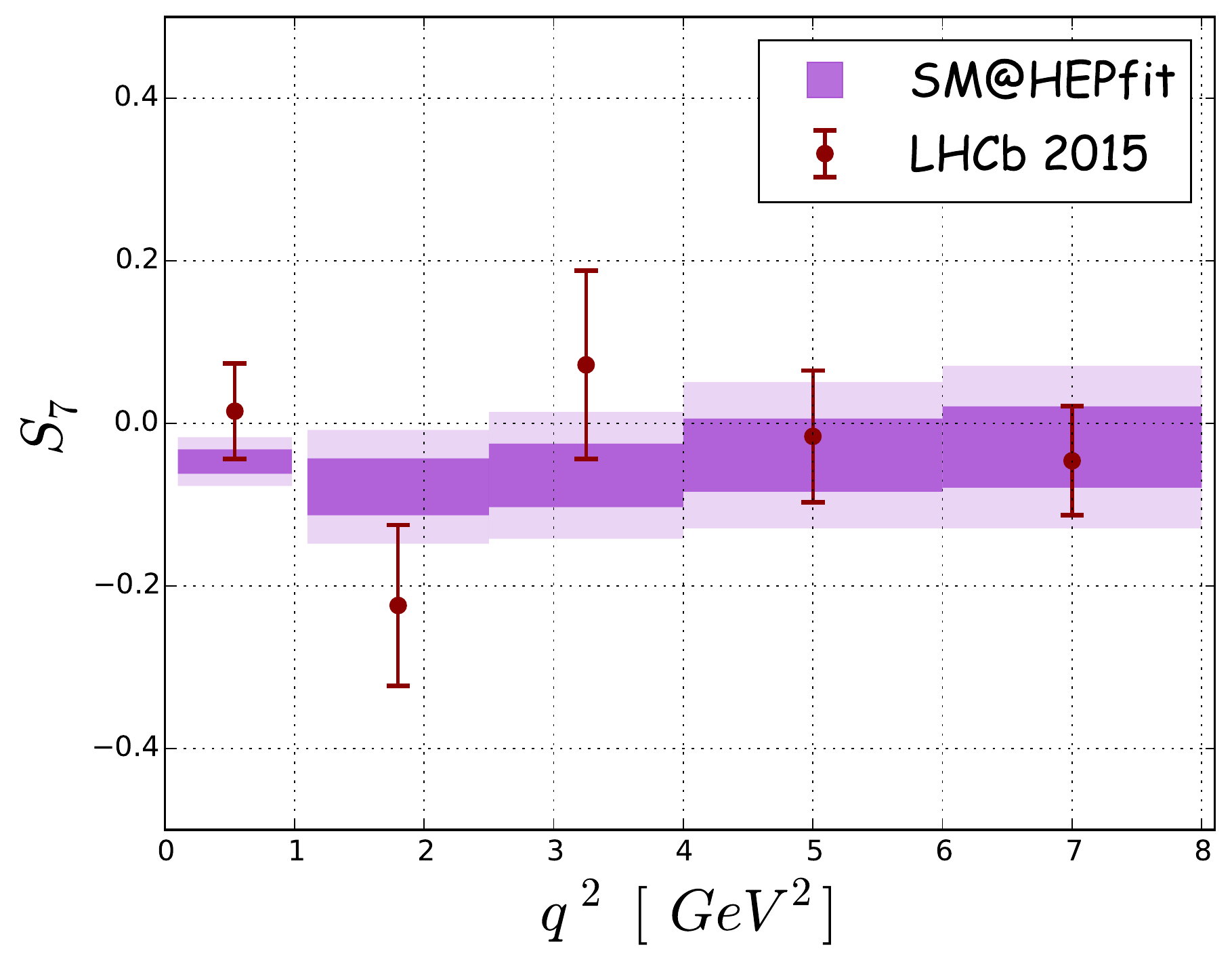}}
\subfigure{\includegraphics[width=.4\textwidth]{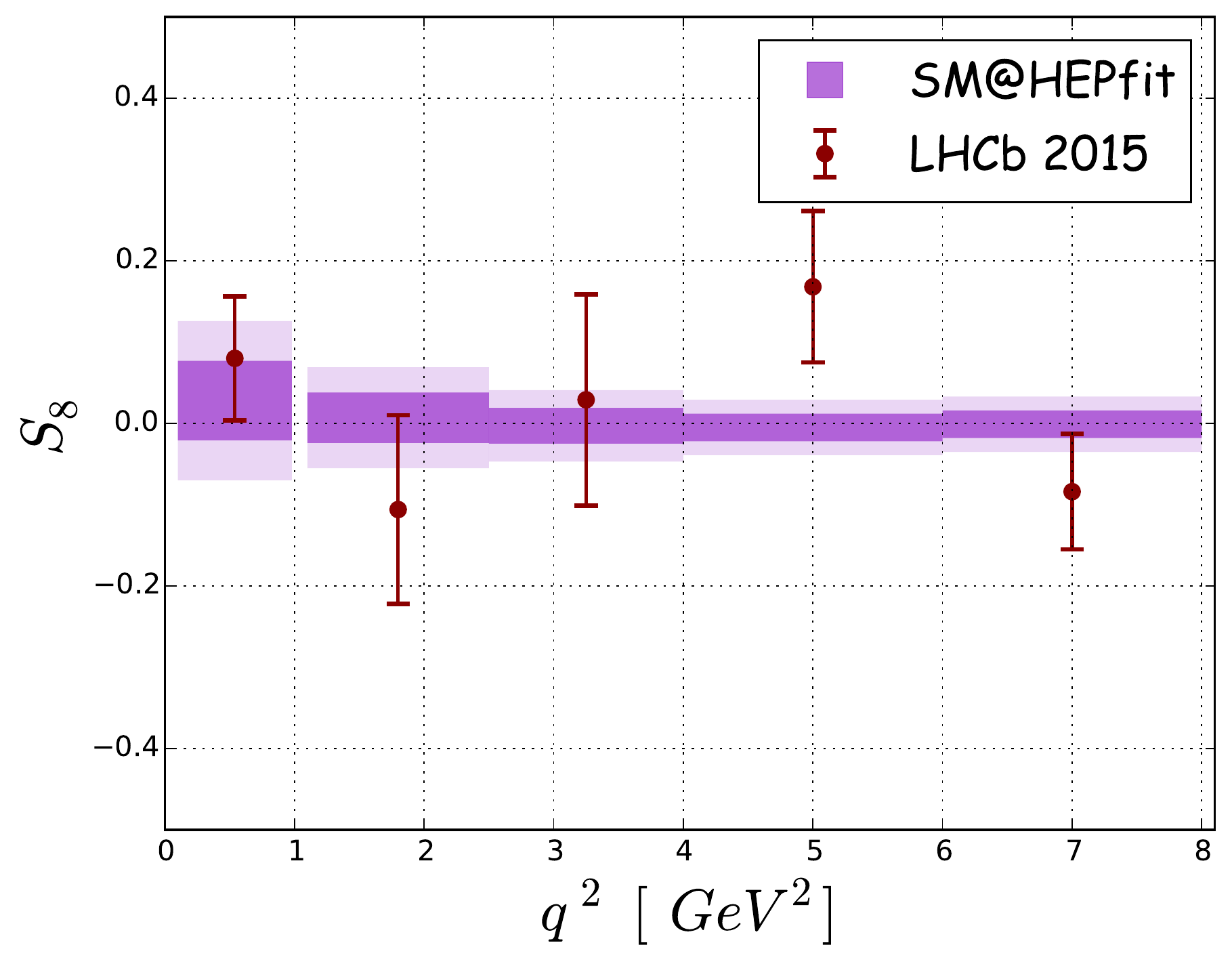}}
\subfigure{\includegraphics[width=.4\textwidth]{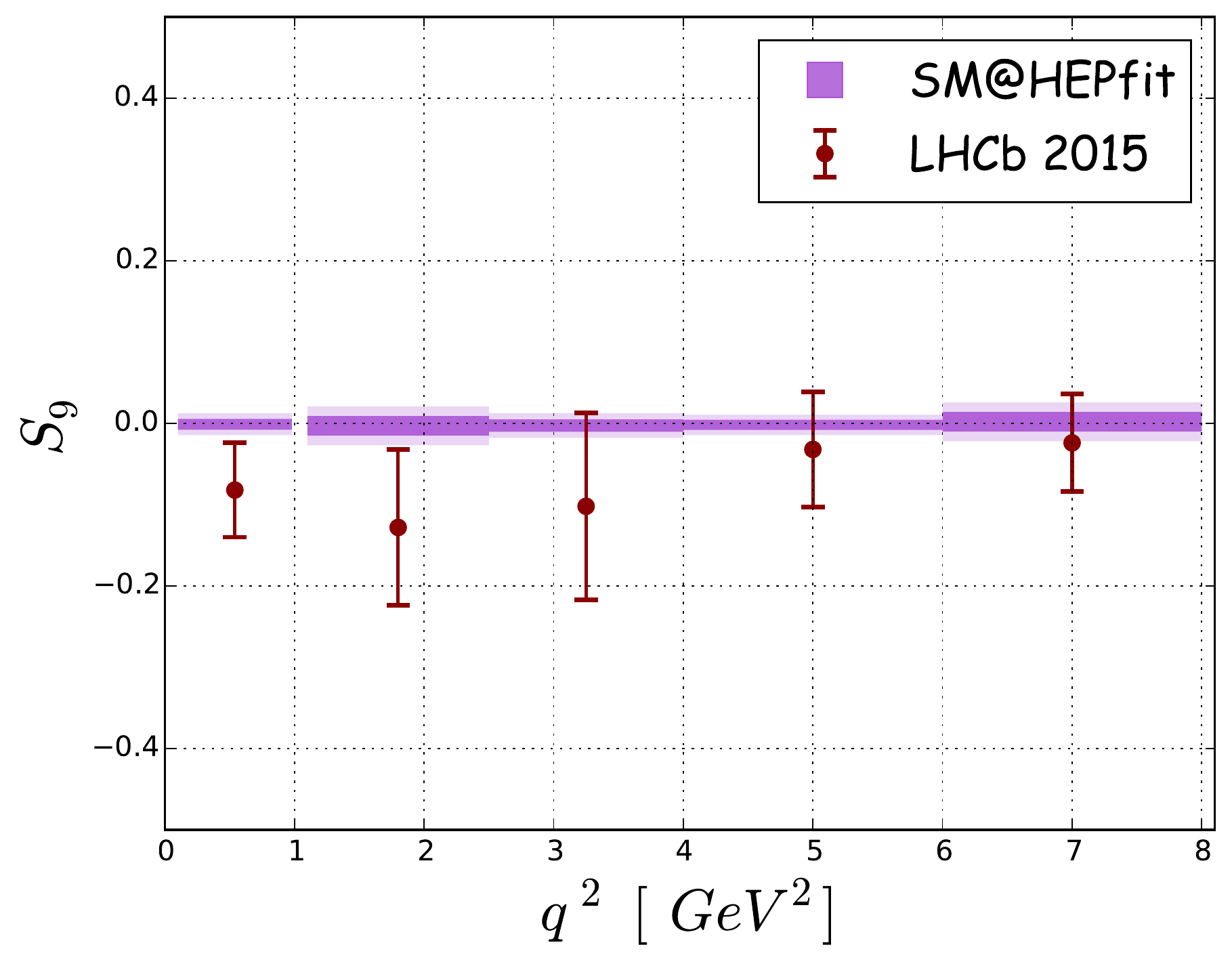}}

\caption{\textit{Results of the full fit and experimental results for
    the $B \to K^* \mu^+ \mu^-$ angular observables obtained assuming
    vanishing $h_\lambda^{(2)}$, i.e.  hadronic corrections fully
    equivalent to a shift in $C_{7,9}$.}}
\label{fig:fullfitnoq4}
\end{figure}

Finally, we performed our fit assuming vanishing $h_\lambda^{(2)}$,
i.e.\ hadronic corrections fully equivalent to a shift in
$C_{7,9}$. The results for the $B \to K^* \mu^+ \mu^-$ observables are
reported in table~\ref{tab:mumunoq4}, while the ones for the
$B \to K^* e^+ e^-$ observables are in table~\ref{tab:eenoq4}. Plots for the
$B \to K^* \mu^+ \mu^-$ angular observables are shown in
figure~\ref{fig:fullfitnoq4}. 

See section~\ref{sec:fit} for a discussion of the physical implications
of the results reported here.

\begin{table}[!htbp]
\fontsize{8}{8}\selectfont
\centering
\begin{tabular}{|c|c|c|c|c|c|}
\hline
&&&&&\\[-1mm]
\textbf{$q^2$ bin [GeV$^2$]} & \textbf{Observable} & \textbf{measurement} & \textbf{full fit} & \textbf{prediction} & $\mathbf{p-value}$ \\[1mm]
\hline
&&&&&\\[-2mm]
 \multirow{9}{*}{\normalsize $ [0.1,0.98] $}
& $F_L
$ & $
\phantom{-}
0.264 \pm 0.048$ & $
\phantom{-}
0.274 \pm 0.036$ & $
\phantom{-}
0.255 \pm 0.037$ & \multirow{8}{*}{0.14}
\\
& $S_3
$ & $
-0.036 \pm 0.063$ & $
-0.017 \pm 0.026$ & $
-0.021 \pm 0.037$ & 
\\
& $S_4
$ & $
\phantom{-}
0.082 \pm 0.069$ & $
\phantom{-}
0.033 \pm 0.045$ & $
-0.032 \pm 0.049$ & 
\\
& $S_5
$ & $
\phantom{-}
0.170 \pm 0.061$ & $
\phantom{-}
0.259 \pm 0.029$ & $
\phantom{-}
0.261 \pm 0.040$ & 
\\
& $ A_{FB}
$ & $
-0.003 \pm 0.058$ & $
-0.098 \pm 0.009$ & $
-0.092 \pm 0.012$ & 
\\
& $S_7
$ & $
\phantom{-}
0.015 \pm 0.059$ & $
-0.031 \pm 0.055$ & $
-0.119 \pm 0.072$ & 
\\
& $S_8
$ & $
\phantom{-}
0.080 \pm 0.076$ & $
\phantom{-}
0.030 \pm 0.049$ & $
-0.008 \pm 0.045$ & 
\\
& $S_9
$ & $
-0.082 \pm 0.058$ & $
-0.020 \pm 0.026$ & $
-0.007 \pm 0.028$ & 
\\
\cline{2-6}
&&&&&\\[-2mm]
& $ P_5'
$ & $
\phantom{-}
0.387 \pm 0.142$ & $
\phantom{-}
0.740 \pm 0.096$ & $
\phantom{-}
0.760 \pm 0.120$ & 0.045
\\[.5mm]
\hline
&&&&&\\[-2mm]
 \multirow{9}{*}{\normalsize $ [1.1,2.5] $}
& $F_L
$ & $
\phantom{-}
0.663 \pm 0.083$ & $
\phantom{-}
0.668 \pm 0.039$ & $
\phantom{-}
0.664 \pm 0.046$ & \multirow{8}{*}{0.52}
\\
& $S_3
$ & $
-0.086 \pm 0.096$ & $
-0.017 \pm 0.032$ & $
-0.009 \pm 0.033$ & 
\\
& $S_4
$ & $
-0.078 \pm 0.112$ & $
-0.055 \pm 0.037$ & $
-0.061 \pm 0.039$ & 
\\
& $S_5
$ & $
\phantom{-}
0.140 \pm 0.097$ & $
\phantom{-}
0.170 \pm 0.052$ & $
\phantom{-}
0.186 \pm 0.064$ & 
\\
& $ A_{FB}
$ & $
-0.197 \pm 0.075$ & $
-0.195 \pm 0.023$ & $
-0.194 \pm 0.026$ & 
\\
& $S_7
$ & $
-0.224 \pm 0.099$ & $
-0.077 \pm 0.063$ & $
\phantom{-}
0.020 \pm 0.078$ & 
\\
& $S_8
$ & $
-0.106 \pm 0.116$ & $
\phantom{-}
0.014 \pm 0.040$ & $
\phantom{-}
0.034 \pm 0.040$ & 
\\
& $S_9
$ & $
-0.128 \pm 0.096$ & $
-0.028 \pm 0.036$ & $
-0.014 \pm 0.036$ & 
\\
\cline{2-6}
&&&&&\\[-2mm]
& $ P_5'
$ & $
\phantom{-}
0.298 \pm 0.212$ & $
\phantom{-}
0.370 \pm 0.110$ & $
\phantom{-}
0.410 \pm 0.140$ & 0.66
\\[.5mm]
\hline
&&&&&\\[-2mm]
 \multirow{9}{*}{\normalsize $ [2.5,4] $}
& $F_L
$ & $
\phantom{-}
0.882 \pm 0.104$ & $
\phantom{-}
0.725 \pm 0.033$ & $
\phantom{-}
0.700 \pm 0.041$ & \multirow{8}{*}{0.72}
\\
& $S_3
$ & $
\phantom{-}
0.040 \pm 0.094$ & $
-0.016 \pm 0.017$ & $
-0.024 \pm 0.023$ & 
\\
& $S_4
$ & $
-0.242 \pm 0.136$ & $
-0.167 \pm 0.029$ & $
-0.167 \pm 0.033$ & 
\\
& $S_5
$ & $
-0.019 \pm 0.107$ & $
-0.055 \pm 0.054$ & $
-0.066 \pm 0.066$ & 
\\
& $ A_{FB}
$ & $
-0.122 \pm 0.086$ & $
-0.093 \pm 0.031$ & $
-0.091 \pm 0.037$ & 
\\
& $S_7
$ & $
\phantom{-}
0.072 \pm 0.116$ & $
-0.066 \pm 0.059$ & $
-0.113 \pm 0.072$ & 
\\
& $S_8
$ & $
\phantom{-}
0.029 \pm 0.130$ & $
\phantom{-}
0.005 \pm 0.032$ & $
\phantom{-}
0.005 \pm 0.034$ & 
\\
& $S_9
$ & $
-0.102 \pm 0.115$ & $
-0.011 \pm 0.018$ & $
-0.015 \pm 0.023$ & 
\\
\cline{2-6}
&&&&&\\[-2mm]
& $ P_5'
$ & $
-0.077 \pm 0.354$ & $
-0.130 \pm 0.120$ & $
-0.150 \pm 0.150$ & 0.85
\\[.5mm]
\hline
&&&&&\\[-2mm]
 \multirow{9}{*}{\normalsize $ [4,6] $}
& $F_L
$ & $
\phantom{-}
0.610 \pm 0.055$ & $
\phantom{-}
0.652 \pm 0.031$ & $
\phantom{-}
0.667 \pm 0.036$ & \multirow{8}{*}{0.56}
\\
& $S_3
$ & $
\phantom{-}
0.036 \pm 0.069$ & $
-0.027 \pm 0.017$ & $
-0.028 \pm 0.018$ & 
\\
& $S_4
$ & $
-0.218 \pm 0.085$ & $
-0.235 \pm 0.017$ & $
-0.232 \pm 0.020$ & 
\\
& $S_5
$ & $
-0.146 \pm 0.078$ & $
-0.182 \pm 0.044$ & $
-0.204 \pm 0.052$ & 
\\
& $ A_{FB}
$ & $
\phantom{-}
0.024 \pm 0.052$ & $
\phantom{-}
0.042 \pm 0.030$ & $
\phantom{-}
0.060 \pm 0.037$ & 
\\
& $S_7
$ & $
-0.016 \pm 0.081$ & $
-0.039 \pm 0.049$ & $
-0.047 \pm 0.062$ & 
\\
& $S_8
$ & $
\phantom{-}
0.168 \pm 0.093$ & $
-0.005 \pm 0.030$ & $
-0.023 \pm 0.030$ & 
\\
& $S_9
$ & $
-0.032 \pm 0.071$ & $
-0.006 \pm 0.015$ & $
-0.011 \pm 0.016$ & 
\\
\cline{2-6}
&&&&&\\[-2mm]
& $ P_5'
$ & $
-0.301 \pm 0.160$ & $
-0.386 \pm 0.093$ & $
-0.440 \pm 0.110$ & 0.47
\\[.5mm]
\hline
&&&&&\\[-2mm]
 \multirow{9}{*}{\normalsize $ [6,8] $}
& $F_L
$ & $
\phantom{-}
0.579 \pm 0.048$ & $
\phantom{-}
0.569 \pm 0.035$ & $
\phantom{-}
0.516 \pm 0.075$ & \multirow{8}{*}{0.74}
\\
& $S_3
$ & $
-0.042 \pm 0.060$ & $
-0.046 \pm 0.031$ & $
\phantom{-}
0.005 \pm 0.060$ & 
\\
& $S_4
$ & $
-0.298 \pm 0.066$ & $
-0.262 \pm 0.018$ & $
-0.213 \pm 0.040$ & 
\\
& $S_5
$ & $
-0.250 \pm 0.061$ & $
-0.238 \pm 0.050$ & $
-0.160 \pm 0.110$ & 
\\
& $ A_{FB}
$ & $
\phantom{-}
0.152 \pm 0.041$ & $
\phantom{-}
0.148 \pm 0.036$ & $
\phantom{-}
0.107 \pm 0.080$ & 
\\
& $S_7
$ & $
-0.046 \pm 0.067$ & $
-0.028 \pm 0.056$ & $
\phantom{-}
0.040 \pm 0.120$ & 
\\
& $S_8
$ & $
-0.084 \pm 0.071$ & $
-0.017 \pm 0.040$ & $
\phantom{-}
0.043 \pm 0.058$ & 
\\
& $S_9
$ & $
-0.024 \pm 0.060$ & $
-0.013 \pm 0.033$ & $
\phantom{-}
0.020 \pm 0.055$ & 
\\
\cline{2-6}
&&&&&\\[-2mm]
& $ P_5'
$ & $
-0.505 \pm 0.124$ & $
-0.490 \pm 0.100$ & $
-0.320 \pm 0.230$ & 0.48
\\[.5mm]
\hline
\hline&&&&&\\[-2mm]
 $ [0.1,2] $ & 
\multirow{3}{*}{\normalsize $ {\rm BR } \cdot 10^7 $}
 & $ 0.58 \pm 0.09$ & $
0.67 \pm 0.04$ & $
0.70 \pm 0.06$ & 0.27
\\
 $ [2,4.3] $ & 
 & $ 0.29 \pm 0.05$ & $
0.34 \pm 0.03$ & $
0.37 \pm 0.05$ & 0.26
\\
 $ [4.3,8.68] $ & 
 & $ 0.47 \pm 0.07$ & $
0.46 \pm 0.06$ & $
0.49 \pm 0.13$ & 0.89
\\
\hline

&&&&&\\[-2mm]
& { \normalsize $ {\rm BR }_{B \to K^* \gamma} \cdot 10^5 $}
& $ 4.33 \pm 0.15$ & $
4.34 \pm 0.15$ & $
4.59 \pm 0.77$ & 0.74
\\[1mm]
\hline
\end{tabular}
\caption{\textit{Experimental results, results from the full fit,
    predictions and $p$-values for $B \to K^* \mu^+ \mu^-$ BR's and angular
  observables obtained without using the numerical information from ref.~\cite{Khodjamirian:2010vf}.
  The predictions for the  BR's (angular observables) are obtained removing the 
  corresponding observable (the experimental information in one bin at a time) from the fit. 
  We also report the results for BR$(B \to K^* \gamma)$
  (including the experimental value from refs.~\cite{Coan:1999kh,Nakao:2004th,Aubert:2009ak,Agashe:2014kda})
  and for the optimized observable $P_5^\prime$. The latter is however
  not explicitly used in the fit as a constraint, since it is not
  independent of $F_L$ and $S_5$. }}
\label{tab:mumunoK}
\end{table}

\FloatBarrier

\begin{table}[!htbp]
\fontsize{8}{8}\selectfont
\centering
\begin{tabular}{|c|c|c|c|c|c|}
\hline
&&&&&\\[-1mm]
\textbf{$q^2$ bin [GeV$^2$]} & \textbf{Observable} & \textbf{measurement} & \textbf{full fit} & \textbf{prediction} & $\mathbf{p-value}$ \\[1mm]
\hline
&&&&&\\[-2mm]
 \multirow{9}{*}{\normalsize $ [0.1,0.98] $}
& $F_L
$ & $
\phantom{-}
0.264 \pm 0.048$ & $
\phantom{-}
0.270 \pm 0.032$ & $
\phantom{-}
0.257 \pm 0.025$ & \multirow{8}{*}{0.056}
\\
& $S_3
$ & $
-0.036 \pm 0.063$ & $
\phantom{-}
0.004 \pm 0.004$ & $
\phantom{-}
0.004 \pm 0.004$ & 
\\
& $S_4
$ & $
\phantom{-}
0.082 \pm 0.069$ & $
\phantom{-}
0.010 \pm 0.048$ & $
-0.047 \pm 0.035$ & 
\\
& $S_5
$ & $
\phantom{-}
0.170 \pm 0.061$ & $
\phantom{-}
0.293 \pm 0.024$ & $
\phantom{-}
0.314 \pm 0.015$ & 
\\
& $ A_{FB}
$ & $
-0.003 \pm 0.058$ & $
-0.101 \pm 0.005$ & $
-0.102 \pm 0.005$ & 
\\
& $S_7
$ & $
\phantom{-}
0.015 \pm 0.059$ & $
-0.046 \pm 0.015$ & $
-0.041 \pm 0.014$ & 
\\
& $S_8
$ & $
\phantom{-}
0.080 \pm 0.076$ & $
\phantom{-}
0.023 \pm 0.045$ & $
-0.005 \pm 0.036$ & 
\\
& $S_9
$ & $
-0.082 \pm 0.058$ & $
-0.001 \pm 0.003$ & $
-0.001 \pm 0.003$ & 
\\
\cline{2-6}
&&&&&\\[-2mm]
& $ P_5'
$ & $
\phantom{-}
0.387 \pm 0.142$ & $
\phantom{-}
0.840 \pm 0.088$ & $
\phantom{-}
0.919 \pm 0.051$ & 0.0004
\\[.5mm]
\hline
&&&&&\\[-2mm]
 \multirow{9}{*}{\normalsize $ [1.1,2.5] $}
& $F_L
$ & $
\phantom{-}
0.663 \pm 0.083$ & $
\phantom{-}
0.711 \pm 0.024$ & $
\phantom{-}
0.711 \pm 0.027$ & \multirow{8}{*}{0.58}
\\
& $S_3
$ & $
-0.086 \pm 0.096$ & $
\phantom{-}
0.001 \pm 0.003$ & $
\phantom{-}
0.001 \pm 0.003$ & 
\\
& $S_4
$ & $
-0.078 \pm 0.112$ & $
-0.073 \pm 0.020$ & $
-0.078 \pm 0.022$ & 
\\
& $S_5
$ & $
\phantom{-}
0.140 \pm 0.097$ & $
\phantom{-}
0.190 \pm 0.039$ & $
\phantom{-}
0.201 \pm 0.043$ & 
\\
& $ A_{FB}
$ & $
-0.197 \pm 0.075$ & $
-0.185 \pm 0.016$ & $
-0.186 \pm 0.017$ & 
\\
& $S_7
$ & $
-0.224 \pm 0.099$ & $
-0.061 \pm 0.030$ & $
-0.050 \pm 0.032$ & 
\\
& $S_8
$ & $
-0.106 \pm 0.116$ & $
-0.010 \pm 0.021$ & $
-0.014 \pm 0.021$ & 
\\
& $S_9
$ & $
-0.128 \pm 0.096$ & $
-0.002 \pm 0.003$ & $
-0.001 \pm 0.003$ & 
\\
\cline{2-6}
&&&&&\\[-2mm]
& $ P_5'
$ & $
\phantom{-}
0.298 \pm 0.212$ & $
\phantom{-}
0.434 \pm 0.082$ & $
\phantom{-}
0.458 \pm 0.090$ & 0.49
\\[.5mm]
\hline
&&&&&\\[-2mm]
 \multirow{9}{*}{\normalsize $ [2.5,4] $}
& $F_L
$ & $
\phantom{-}
0.882 \pm 0.104$ & $
\phantom{-}
0.770 \pm 0.020$ & $
\phantom{-}
0.767 \pm 0.021$ & \multirow{8}{*}{0.79}
\\
& $S_3
$ & $
\phantom{-}
0.040 \pm 0.094$ & $
-0.016 \pm 0.004$ & $
-0.017 \pm 0.004$ & 
\\
& $S_4
$ & $
-0.242 \pm 0.136$ & $
-0.187 \pm 0.012$ & $
-0.188 \pm 0.013$ & 
\\
& $S_5
$ & $
-0.019 \pm 0.107$ & $
-0.112 \pm 0.029$ & $
-0.119 \pm 0.030$ & 
\\
& $ A_{FB}
$ & $
-0.122 \pm 0.086$ & $
-0.034 \pm 0.011$ & $
-0.032 \pm 0.012$ & 
\\
& $S_7
$ & $
\phantom{-}
0.072 \pm 0.116$ & $
-0.029 \pm 0.030$ & $
-0.035 \pm 0.030$ & 
\\
& $S_8
$ & $
\phantom{-}
0.029 \pm 0.130$ & $
-0.013 \pm 0.006$ & $
-0.012 \pm 0.006$ & 
\\
& $S_9
$ & $
-0.102 \pm 0.115$ & $
-0.002 \pm 0.001$ & $
-0.002 \pm 0.001$ & 
\\
\cline{2-6}
&&&&&\\[-2mm]
& $ P_5'
$ & $
-0.077 \pm 0.354$ & $
-0.271 \pm 0.072$ & $
-0.287 \pm 0.074$ & 0.56
\\[.5mm]
\hline
&&&&&\\[-2mm]
 \multirow{9}{*}{\normalsize $ [4,6] $}
& $F_L
$ & $
\phantom{-}
0.610 \pm 0.055$ & $
\phantom{-}
0.679 \pm 0.024$ & $
\phantom{-}
0.682 \pm 0.026$ & \multirow{8}{*}{0.025}
\\
& $S_3
$ & $
\phantom{-}
0.036 \pm 0.069$ & $
-0.036 \pm 0.008$ & $
-0.035 \pm 0.009$ & 
\\
& $S_4
$ & $
-0.218 \pm 0.085$ & $
-0.249 \pm 0.008$ & $
-0.247 \pm 0.009$ & 
\\
& $S_5
$ & $
-0.146 \pm 0.078$ & $
-0.295 \pm 0.021$ & $
-0.312 \pm 0.021$ & 
\\
& $ A_{FB}
$ & $
\phantom{-}
0.024 \pm 0.052$ & $
\phantom{-}
0.139 \pm 0.014$ & $
\phantom{-}
0.146 \pm 0.016$ & 
\\
& $S_7
$ & $
-0.016 \pm 0.081$ & $
-0.002 \pm 0.026$ & $
-0.002 \pm 0.026$ & 
\\
& $S_8
$ & $
\phantom{-}
0.168 \pm 0.093$ & $
-0.006 \pm 0.005$ & $
-0.006 \pm 0.005$ & 
\\
& $S_9
$ & $
-0.032 \pm 0.071$ & $
-0.002 \pm 0.002$ & $
-0.002 \pm 0.002$ & 
\\
\cline{2-6}
&&&&&\\[-2mm]
& $ P_5'
$ & $
-0.301 \pm 0.160$ & $
-0.637 \pm 0.047$ & $
-0.676 \pm 0.047$ & 0.025
\\[.5mm]
\hline
&&&&&\\[-2mm]
 \multirow{9}{*}{\normalsize $ [6,8] $}
& $F_L
$ & $
\phantom{-}
0.579 \pm 0.048$ & $
\phantom{-}
0.585 \pm 0.029$ & $
\phantom{-}
0.561 \pm 0.038$ & \multirow{8}{*}{0.058}
\\
& $S_3
$ & $
-0.042 \pm 0.060$ & $
-0.054 \pm 0.011$ & $
-0.053 \pm 0.013$ & 
\\
& $S_4
$ & $
-0.298 \pm 0.066$ & $
-0.271 \pm 0.007$ & $
-0.271 \pm 0.007$ & 
\\
& $S_5
$ & $
-0.250 \pm 0.061$ & $
-0.383 \pm 0.017$ & $
-0.392 \pm 0.019$ & 
\\
& $ A_{FB}
$ & $
\phantom{-}
0.152 \pm 0.041$ & $
\phantom{-}
0.264 \pm 0.019$ & $
\phantom{-}
0.286 \pm 0.025$ & 
\\
& $S_7
$ & $
-0.046 \pm 0.067$ & $
-0.000 \pm 0.036$ & $
\phantom{-}
0.021 \pm 0.039$ & 
\\
& $S_8
$ & $
-0.084 \pm 0.071$ & $
-0.001 \pm 0.010$ & $
\phantom{-}
0.005 \pm 0.011$ & 
\\
& $S_9
$ & $
-0.024 \pm 0.060$ & $
-0.001 \pm 0.004$ & $
-0.000 \pm 0.005$ & 
\\
\cline{2-6}
&&&&&\\[-2mm]
& $ P_5'
$ & $
-0.505 \pm 0.124$ & $
-0.783 \pm 0.038$ & $
-0.797 \pm 0.041$ & 0.025
\\[.5mm]
\hline
\hline&&&&&\\[-2mm]
 $ [0.1,2] $ & 
\multirow{3}{*}{\normalsize $ {\rm BR } \cdot 10^7 $}
 & $ 0.58 \pm 0.09$ & $
0.64 \pm 0.03$ & $
0.65 \pm 0.04$ & 0.48
\\
 $ [2,4.3] $ & 
 & $ 0.29 \pm 0.05$ & $
0.33 \pm 0.03$ & $
0.35 \pm 0.03$ & 0.30
\\
 $ [4.3,8.68] $ & 
 & $ 0.47 \pm 0.07$ & $
0.48 \pm 0.04$ & $
0.49 \pm 0.05$ & 0.82
\\
\hline

&&&&&\\[-2mm]
& { \normalsize $ {\rm BR }_{B \to K^* \gamma} \cdot 10^5 $}
& $ 4.33 \pm 0.15$ & $
4.35 \pm 0.14$ & $
4.69 \pm 0.53$ & 0.51
\\[1mm]
\hline
\end{tabular}
\caption{\textit{Experimental results, results from the full fit,
    predictions and $p$-values for $B \to K^* \mu^+ \mu^-$ BR's and angular
  observables obtained using the phenomenological model from
  ref.~\cite{Khodjamirian:2010vf}. The predictions for the  BR's (angular observables) are obtained removing the 
  corresponding observable (the experimental information in one bin at a time) from the fit. 
  We also report the results for BR$(B \to K^* \gamma)$
  (including the experimental value from refs.~\cite{Coan:1999kh,Nakao:2004th,Aubert:2009ak,Agashe:2014kda})
  and for the optimized observable $P_5^\prime$. The latter is however
  not explicitly used in the fit as a constraint, since it is not
  independent of $F_L$ and $S_5$. }}
\label{tab:mumuallK}
\end{table}

\FloatBarrier

\begin{table}[!htbp]
\fontsize{8}{8}\selectfont
\centering
\begin{tabular}{|c|c|c|c|c|c|}
\hline
&&&&&\\[-1mm]
\textbf{$q^2$ bin [GeV$^2$]} & \textbf{Observable} & \textbf{measurement} & \textbf{full fit} & \textbf{prediction} & $\mathbf{p-value}$ \\[1mm]
\hline
&&&&&\\[-2mm]
 \multirow{9}{*}{\normalsize $ [0.1,0.98] $}
& $F_L
$ & $
\phantom{-}
0.264 \pm 0.048$ & $
\phantom{-}
0.272 \pm 0.034$ & $
\phantom{-}
0.251 \pm 0.033$ & \multirow{8}{*}{0.11}
\\
& $S_3
$ & $
-0.036 \pm 0.063$ & $
\phantom{-}
0.004 \pm 0.007$ & $
\phantom{-}
0.004 \pm 0.008$ & 
\\
& $S_4
$ & $
\phantom{-}
0.082 \pm 0.069$ & $
\phantom{-}
0.039 \pm 0.040$ & $
-0.024 \pm 0.045$ & 
\\
& $S_5
$ & $
\phantom{-}
0.170 \pm 0.061$ & $
\phantom{-}
0.274 \pm 0.027$ & $
\phantom{-}
0.305 \pm 0.023$ & 
\\
& $ A_{FB}
$ & $
-0.003 \pm 0.058$ & $
-0.104 \pm 0.006$ & $
-0.106 \pm 0.006$ & 
\\
& $S_7
$ & $
\phantom{-}
0.015 \pm 0.059$ & $
-0.047 \pm 0.015$ & $
-0.041 \pm 0.016$ & 
\\
& $S_8
$ & $
\phantom{-}
0.080 \pm 0.076$ & $
\phantom{-}
0.028 \pm 0.049$ & $
-0.003 \pm 0.046$ & 
\\
& $S_9
$ & $
-0.082 \pm 0.058$ & $
-0.001 \pm 0.007$ & $
-0.002 \pm 0.007$ & 
\\
\cline{2-6}
&&&&&\\[-2mm]
& $ P_5'
$ & $
\phantom{-}
0.387 \pm 0.142$ & $
\phantom{-}
0.782 \pm 0.093$ & $
\phantom{-}
0.896 \pm 0.080$ & 0.0018
\\[.5mm]
\hline
&&&&&\\[-2mm]
 \multirow{9}{*}{\normalsize $ [1.1,2.5] $}
& $F_L
$ & $
\phantom{-}
0.663 \pm 0.083$ & $
\phantom{-}
0.662 \pm 0.029$ & $
\phantom{-}
0.656 \pm 0.033$ & \multirow{8}{*}{0.53}
\\
& $S_3
$ & $
-0.086 \pm 0.096$ & $
\phantom{-}
0.005 \pm 0.011$ & $
\phantom{-}
0.006 \pm 0.012$ & 
\\
& $S_4
$ & $
-0.078 \pm 0.112$ & $
-0.048 \pm 0.023$ & $
-0.060 \pm 0.027$ & 
\\
& $S_5
$ & $
\phantom{-}
0.140 \pm 0.097$ & $
\phantom{-}
0.214 \pm 0.040$ & $
\phantom{-}
0.238 \pm 0.046$ & 
\\
& $ A_{FB}
$ & $
-0.197 \pm 0.075$ & $
-0.216 \pm 0.019$ & $
-0.221 \pm 0.021$ & 
\\
& $S_7
$ & $
-0.224 \pm 0.099$ & $
-0.078 \pm 0.035$ & $
-0.064 \pm 0.038$ & 
\\
& $S_8
$ & $
-0.106 \pm 0.116$ & $
\phantom{-}
0.007 \pm 0.031$ & $
\phantom{-}
0.010 \pm 0.032$ & 
\\
& $S_9
$ & $
-0.128 \pm 0.096$ & $
-0.003 \pm 0.012$ & $
\phantom{-}
0.001 \pm 0.013$ & 
\\
\cline{2-6}
&&&&&\\[-2mm]
& $ P_5'
$ & $
\phantom{-}
0.298 \pm 0.212$ & $
\phantom{-}
0.468 \pm 0.085$ & $
\phantom{-}
0.519 \pm 0.097$ & 0.34
\\[.5mm]
\hline
&&&&&\\[-2mm]
 \multirow{9}{*}{\normalsize $ [2.5,4] $}
& $F_L
$ & $
\phantom{-}
0.882 \pm 0.104$ & $
\phantom{-}
0.731 \pm 0.023$ & $
\phantom{-}
0.721 \pm 0.025$ & \multirow{8}{*}{0.79}
\\
& $S_3
$ & $
\phantom{-}
0.040 \pm 0.094$ & $
-0.010 \pm 0.007$ & $
-0.011 \pm 0.007$ & 
\\
& $S_4
$ & $
-0.242 \pm 0.136$ & $
-0.166 \pm 0.017$ & $
-0.166 \pm 0.018$ & 
\\
& $S_5
$ & $
-0.019 \pm 0.107$ & $
-0.023 \pm 0.041$ & $
-0.021 \pm 0.045$ & 
\\
& $ A_{FB}
$ & $
-0.122 \pm 0.086$ & $
-0.113 \pm 0.024$ & $
-0.119 \pm 0.025$ & 
\\
& $S_7
$ & $
\phantom{-}
0.072 \pm 0.116$ & $
-0.064 \pm 0.039$ & $
-0.080 \pm 0.041$ & 
\\
& $S_8
$ & $
\phantom{-}
0.029 \pm 0.130$ & $
-0.003 \pm 0.022$ & $
-0.005 \pm 0.023$ & 
\\
& $S_9
$ & $
-0.102 \pm 0.115$ & $
-0.003 \pm 0.008$ & $
-0.004 \pm 0.008$ & 
\\
\cline{2-6}
&&&&&\\[-2mm]
& $ P_5'
$ & $
-0.077 \pm 0.354$ & $
-0.054 \pm 0.095$ & $
-0.050 \pm 0.100$ & 0.94
\\[.5mm]
\hline
&&&&&\\[-2mm]
 \multirow{9}{*}{\normalsize $ [4,6] $}
& $F_L
$ & $
\phantom{-}
0.610 \pm 0.055$ & $
\phantom{-}
0.662 \pm 0.025$ & $
\phantom{-}
0.678 \pm 0.028$ & \multirow{8}{*}{0.44}
\\
& $S_3
$ & $
\phantom{-}
0.036 \pm 0.069$ & $
-0.031 \pm 0.009$ & $
-0.033 \pm 0.010$ & 
\\
& $S_4
$ & $
-0.218 \pm 0.085$ & $
-0.238 \pm 0.013$ & $
-0.236 \pm 0.015$ & 
\\
& $S_5
$ & $
-0.146 \pm 0.078$ & $
-0.193 \pm 0.043$ & $
-0.234 \pm 0.049$ & 
\\
& $ A_{FB}
$ & $
\phantom{-}
0.024 \pm 0.052$ & $
\phantom{-}
0.049 \pm 0.028$ & $
\phantom{-}
0.076 \pm 0.035$ & 
\\
& $S_7
$ & $
-0.016 \pm 0.081$ & $
-0.039 \pm 0.045$ & $
-0.045 \pm 0.054$ & 
\\
& $S_8
$ & $
\phantom{-}
0.168 \pm 0.093$ & $
-0.005 \pm 0.017$ & $
-0.012 \pm 0.018$ & 
\\
& $S_9
$ & $
-0.032 \pm 0.071$ & $
-0.002 \pm 0.006$ & $
-0.004 \pm 0.007$ & 
\\
\cline{2-6}
&&&&&\\[-2mm]
& $ P_5'
$ & $
-0.301 \pm 0.160$ & $
-0.413 \pm 0.093$ & $
-0.510 \pm 0.110$ & 0.28
\\[.5mm]
\hline
&&&&&\\[-2mm]
 \multirow{9}{*}{\normalsize $ [6,8] $}
& $F_L
$ & $
\phantom{-}
0.579 \pm 0.048$ & $
\phantom{-}
0.574 \pm 0.030$ & $
\phantom{-}
0.552 \pm 0.043$ & \multirow{8}{*}{0.72}
\\
& $S_3
$ & $
-0.042 \pm 0.060$ & $
-0.054 \pm 0.015$ & $
-0.051 \pm 0.018$ & 
\\
& $S_4
$ & $
-0.298 \pm 0.066$ & $
-0.268 \pm 0.010$ & $
-0.261 \pm 0.017$ & 
\\
& $S_5
$ & $
-0.250 \pm 0.061$ & $
-0.302 \pm 0.037$ & $
-0.311 \pm 0.055$ & 
\\
& $ A_{FB}
$ & $
\phantom{-}
0.152 \pm 0.041$ & $
\phantom{-}
0.200 \pm 0.029$ & $
\phantom{-}
0.245 \pm 0.043$ & 
\\
& $S_7
$ & $
-0.046 \pm 0.067$ & $
-0.029 \pm 0.050$ & $
\phantom{-}
0.014 \pm 0.077$ & 
\\
& $S_8
$ & $
-0.084 \pm 0.071$ & $
-0.001 \pm 0.017$ & $
\phantom{-}
0.010 \pm 0.023$ & 
\\
& $S_9
$ & $
-0.024 \pm 0.060$ & $
\phantom{-}
0.002 \pm 0.012$ & $
\phantom{-}
0.006 \pm 0.015$ & 
\\
\cline{2-6}
&&&&&\\[-2mm]
& $ P_5'
$ & $
-0.505 \pm 0.124$ & $
-0.616 \pm 0.077$ & $
-0.630 \pm 0.110$ & 0.45
\\[.5mm]
\hline
\hline&&&&&\\[-2mm]
 $ [0.1,2] $ & 
\multirow{3}{*}{\normalsize $ {\rm BR } \cdot 10^7 $}
 & $ 0.58 \pm 0.09$ & $
0.69 \pm 0.04$ & $
0.71 \pm 0.04$ & 0.19
\\
 $ [2,4.3] $ & 
 & $ 0.29 \pm 0.05$ & $
0.34 \pm 0.02$ & $
0.36 \pm 0.03$ & 0.23
\\
 $ [4.3,8.68] $ & 
 & $ 0.47 \pm 0.07$ & $
0.44 \pm 0.04$ & $
0.43 \pm 0.04$ & 0.62
\\
\hline

&&&&&\\[-2mm]
& { \normalsize $ {\rm BR }_{B \to K^* \gamma} \cdot 10^5 $}
& $ 4.33 \pm 0.15$ & $
4.32 \pm 0.14$ & $
4.30 \pm 0.48$ & 0.95
\\[1mm]
\hline
\end{tabular}
\caption{\textit{Experimental results, results from the full fit,
    predictions and $p$-values for $B \to K^* \mu^+ \mu^-$ BR's and angular
    observables obtained assuming vanishing $h_\lambda^{(2)}$,
    i.e.  hadronic corrections fully equivalent to a shift in
    $C_{7,9}$. The predictions for the  BR's (angular observables) are obtained removing the 
    corresponding observable (the experimental information in one bin at a time) from the fit. 
    We also report the results for BR$(B \to K^* \gamma)$
    (including the experimental value from refs.~\cite{Coan:1999kh,Nakao:2004th,Aubert:2009ak,Agashe:2014kda})
    and for the optimized observable $P_5^\prime$. The latter is however
    not explicitly used in the fit as a constraint, since it is not
    independent of $F_L$ and $S_5$. }}
\label{tab:mumunoq4}
\end{table}

\FloatBarrier

\FloatBarrier

\begin{table}[!htbp]
\centering
\begin{tabular}{|c|c|c|c|c|}
\hline
&&&&\\[-4mm]
\textbf{Observable} & \textbf{measurement} & \textbf{full fit} & \textbf{prediction} & \textbf{p-value} \\[1mm]
\hline
$ {P_1}$
 & $
-0.23 \pm 0.24$ & $
-0.040 \pm 0.07$ & $
-0.03 \pm 0.07$ & 0.42
\\
$ {P_2}$
 & $
\phantom{-}0.05 \pm 0.09$ & $
-0.040 \pm 0.00$ & $
-0.040 \pm 0.00$ & 0.32
\\
$ {P_3}$
 & $
-0.07 \pm 0.11$ & $
\phantom{-}0.02 \pm 0.03$ & $
\phantom{-}0.03 \pm 0.04$ & 0.39
\\
$ {F_L}$
 & $
\phantom{-}0.16 \pm 0.08$ & $
\phantom{-}0.170 \pm 0.04$ & $
\phantom{-}0.18 \pm 0.05$ & 0.82
\\
$ {\rm BR }\cdot 10^7 $
 & $\phantom{-} 3.1 \pm 1.0$ & $
\phantom{-}1.4 \pm 0.1$ & $
\phantom{-}1.4 \pm 0.1$ & 0.06
\\
\hline
\end{tabular}
\caption{\textit{Experimental results (with symmetrized errors), results from the full fit,
    predictions and $p$-values for $B \to K^* e^+ e^-$ BR and angular
    observables obtained without using the numerical information from ref.~\cite{Khodjamirian:2010vf}. The predictions are obtained removing the corresponding
    observable from the fit.}}
\label{tab:eenoK}
\end{table}

\begin{table}[!htbp]
\centering
\begin{tabular}{|c|c|c|c|c|}
\hline
&&&&\\[-4mm]
\textbf{Observable} & \textbf{measurement} & \textbf{full fit} & \textbf{prediction} & \textbf{p-value} \\[1mm]
\hline
$ {P_1}$
 & $
-0.23 \pm 0.24$ & $
\phantom{-}0.01 \pm 0.01$ & $
\phantom{-}0.01 \pm 0.01$ & 0.32
\\
$ {P_2}$
 & $
\phantom{-}0.05 \pm 0.09$ & $
-0.040 \pm 0.00$ & $
-0.040 \pm 0.00$ & 0.32
\\
$ {P_3}$
 & $
-0.07 \pm 0.11$ & $
\phantom{-}0.00 \pm 0.01$ & $
\phantom{-}0.00 \pm 0.01$ & 0.53
\\
$ {F_L}$
 & $
\phantom{-}0.16 \pm 0.08$ & $
\phantom{-}0.18 \pm 0.04$ & $
\phantom{-}0.20 \pm 0.060$ & 0.66
\\
$ {\rm BR }\cdot 10^7 $
 & $ \phantom{-}3.1 \pm 1.0$ & $
\phantom{-}1.4 \pm 0.1$ & $
\phantom{-}1.4 \pm 0.1$ & 0.06
\\
\hline
\end{tabular}
\caption{\textit{Experimental results (with symmetrized errors), results from the full fit,
    predictions and $p$-values for $B \to K^* e^+ e^-$ BR and angular
    observables obtained using the phenomenological model from
  ref.~\cite{Khodjamirian:2010vf}. The predictions are obtained removing the corresponding
    observable from the fit.}}
\label{tab:eeallK}
\end{table}

\begin{table}[!htbp]
\centering
\begin{tabular}{|c|c|c|c|c|}
\hline
&&&&\\[-4mm]
\textbf{Observable} & \textbf{measurement} & \textbf{full fit} & \textbf{prediction} & \textbf{p-value} \\[1mm]
\hline
$ {P_1}$
 & $
-0.23 \pm 0.24$ & $
\phantom{-}0.00 \pm 0.02$ & $
\phantom{-}0.01 \pm 0.02$ & 0.32
\\
$ {P_2}$
 & $
\phantom{-} 0.05 \pm 0.09$ & $
-0.05 \pm 0.00$ & $
-0.05 \pm 0.00$ & 0.27
\\
$ {P_3}$
 & $
-0.07 \pm 0.11$ & $
\phantom{-}0.00 \pm 0.01$ & $
\phantom{-}0.00 \pm 0.01$ & 0.53
\\
$ {F_L}$
 & $
\phantom{-}0.16 \pm 0.08$ & $
\phantom{-}0.170 \pm 0.04$ & $
\phantom{-}0.170 \pm 0.05$ & 0.91
\\
$ {\rm BR }\cdot 10^7 $
 & $ \phantom{-}3.1 \pm 1.0$ & $
\phantom{-}1.4 \pm 0.1$ & $
\phantom{-}1.4 \pm 0.1$ & 0.06
\\
\hline
\end{tabular}
\caption{\textit{Experimental results (with symmetrized errors), results from the full fit,
    predictions and $p$-values for $B \to K^* e^+ e^-$ BR and angular
    observables obtained assuming vanishing $h_\lambda^{(2)}$,
i.e.  hadronic corrections fully equivalent to a shift in
$C_{7,9}$. The predictions are obtained removing the corresponding
    observable from the fit.}}
\label{tab:eenoq4}
\end{table}

\bibliographystyle{JHEP}
\bibliography{DB1}

\end{document}